\documentclass[a4paper,12pt]{article}

\usepackage{amssymb,amsmath,mathbbol,mathrsfs}
\usepackage[usenames,dvipsnames]{color}
\usepackage{hyperref}
\usepackage{xcolor}
\usepackage{stmaryrd}
\usepackage{authblk}
\usepackage{framed}
\usepackage{empheq}
\usepackage{slashed}
\usepackage{hyphenat}
\usepackage{cite}

\usepackage{marginnote}    

\usepackage[left=.8in,right=.8in,top=.8in,bottom=.8in]{geometry}                  % For good copy.

\usepackage{sectsty}
\allsectionsfont{\sffamily\mdseries\upshape} % (See the fntguide.pdf for font help)
\usepackage{tocloft}



\numberwithin{equation}{section}
5

\hypersetup{
	colorlinks=true,         
	linkcolor=MidnightBlue,          
	citecolor=BrickRed,        
	urlcolor=MidnightBlue            
}

\newcommand\mathC{\mkern1mu\raise2.2pt\hbox{$\scriptscriptstyle|$}
        {\mkern-7mu\rm C}}              %%% The complex  numbers
\newcommand{\mathR}{{\rm I\! R}}         % The real numbers

\newcommand{\be}{\begin{equation}}
\newcommand{\ee}{\end{equation}}

\newcommand{\cint}{{\int\kern-.87em{<}}}
\newcommand{\sint}{{\int\kern-.75em{\sim}}}
\newcommand{\fint}{{\int\kern-1.00em{\int}}}

\let\oldmarginpar\marginpar
\renewcommand\marginpar[1]{\oldmarginpar{\color{red}\raggedright\footnotesize #1}}

\date{July 23, 2020}
\title{Functionalism as a Species of Reduction}
\author{Jeremy Butterfield and Henrique Gomes%\footnote{\href{mailto:gomes.ha@gmail.com}{gomes.ha@gmail.com}} \\\it Perimeter Institute for Theoretical Physics\\ \it 31 Caroline Street, ON, N2L 2Y5, Canada} 
}

\begin{document}

\maketitle

\begin{center}
Submitted to {\em Current Debates in Philosophy of Science: In honor of Roberto Torretti}, edited by C. Soto. 
\end{center} 

\noindent Abstract: This is the first of four papers prompted by a recent literature about a doctrine dubbed {\em spacetime functionalism}. This paper gives our general framework for discussing functionalism. Following Lewis, we take it as a species of {\em reduction}. We start by expounding reduction in a broadly Nagelian sense. Then we argue that Lewis' functionalism is an {\em improvement} on Nagelian reduction. 

This paper thereby sets the scene for the other papers, which will apply our framework to theories of space and time. (So those papers address the space and time literature: both recent and older, and physical as well as philosophical literature. But the four papers can be read independently.)

Overall, we come to praise  spacetime functionalism, not to bury it. But we criticize the recent  philosophical  literature for failing to stress:\\
\indent \indent (i) functionalism's being a species of reduction (in particular: reduction of chrono-geometry to the physics of matter and radiation); \\
\indent \indent (ii) functionalism's  idea, not just of specifying a concept by its functional role, but of specifying several concepts {\em simultaneously} by their roles; \\
\indent \indent (iii) functionalism's providing bridge laws that are mandatory, not optional: they are statements of identity (or co-extension) that are conclusions of a deductive argument, rather than contingent guesses  or verbal stipulations; and once we infer them, we have a reduction in a Nagelian sense.  

On the other hand, some of the older philosophical  literature, and the mathematical physics literature, {\em is} faithful to these ideas (i) to (iii)---as are Torretti's writings. (But of course, the word `functionalism' is not used; and themes like simultaneous unique definition are not articulated.) Thus in various papers, falling under various research programmes, the unique definability of a chrono-geometric concept (or concepts) in terms of matter and radiation, and a corresponding bridge law and reduction, is secured by a precise theorem. Hence our desire to celebrate these results as rigorous renditions of spacetime functionalism. 

\newpage
  
\tableofcontents

\newpage

\section{Introduction}\label{intro}

This is the first of four papers prompted by a recent literature about a doctrine dubbed {\em spacetime functionalism}. This paper gives our general framework for discussing functionalism. Following Lewis, we take it as a species of reduction.  In this, our views are close to the `Canberra Plan’.\footnote{\label{Canberra}{This programme (surveyed in Braddon-Mitchell and Nola 2009) is familiar in metaphysics, philosophy of mind and ethics; but unfortunately little-known in philosophy of science. But our views are only `close to Canberra’: in particular, we will not require that the functional role of a concept, extracted from the given theory, gives a conceptual analysis of it.}}

We first expound reduction in the broadly Nagelian sense of deduction enabled by extra premises, usually called `bridge laws' (Sections \ref{homered} and \ref{rednobstacle}). We start with the general idea of reduction and problems it faces (Sections \ref{problegi} and \ref{pleniscarce}). Then we specialize to Nagelian reduction (Section \ref{rednobstacle}). Then we describe how Lewis' functionalism is a cousin of Nagelian reduction: and, we  argue, an {\em improvement} on it (Sections \ref{homefunc} and \ref{DKL4}). Finally,  we discuss connections with Torretti's writings especially about reduction and physical geometry (Section \ref{roberto}). Section \ref{concl} concludes.
 
This paper thereby sets the scene for the other papers, which will apply our framework to theories of space and time. So those papers address the literature about space and time---both recent and older, and physical as well as philosophical; (but they can be read independently).  In the second paper, we present four older examples, drawn from geometry and dynamics (including special relativity), that give attractively precise examples of spacetime functionalism---in our reductive and Lewisian sense of `functionalism'. Namely: results by Lie (building on Helmholtz), by  Malament (building on Robb), by Mundy, and  by Barbour and Bertotti. So these examples span some hundred years, from the 1880s to the 1980s. Happily for us, with our wish to honour Roberto Torretti: he has written in detail about the first two of these examples, and our verdicts about them mesh with his discussions.  In Section \ref{roberto} below, we will get a  glimpse of this meshing.

The  third  and fourth papers address other more recent  literature. The third paper discusses some recent philosophical literature (by authors such as Belot, Callender and Knox). The fourth discusses some physics literature, about the foundations of general relativity and kindred theories. (There, our first example will be a classic 1976 paper by Hojman et al. which, unfortunately, the philosophical literature has ignored---but which, happily for us, Torretti commends.)

Our overall position is that we come to praise  spacetime functionalism, not to bury it.
But we criticise the recent philosophical literature: criticisms which, it will be clear, happily do {\em not} apply to Torretti's views. 

In short: our criticism is that although this literature picks up on the idea of {\em functional role},  it fails to stress three significant ways that functionalism, especially in Lewis’ hands, develops that idea. Namely:\\
\indent \indent (i) functionalism's being a species of reduction (in this context: reduction of chrono-geometry to the physics of matter and radiation); \\
\indent \indent (ii)  functionalism's  idea, not just of specifying a concept by its functional role, but of specifying several concepts {\em simultaneously} by their roles; \\
\indent \indent (iii) functionalism's providing bridge laws that are mandatory, not optional: they are statements of identity (or co-extension) that are conclusions of a deductive argument, rather than contingent guesses  or verbal stipulations; and once we infer them, we have a reduction in a Nagelian sense; i.e. the reduction is a deduction.   

These lacunae are a missed opportunity, in two ways.

First, it turns out that some of the older philosophy of physics literature, and the mathematical physics literature, {\em is} faithful to the ideas (i) to (iii)---as are Torretti's writings. (But of course, the word `functionalism' is not used; and themes like simultaneous unique definition are not articulated.) Thus in various papers, falling under various research programmes, the unique definability of a chrono-geometric concept (or concepts) in terms of matter and radiation, and a corresponding bridge law and reduction, is secured by a precise  theorem. Witness the four older examples in our second paper, and the physics examples in our fourth paper. In short: the recent literature has missed some {\em golden oldies} that illustrate {\em real} functionalism, and not merely the idea of functional role.

The second missed opportunity concerns the wider literature in philosophy of science about reduction. We believe it has missed the power and plausibility of the ideas (i) to (iii). The reason is that the literature works predominantly with a {\em binary} contrast between ``reduced'' and ``reducing'': either of vocabularies or of theories, or both. (Of course, this contrast derives in part from the traditional positivist project of reducing theory to observation, as a matter of defining theoretical terms and then deducing theoretical claims.) On the other hand, functionalism especially in Lewis' hands---and more generally, the Canberra Plan---works with a {\em ternary} contrast. It is this ternary contrast that yields (i) to (iii). 

For there is, first, within a {\em single} theory (the `higher-level' theory), a binary division of its terms, with each of the terms of the first sort having a functional definition—i.e. a unique specification---using all the terms of the second sort: idea (ii) above.  Here, `functional definition' will mean `specification' (i.e. `uniquely picking out'), but need not give the term's conceptual analysis or meaning.

But there is also a {\em second} (`lower-level') theory that specifies each {\em definiendum}---the unique {\em realizer} or {\em occupant} of each functional role---{\em independently} of the first theory. Here,  `independently' means using terms (or concepts or properties) that are not in the first theory. So a single entity (extension) is picked out in two independent ways: \\
\indent \indent (a) as the unique occupant of a functional role extracted from the first theory, and \\
\indent \indent (b) as specified by the second theory. \\
This gives a statement of identity (coextension) that is a {\em derived bridge law}. It is mandatory, not optional, because it is deduced from the two theories. Besides, one can show that the collection of such statements gives a deduction of the {\em first} theory. The reason, in short, is that the terms in the bridge laws implicitly contain the rich information of the functional roles. Thus we obtain idea (iii) above.  

So we stress that here, one must distinguish three items: \\
\indent \indent (a) the specification extracted from the first theory (using a binary division of its terms); from \\
\indent \indent (b) the specification using the second theory (so invoking a ternary contrast); and from \\
\indent \indent (c) the derived bridge law, that comes from combining (a) and (b).

Thus this paper is mostly about the second missed opportunity. We want to expound and defend what we will label as {\em (Lewisian) functionalist reduction}. It is another {\em golden oldie} that much of the  literature  in philosophy of science has sadly missed. Though simple, we will argue that it is not too simple. It is powerful and flexible enough to address issues that beset reduction: such as multiple realizability, meaning variance, and most real-life reductions being, in one or more ways, partial. 

But even if you end up rejecting our advertisement as regards reduction in general, we in any case commend functionalist reduction for our several examples of the physics of matter and radiation contributing to determining, or explaining, some chrono-geometric concepts.  As we will see in our other papers, it fits those examples very well.  

In the rest of this Introduction, we will add some details about: functionalism in general (Section \ref{111}); functionalism about spacetime (Section \ref{112}); and the connection to Torretti (Section \ref{113}).

\subsection{Introducing functionalist reduction}\label{111}
In philosophy, the word `functionalism' is mostly associated  with proposals in the philosophy of mind by Armstrong, Lewis and Putnam in the mid-1960s. It will be clearest to introduce functionalism, and Lewisian functionalist reduction, via this example (Section \ref{1111}); and then make three general comments (Section \ref{1112}).

\subsubsection{... in the philosophy of mind}\label{1111}
  These authors' initial idea  was that each mental state, or concept, could be {\em  uniquely specified} by its characteristic  pattern (`web') of relations to various other states, or concepts, both mental and physical. Such  a characteristic pattern, dubbed {\em functional role}, was spelt out in terms of nomic and-or causal relations: relations that some appropriate body of knowledge or belief (often called a `theory'), whether everyday or scientific, claims to hold between the  various states or concepts. 
  
  This idea of functional role has indeed been taken up by recent articles on spacetime functionalism (cf. Sec \ref{112}). They focus on the idea that spacetime, or specific spatiotemporal concepts such as the metric, or connection, or inertial frame, are specified by such a  pattern.\footnote{{\label{fnstateconcept}}
We will mostly use both `state' and `concept' throughout, although: (i) they are used more in philosophy of mind and metaphysics than in philosophy of physics (whose main analogues are `physical state'  and `quantity' i.e. `physical magnitude'); and  (ii) they are terms of art, without widely agreed criteria of individuation. But  we shall not need to be precise about such criteria. Indeed, almost all our points are unaffected by what exactly these criteria are; and we will signal the exceptions. 

It suffices for us to note that: (a) for most authors, a state is a localized fact or state of affairs; (b) some authors distinguish type-states, e.g. being in pain, or being in love, or seeing yellow in the top-left of the visual field, from token-states e.g. Fred's being in pain at $t$, John's loving Mary now, and these authors often go on to individuate token-states as $n$-tuples of the objects and times  involved and the  properties or relations attributed e.g. $\langle {\rm{John, Mary,  now, ... loves ...}} \rangle$ (such $n$-tuples are often called `Russellian propositions'); (c) for most authors, a concept is the same, or pretty much the same, as a type-state. It is also the same, or pretty much the same, as a property or relation. For it is general, not localized, and is individuated more finely than by its set of instances; for example, it is individuated by a Fregean sense or Carnapian intension rather than by extension. 

We can go along with (a) to (c), both in this paper and the others. We will simply note that some of the debates,  e.g. in philosophy of mind  about whether the mental state or concept of being in pain is reducible to material states or concepts, turn on  controversies about the criteria of identity of properties---indeed a misty topic.

We can be sanguine in this way because, happily, for our examples for philosophy of physics, the state and concepts in question---the properties specified by their functional roles--- are much less vague and controversial than e.g. pain. For they are physical properties: often, familiar physical quantities. Here are examples from our second paper: being freely mobile; being simultaneous with (as a relation between events); being congruent  (as a relation between spatial intervals); and being isochronous, i.e. of equal duration (as a relation between temporal intervals).}

But there is much more to functionalism than just the idea of a state or concept having such a characteristic pattern (functional role). For after all: it is plausible that many states and concepts each have a  pattern of relations to other states or concepts that is characteristic, i.e. idiosyncratic, enough, that the state or concept can be uniquely specified by that pattern. So in this minimal sense,  almost every concept is functional. Cf. footnote \ref{fnstateconcept}.

To say what more there is to functionalism, we  begin by recalling that  functionalism, like its forebears, viz. logical behaviourism and mind-brain identity theory, accepted a basic contrast between `mind’ and `body’; or better, between `mental’ and `material’  states and concepts. The general theme, of both these forebears and of functionalism, was of course that mental discourse is {\em problematic}; and is to be vindicated or legitimized by showing how it is {\em suitably related to unproblematic} material discourse. Besides, the `suitable relation' was, in effect, `being a part of'.  That is, mental discourse was to be {\em reduced} to material discourse. 

But functionalism was distinctive in having this reduction proceed in two stages: stages that reflect the {\em ternary} contrast we stressed in this Section's preamble. In other words, functionalism makes two proposals. Or as we advocates would say: it has two insights about the reduction of mental to material. 

The first is about the binary division, mental vs. material, among the terms used in everyday knowledge and belief about the mental and material---in what came to be called `folk psychology'. The proposal is that this body of knowledge and belief is sufficiently rich that each of the many mental states and concepts can be uniquely specified by its functional role. And this can be done {\em simultaneously} for all of them, even though a mental state's or concept's functional role invariably mentions, not just material states and concepts, but also other mental ones.  A standard pedagogic example is that the functional role of the mental state, belief that it is raining, cannot mention only bodily and-or behavioural states and concepts, such as being disposed to pick up an umbrella when going outdoors. For if someone believes it is raining, they have that disposition {\em only if} they also desire to stay dry---a mental state. That the functional role of a mental state needs to mention other mental states  apparently implies a threat of a logical circle. But functionalism shows that the threat can be overcome. We will see later how  Lewis consistently extracts  simultaneous unique specifications of many terms from a theory---here, folk psychology.

The second proposal is that each mental state, each unique occupant of a corresponding functional role, is also specified by {\em another} body of doctrine---a second theory---that we accept, independently of (perhaps after) our acceptance of the first, i.e. folk psychology. For Armstrong, Lewis and Putnam, this second theory was of course neurophysiology. 

Then the two proposals taken together yield deductions of bridge laws as statements of identity (or co-extension of predicates). By specifying a mental state in two independent ways, we can infer such a statement.  These bridge laws, taken together, then yield a reduction, i.e. a deduction of the first theory from the second: details in Section \ref{DKL4}. 

This is vividly illustrated by the time-honoured example of inferring that the mental state of pain {\em is} a neurophysiological state. As an inference, it is trivially simple. For it is merely a case of the transitivity of identity.  But since this inference will be a {\em leitmotif} in what follows, and will have striking parallels in our spacetime examples in other papers, it is worth rehearsing it at the outset of our discussion, albeit briefly. The classic, crystal-clear expositions are Lewis (1966; 1972, Introduction and Section 3).   

Thus consider:---\\
\indent \indent (i) Accepting the characterization of pain given by folk psychology, we endorse the premise: pain is the unique occupant of so-and-so role. \\
\indent \indent  (ii) Accepting neurophysiology, we endorse the premise: C-fibre firing  is the unique occupant of so-and-so role. (Here, `C-fibre firing' is the ignorant philosopher’s catch-all for a technical, maybe long, specification in the language of neurophysiology.) \\
\indent \indent  (iii) So by the transitivity of identity, we must infer: pain {\em is} C-fibre firing.

Thus we see, in summary form, how  the functionalism of Armstrong, Lewis and Putnam clarified and unified its forebears, logical behaviourism and mind-brain identity theory. In terms of the inference just rehearsed: \\
\indent \indent  (a) logical behaviourism---roughly speaking: conceptual analysis of mental terms as they occur in folk psychology---gives the warrant for premise (i); \\
\indent \indent  (b) neurophysiology---i.e. contingent empirical discoveries---gives the warrant for premise (ii); \\
\indent \indent  (c) so we infer (iii), a derived so-called ``bridge law'': a statement of identity between pain as specified by folk psychology and pain as specified by neurophysiology.\\
This statement of identity is an instance of mind-brain identity theory. But the identity is not merely recommended as a hypothesis that is attractive because ontologically parsimonious (as earlier advocates of mind-brain identity theory, such as Place and Smart, had said). It is the conclusion of a valid argument from accepted premises: viz. premises that describe the unique occupant of a functional role in two ways. 

Note also that we must of course expect this statement of identity, and other such identities of mental and neural states, to be relative to a {\em kind}. That is: which neural state is identified with a given mental state of course varies between organisms. Feeling pain might well be a very different state of a mollusc brain than of a human brain. Besides, the kind relative to which such a so-called ``type-type'' identity holds might well be narrower than a species (cf. Lewis 1969, p. 25). So this kind of mind-brain identity theory has no conflict with the multiple realizability of mental states.

So overall, we have a reduction of `mental' to `material'. But thanks to the inference's premise (ii), `material' now involves more than just behavioural and non-technical concepts, as in the  example of being disposed to pick up an umbrella when going outdoors. Agreed, each mental state or concept has a unique functional role in terms of such behavioural and non-technical concepts. That is what we learn from behaviourism, i.e. from conceptually analysing folk psychology. But each mental state or concept is {\em also} uniquely specified in the language of neurophysiology. {\em That} is what we learn from contingent empirical enquiry.  In short, as we said above: there is a {\em ternary}, not binary, contrast of vocabulary: `mental', `behavioural/non-technical' and `neurophysiological'. 

So much by way of using the familiar example from the philosophy of mind to sketch functionalist reduction. We now complete this sketch with three further comments, not tied to this example. 

\subsubsection{... in general}\label{1112}
These comments are about:\\
\indent \indent (1) using the word `definition', instead of `specification'; \\
\indent \indent (2) how functionalist reduction avoids problems that beset other kinds of reduction; \\
\indent \indent (3) the state of the literature about functionalism and reduction.\\

\noindent  (1): We have so far mostly said `specified' and `specification'. We will also say, since the literature often does:  `defined' and `definition'. Beware: these words of course connote both: (a) arbitrary verbal stipulation (cf. Humpty Dumpty in {\em Through the Looking Glass}: `When I use a word, it means just what I choose it to mean---neither more nor less'); and (b) staying faithful to a given meaning of the word (cf. lexicography). But as we will stress: these connotations are often misleading for our cases in philosophy of mind, of science in general, and of physics. The `definition' of a state or concept by its functional role is very often neither a stipulation, nor faithful to some given meaning.

Agreed, logical behaviourism did aim to define (the words for) mental states and concepts in behavioural terms, faithfully to the words' meanings. Before functionalism (and especially Lewis' work), this endeavour had long been thought to be beset by problems of logical circularity. Again, a standard pedagogic example is that  the logical behaviourist wants to define belief that it is raining along the lines of being disposed to pick up an umbrella if going outdoors. But, as we said, this seems to stumble on the apparent need to also assume that the agent desires to stay dry---a mental state. And since the logical behaviourist's natural strategy for defining  desire to stay dry is, again, along the lines of being disposed to pick up an umbrella if going outdoors, and this strategy apparently needs to also assume that the agent   believes it is raining ... clearly, a vicious logical circle of definition looms. It is this sort of logical circle that Lewis' idea of {\em  simultaneous unique definition} avoids at a stroke. 

Besides, as we will explain: this proposal for avoiding circularity does not at all depend on taking a definition to be (a) stipulative or (b) faithful to a given meaning. So one can equally well speak of {\em  simultaneous unique specification}. \\

\noindent (2): In Sections \ref{homered} and \ref{rednobstacle}, we will prepare for our advocacy of functionalist reduction (Sections \ref{homefunc} and \ref{DKL4}) by discussing the enterprise of reduction: first in general (Section \ref{homered}) and then {\em a la} Nagel (Section \ref{rednobstacle}). The general idea of reduction will be that a problematic discourse (or theory) is shown to be part of an unproblematic discourse (or theory), by: (i) suitably specifying or defining, in terms of the unproblematic discourse, the words (or concepts or properties) of the problematic discourse; and thereby (ii) recovering the problematic discourse's claims, usually by deduction. Note that this idea is much less specific than the functionalist reduction we advocate. For (i) is less specific than the  idea of unique specification by a functional role; and the deduction in (ii) is not underwritten by {\em mandatory} i.e. inferred bridge laws.

 In this enterprise there are three kinds of problem, or at least objection,  that a programme of reduction, is liable to face. We spell them out in Section \ref{pleniscarce}. But since in later Sections, and in our other papers, they will  be a template for discussing and assessing Nagelian reductions and functionalist reductions, it will help set the stage if we announce them here, as follows.\\
\indent \indent (1): {\em Faithlessness}: The proposed reduction is faithless to the original problematic discourse. That is, using just the concepts of the unproblematic discourse, one cannot faithfully specify some concept of the problematic discourse.\\
\indent \indent  (2): {\em  Plenitude}: The unproblematic discourse provides many specifications of some concept(s) of the problematic discourse: many specifications that are equally good, so that one cannot choose between them in a non-arbitrary way---so that they are also equally bad. In effect, the objection is: `existence of a specification is all too easy, but uniqueness is unobtainable'.\\
\indent \indent  (3): {\em  Scarcity}: This is the opposite of Plenitude: the unproblematic discourse cannot provide even one good specification of some concept(s) of the problematic discourse. In effect, the objection is: `uniqueness  of the specification is all too easy, but existence is unobtainable'.

These labels will recur. In this paper and its companions, we will see both: \\
\indent \indent (a) examples of these problems, for both philosophical and physical programmes of reduction; and \\
\indent \indent  (b) more positively: how functionalist reduction claims the existence and uniqueness of a concept's specification, thus avoiding the objections of Plenitude and Scarcity. \\
Besides, as we announced in this Section's preamble: in our other papers' examples of (b), the existence and uniqueness claim is indeed a {\em theorem}.

Also, beware on two counts. (i): We have for simplicity stated these problems, (1) to (3), as if they were mutually exclusive; but we shall see that in fact they mingle with each other. (ii): We have stated these problems as about {\em specifications}. And `specification’ is a word that, we admit, tends to be ambiguous between the role, and the realizer i.e. occupant of the role. (`Definition’, on the other hand, lacks this ambiguity: one naturally hears it as referring only to the role, or the words that express the role.) But this ambiguity is a convenience, indeed an advantage, for us. For reduction and functionalism will face objections of both sorts: about many, or no, roles; and about many, or no, realizers. So we have here used `specification’ for simplicity, i.e. so as to concisely cover the various cases.  \\

\noindent  (3): We believe the literature  in philosophy of science about functionalism and reduction has tended to forget the way they can be combined, as above, in functionalist reduction. Recall the preamble of this Section, where we lamented what we dubbed a `second missed opportunity’ and the tendency of philosophers of science to ignore the Canberra Plan. The questions arise: (i) whether our contention is right; and (ii) if so, why this has happened. 

As to the first question (i):--- We admit that we have not attempted a systematic survey. But we note that a recent comprehensive survey of `scientific reduction' (van Riel and van Gulick 2019)---and more important, the great majority of the literature it cites---works with what we called the {\em binary contrast} between reduced and reducing: not the {\em ternary contrast} of functionalist reduction, that we sketched above  (and that matches what has been called the {\em two steps} of the Canberra Plan; cf. Braddon-Mitchell and Nola (2009: 7-9, 185-190, 267-269)). 

Much of the literature also sees functionalism as offering a {\em non}-reductive relation between the higher and lower levels (or theories). The idea here is that the fact that the words (or concepts or properties) at the higher level/theory are functional, i.e. are each specified by a functional role, makes the higher level/theory suitably dependent on, or rooted in, the lower level/theory---but {\em without} being reduced to it. This is broadly similar to, indeed sometimes combined with, another widespread strategy for securing a {\em non}-reductive relation between higher and lower levels (or theories): namely to say that the higher level supervenes on, is determined by, the lower.\footnote{Beware: this strategy stumbles on Beth's theorem. Namely: for first-order languages, supervenience i.e. determination is, surprisingly, equivalent to explicit definability, and so to reduction. In philosophy of science and mind, this was first pointed out by Hellman and Thompson (1975), and later stressed by various authors, e.g. Butterfield (2011: Section 5.1, pp. 948-951). Cf. Dewar (2019) and footnotes \ref {rigorlogic} and \ref {BethRaum}.}     

As we have said: following Lewis, we will reject the claim of non-reduction. Besides, the sense of reduction given by functionalist reduction will be a deduction of the reduced theory. It will be like Nagelian reduction in using bridge laws---but with the difference that the bridge laws are themselves deduced! The details will be in Section \ref{DKL4}.

As to the second question (ii):--- Of course, one can only speculate. But we surmise that the facts just mentioned---i.e. the binary contrast; and functionalism being, like supervenience, widely thought to articulate non-reductionist relations between levels---have been influential. We suspect also that there is a tendency to think that functionalism is a doctrine {\em confined to}, or at least plausible only in, the philosophy of mind. A case in point is that van Riel and van Gulick, for all their merits, discuss functionalism only within their Sections on philosophy of mind (viz.  Sections 3.3, 3.4, 4.5). Besides, they discuss only the idea of functional role: not the crucial consequential ideas of unique simultaneous specification, and of---again, the need for a ternary contrast!--- derived bridge laws. Indeed, of Lewis’ articles on the topic, they cite two, both on the philosophy of mind (viz. his 1969 (a brief discussion of multiple realisability), and 1972). But they do not cite his definitive 1970: which discusses theories in general (it does not mention philosophy of mind), and which presents with full rigour both: (i) how to extract from a theory unique simultaneous specifications of several, even many, terms, and (ii) how to derive bridge laws, and so deduce the reduced theory.\footnote{\label{Lewis94}{Our criticism is not meant to single out van Riel and van Gulick (2019): it has many merits. Our point is just that its emphases are part of a pattern: witness the fact that Lewis (1970), despite being definitive and general, has fewer citations than his (1972).

For both our first and second questions, we also recommend Lewis' `Reduction of mind' (1994). Its second half (pp. 421f.) is specialist: Lewis rebuts fashionable views, e.g. the `language of thought' hypothesis, about Brentano's problem of intentionality, i.e. the question what determines the contents of mental states like beliefs and desires. But the first half summarizes---and updates---the position we have reported in Sections \ref{1111} and \ref{1112}. We especially commend: Lewis' advocacy of supervenience as reductive, his answer to the objection from {\em qualia}, his use of two-dimensional semantics to analyse the necessary {\em a posteriori}, and his answer to the idea that the term `pain' is a rigid designator (pp. 412-421). 

What Lewis says about the last topic (pp. 420-421) is also relevant to our point that, unfortunately, `functionalism' is often understood as non-reductive. Thus Lewis writes: `It is unfortunate that this superficial question [in effect, a question about the best nomenclature for the occupant or realizer of a role]  has sometimes been taken to mark the boundary of `functionalism’. Sometimes so and sometimes not - and that's why I have no idea whether I am a functionalist' (p. 421). We will briefly return to this in Section \ref{DKLsentences}.      

We also suspect, perhaps cheekily, that people have not sufficiently taken up functionalist reduction simply because its exposition in Lewis (1970) occurs only in the last two Sections (p. 441 et seq.). For most of the paper is taken up with a rigorous logical (though beautifully clear!) exposition  of simultaneous unique definition within the first, i.e. reduced, theory. In short: we suspect the last two Sections have been ignored.}}\\

So much by way of introducing how functionalist reduction  has much more to it than just the idea of functional roles. Of course, for our papers' purposes, we do not need to endorse the functionalist philosophy of mind that served in Section \ref{1111} as our main illustration. (But as it happens, we do endorse it.) All we really need is that functionalist reduction fits cases, indeed many cases, in spacetime theories; and so it gives a good sense of `spacetime functionalism'. 

But in this paper  we also want to urge the more general thesis that in the general enterprise of reduction, it is a powerful and plausible model, well able to address issues such as multiple realizability, meaning variance, reductions being partial, and the three problems of Faithlessness etc. listed in (2) above. We will take up this thesis from Section \ref{homered} onwards. But first, we set the stage  by sketching: how functionalist reduction  applies to space and time (Section \ref{112}), and how this relates to the work of Torretti (Section \ref{113}).

\subsection{Functionalism about spacetime}\label{112}
As we said in this Section's preamble and Section \ref{111}, our over-arching message is that functionalist reduction applies well to space and time. Both its  main claims (i.e. (i) to (iii) of this Section's preamble), and the various issues and possible objections that these claims raise (e.g. those we  labelled Faithlessness, Plenitude and Scarcity), apply well to several much-studied cases of theories of space and time. But since, unfortunately, the recent literature on spacetime functionalism has  not articulated or assessed these applications, we aim do so---in our other papers.

But to complete this announcement of our general aims, we should say here what is the analogue, for spacetime, of the contrast between `mental’ and `material’ that, as we saw, functionalism in the philosophy of mind assumed at the outset, so as to make its claims.

That is; we should say what is the problematic/unproblematic contrast, with spacetime or spatiotemporal concepts allocated to the problematic side. The recent literature on spacetime functionalism has considered two main choices. 

First, there is the contrast between spacetime and the physics of matter and radiation: (intuitively, the problematic `void’ or `receptacle’, vs. the unproblematic `stuff’). Using this contrast, spacetime functionalism is evidently related to relationism about space and time; and to what has recently been called the `dynamical approach' to chrono-geometry. Thus the focus is on how the physics of matter and radiation contributes to determining, or perhaps even determines, or even explains, chrono-geometry. It is {\em this} contrast that we will be concerned with: our overall point being, again, that the older literature already gave us examples of such a functionalist understanding of chrono-geometry in terms of matter and radiation.\footnote{The `dynamical approach' is especially associated with Harvey Brown (especially his 2006); indeed, it was a member of his school, Eleanor Knox, who first coined the phrase `spacetime functionalism’ as a label for a position she favoured, and saw as close to Brown’s.

Recalling from this Section's preamble, and Section \ref{1111}, that functionalist reduction proceeds in two stages, and uses a ternary contrast, not a binary one, you will ask: what will be our  ternary contrast? That is: what will be our `second theory', accepted independently of (or after) the first one? In short: our answer, in subsequent papers, will vary from case to case; but the text's contrast between spacetime and matter-and-radiation will be the unifying theme.}

Second, there is the contrast between: (a) spacetime taken  to be a continuum (as it is in all our established theories), and (b) some non-continuum underpinning that spacetime is supposed to approximate (or to be a coarse-graining of, or to emerge from)---an underpinning proposed by some speculative framework, usually a programme in quantum gravity. Thus some discussions of spacetime functionalism (especially in the philosophy of quantum gravity literature) invoke this contrast.\footnote{{\label{qgprobl}}{Agreed: here, our labels `problematic’ and `unproblematic’ become  strained. For in the present state of knowledge, it is spacetime, i.e. our established theories positing a spacetime continuum, that should be called `unproblematic’, and the speculative quantum gravity programme that should be called `problematic’. But labels aside, the intended analogy between the mind and spacetime cases is as clear as for the first contrast. Namely: just as mind is best understood in terms of mental concepts’ webs of relations to each other, and to material concepts, so also spacetime is best understood in terms of spatiotemporal concepts’ webs of relations to each other, and to the concepts of a postulated `non-spacetime’ theory.}} 

But we will abjure this second contrast, and stick to the first one.\footnote{Agreed: it might be worthwhile to assess some emergent-spacetime research programmes using the taxonomy of problems that we will develop for our versions of spacetime functionalism. In particular, some of these programmes’ efforts to obtain a spacetime continuum may face the problems we label as Faithlessness, Plenitude and Scarcity (cf Section \ref{pleniscarce}). Thanks to Julius Doboszewski for this point.}  That is: we will take spacetime functionalism to focus on how the physics of matter and radiation contributes to determining, or perhaps even determines or explains, chrono-geometry; and so as closely related to relationism, and the `dynamical approach' to chrono-geometry.\footnote{{\label{DKLsubstvlism}} We should mention here, since we will be advocating Lewis' account of functionalism and reduction, that he himself did not espouse spacetime functionalism in our sense. In fact, he did not work in detail on philosophy of physical geometry. But his main view was substantivalist: spacetime is an object, the mereological fusion of its regions, that have various spatiotemporal relations to each other. Besides, in his mature metaphysical system, these relations are, in his jargon, perfectly natural and external. But we will {\em not} need these doctrines in this paper, or in our others. We only need Lewis’ treatment of functionalism and reduction. }

\subsection{Connections with the work of Torretti}\label{113}
Clearly, the project of this paper and its companions puts us in ``the land of Torretti''. For much of his scholarly writing has been about the philosophy and history of geometry, especially of physical geometry and thus, after the advent of relativity theory, of chrono-geometry. For example, we will see already at the start of our second paper that two of his main books discuss in detail our first two examples of spacetime functionalism, viz. the work of Lie (building on Helmholtz) on the ``problem of space'', and the work of Malament (building on Robb), on simultaneity in special relativity. 

Nor is it just the philosophy and history of physical geometry that links our project to Torretti's work. For of course Torretti has written a lot about reduction, and inter-theoretic relations in general, not just in geometry but across all of physics; and he has advocated the semantic (also called: structural) conception of scientific theories, against the traditional syntactic conception assumed by writers like Nagel. So by the end of this paper, after we finish our advocacy of functionalist reduction, it will be clear that various projects beckon: for example, comparing functionalist reduction with Torretti’s treatments of reduction and related topics.

We will not have space for details about any of these projects. But in Section \ref{roberto}, we will discuss three: the comparison of our and Torretti's treatments of reduction; and two  topics in the philosophy and history of the axiomatic method. Let us also, here at the outset, reassure Torretti that despite our inclining more than he does to `scientific realism' and `reductionism'---witness this paper's invoking Lewis and Nagel---we will agree with him about many details. As always, it is a matter of getting beyond the slogans and `isms'. See also the reassurances at the start of Section \ref{homered}.

\section{The enterprise of reduction}\label{homered}

In Section \ref{problegi}, we first introduce reduction as a strategy for legitimizing a problematic discourse; namely by showing it to be really a part of an unproblematic discourse. 
We then take one discourse `being a part' of another as a matter of the latter giving specifications or definitions of the concepts of the problematic discourse, in such a way that the unproblematic discourse, augmented with these specifications, then implies the claims of the problematic discourse. We  illustrate this with some historically influential programmes of reduction, including from the philosophy of mathematics; and we stress the need for deduction, not mere postulation. 

In Section \ref{pleniscarce}, we introduce three sorts of problem, which we label {\em Faithlessness}, {\em Plenitude} and {\em Scarcity}, that are liable to beset reduction---whatever one's exact conception of it. These  labels are helpful for classifying the various objections that specific conceptions or examples of reduction face; as we will see in  the other  papers' examples of chrono-geometry.

Four initial disclaimers: or perhaps better, reassurances. (1): Obviously, reduction and functionalism  are large and much-contested topics; and we have not the space to fully defend the traditional accounts of them---in short: Nagel's for reduction, Lewis' for functionalism---that this paper, and its companions, will adopt. Although we will of course indicate our defence, not least by citation: it suffices for us that these accounts fit perfectly our examples in our companion papers. After all, there is no point in fighting over the words `reduction' and `functionalism'. 

(2): Note that in this paper and its companions, our favoured word is `reduction', not `reductionism'.  We will {\em not} be concerned with either of two `big-picture' {\em reductionisms}:\\
\indent \indent (a): the `unified science' picture, that reality, or science, is arranged in a sequence or hierarchy of `levels' (levels of scale, or of description): with a physical level (roughly: collection of theories), that individually or collectively reduce a chemical level or collection of theories; that individually or collectively reduce a biological level or collection of theories etc.; or \\
\indent \indent (b): the `emergent spacetime' picture, mentioned at the end of Section \ref{112}, that spacetime  is emergent from a fundamental  level  that is not spatiotemporal.\\
We stress this because: firstly, picture (a) is much discussed (criticised!), both in general philosophy of science, and in our wider culture; and secondly, picture (b) is common in quantum gravity research, and thereby in some  recent philosophical literature on `spacetime functionalism'. But we will have no need to endorse, or even assess, either of these big-picture reductionisms.\footnote{Agreed: we will start our discussion of reduction, in Section \ref{examp}, by mentioning some historically influential proposed reductions that were ambitious and philosophical, like the `unified science' picture. But these are just by way of example.  In the same vein, note that our main claims will not depend on the labels, `problematic’ and `unproblematic’. To be sure: in some cases, the discourse to be reduced is unproblematic, but a reduction remains of interest. Cf. footnote \ref {qgprobl}.} 

(3): Similarly, we will not need to endorse, or even formulate, scientific realism (though as it happens, we do endorse some modest formulations).  Agreed, we assume that the endeavour of interpreting physical theories, especially as regards how matter and radiation ``mesh with'' chrono-geometry, makes sense. But this assumption is entirely compatible with, for example, being a constructive empiricist: witness van Fraassen's discussion of what an interpretation of a physical theory is (1991, pp. 8-12). Our discussion of Torretti will also briefly return to this topic (Section \ref{peace}). 

(4): Nor will we, or the precursor spacetime functionalists we celebrate in other papers, be committed to strongly realist metaphysical views, such as Lewis'. Yes, we will endorse the objectivity of reference and of truth, as part of endorsing Lewis' functionalism with its unique definitions. And yes, this means we  face challenges like Putnam's model-theoretic argument, and its precursors like Newman's objection to Russell’s structural realism: challenges urging that by our realist lights (Russell’s structural realist lights), reference and truth are all too easy to attain. (Button (2013: Part A) is a thorough presentation.)  But there are various cogent replies to these challenges. And some of these are significantly less `gung-ho' or ardent in their realism than  Lewis’ own reply, which invokes a strong doctrine of objective similarity.\footnote{ Lewis proposes that the extent to which a property encodes objective similarity---which he dubs: how natural it is---is a feature the property has across all of modal reality, wholly independent of contingencies such as what are the laws of nature. So this is indeed `limning the true and ultimate structure of reality' (Quine 1960, p. 202). Among less gung-ho replies, Taylor's proposal, which  he calls a `vegetarian substitute' to Lewis, is to relativize similarity and associated notions to a theory (1993, especially Section IVf., p. 88f.); and van Fraassen answers Newman and Putnam from an empiricist and pragmatic perspective (2008, pp 229-235).} So one does not face a forced choice between, say, just Putnam’s views and Lewis’: there are intermediate options. We will say more about this in Section \ref{DKL1A}. But the main point is that in this and our other papers, we will not need to choose between these options. The reason for this flexibility will lie in the specificity of the scientific contexts we are concerned with. Thus consider the property that is centre-stage in our second paper’s first example: free mobility, as a property of rigid bodies. This property is sufficiently close to observation, and sufficiently rich in its relations to other properties, that we can be confident that we grasp it---that very property, {\em pace} Putnam!---according to various of these intermediate replies. And similarly for the other properties, simultaneity, congruence etc., that will figure in our other paper’s examples; cf. the end of footnote \ref{fnstateconcept}.

\subsection{The problematic, and how to legitimize it}\label{problegi}
Philosophical problems and projects often begin with a contrast between: some discourse (roughly: a set of concepts, and claims involving them) that is believed to be problematic; and another  discourse that is believed to be unproblematic. 

Some well-known examples are as follows:---\\
\indent \indent (i):  Mind and matter:  i.e. mental concepts and claims are problematic, while material (or bodily) concepts and claims are not.\\
\indent \indent (ii):  Ethics and factual description: i.e. ethical (and maybe other evaluative) concepts and claims are problematic, while factual descriptive concepts and claims are not.   \\
\indent \indent (iii): Pure mathematics and the empirical: i.e. pure mathematics is problematic, the main question being, since we seem to have no experience of mathematical objects such as numbers: how do we know mathematical truths? On the other hand, empirical concepts and claims seem unproblematic, since rooted in our experience.\\
\indent \indent (iv): The unobservable and the observable: concepts and claims about the unobservable are problematic, while  observable concepts and claims  are not.\\
Of course, in all examples  the concepts at issue in either the problematic or unproblematic discourse are only vaguely delimited; and authors vary about how to be more precise.\footnote{For our use of `concept' and `state', recall footnote \ref{fnstateconcept}. References for each example, out of countless that could be given, are: (i) Smith and Jones (1986); (ii) Harman (1977, Part I); (iii) Benacerraf (1973); (iv) van Fraassen (1980, Chap. 2). For variety, we have here chosen expository references that are {\em not} committed to a reduction. Section \ref{examp} will discuss reductions for these examples.} \\

Of course, responses to each of these contrasts vary greatly. Some reject the contrast as a mistake. For example, they diagnose the mistake as rooted in a simplistic picture of how some part of our language (or other practices) works. So what seemed problematic is, in fact, not. Some accept the contrast and conclude: `So much the worse for the problematic: though it may have meritorious uses, it is, as it stands, cognitive rubbish'. Some accept the contrast and yet conclude: `We cannot dismiss the problematic as rubbish, albeit useful rubbish; so we must revise our previous views about why it is problematic'.\footnote{\label{Moore}{Again, there are many examples of these responses. The `mathematical atheism' of Field (1980) exemplifies the second, eliminativist, response for example (iii) above. The condemnation of the `naturalistic fallacy' by Moore (1903, Chap. 2) exemplifies the third, `{\em sui generis}', response for example (ii).}} 

But we will focus on another more irenic response: that  the problematic discourse must be legitimized by appeal to the unproblematic discourse. Here of course: (a) it may only be some part of the problematic discourse  that gets legitimized; and (b) to win legitimacy, one might also use ingredients (concepts and claims) from discourses other than the  unproblematic one originally delineated. 

Again, there are various versions of this response. One version is {\em conceptual analysis}: each problematic concept is to be analysed in terms of unproblematic concepts---and so legitimized.  It is this version that is espoused by the Canberra Plan, mentioned in Section \ref{intro}; a well-known example being Jackson (1998). Of course, the given problematic concept might be vague, so that its {\em analysans} will either be correspondingly vague, or be an analysis of a `precisification' of the concept. Another version is Carnap's notion of {\em explication} (cf. Stein 1992, 280-282; Beaney 2004). This is like conceptual analysis, but more revisionary: it allows that the given concept is defective in some way, and if so, it provides a precise version aiming to rectify the defective aspect. 

Another version is Russell's idea of {\em logical construction}: (which he also  called `logical fiction'). The problematic concept is to be replaced by one that is precisely defined in terms of unproblematic ingredients, in such as way as to mimic its properties; or again, with some allowance of revision: to mimic its desirable or correct properties (1918, p. 122, p. 144; 1924, p. 160).  This version, with its word `construction’, introduces another theme we have so far been silent about. Namely: whether the unproblematic discourse is assumed to have tools, such as set theory or mereology, with which to {\em construct} entities (whether objects or properties) that it is not initially given, or thought of, as including (as being `ontologically committed to’). If so, the power of the unproblematic discourse to secure a reduction is, in general, increased. That is: provided one is willing to identify the  objects or properties at issue within the problematic discourse with such constructions---a misgiving we will return to, in Section \ref{pleniscarce} and also later. 

We will not need to choose between these  options of analysis, explication and construction (including constructing objects and properties). So we need a word to cover them all. As discussed in (1) of Section \ref{1112},  we will say {\bf define}, {\bf definition} and thus also `{\bf definiens}' and `{\bf definiendum}'. 

This usage is analogous to the practice in logic books of calling a universally quantified bi-conditional with a single predicate $F$ on the left-hand side, $(\forall x)(Fx \equiv \Phi(x))$ where $F$ does not occur in  the open sentence $\Phi(x)$ on the right, a {\bf definition} of $F$ in terms of the vocabulary in $\Phi$. This analogy has three aspects:\\
\indent \indent (i) such a bi-conditional determines the extension of $F$ in terms of the (extensions of) the vocabulary occurring in $\Phi(x)$; but also, \\
\indent \indent (ii) there is no mandatory implication  that the definition must be true, or nearly true, to any pre-existing meaning of $F$; \\
\indent \indent (iii) nor is it implied that $F$ and $\Phi$ are co-extensive in other ``worlds'', i.e. in domains other than the given one.\\
The flexibility stated by (ii) and (iii) will be important for functionalism.  It is also an advantage that the words, {\em definiens} and {\em definiendum}, are established jargon (unlike, say, {\em specificans} and {\em specificandum}). So here, $F$ is the {\em definiendum}, and $\Phi$ is the {\em definiens}. Note: these Latin tags are  often used---and we will also use them---not for the linguistic item, but for what they express or denote. So on this usage, one says, in functionalist jargon: the {\em definiens} is the role, and the {\em definiendum} is the realizer. (In general, both role and realizer are properties, states or concepts: cf. footnote \ref{fnstateconcept}.) 

  But we admit: this analogy of usage is also limited, in three ways. First: one naturally hears  `definition’  as referring to the {\em definiens}, not the {\em definiendum}; or in functionalist jargon: to the role, not the realizer. But we will sometimes want to refer to the realizer; and for this, the word `specification’ is often more natural than `definition’ (as we noted in (ii) at the end of (2) in Section \ref{1112}). Second: the word `definition’ connotes (outside a logic book!) both (a) arbitrary verbal stipulation and (b) faithfulness to a given meaning; and these connotations are stronger than for the alternative word, `specification’. (Cf. (1) in Section \ref{1112}.) Third: `definition’ does not connote---and in logic books, usually excludes---logical construction of the kinds just mentioned; but we will use `definition’ widely, as also covering constructions.  \\

So far, we have gestured at the various versions of the irenic response, by considering how to treat the problematic {\em concepts}: analysis vs. explication vs. logical construction. But there is a corresponding variety in the treatment of the problematic discourse's use of its concepts, i.e. its {\em claims}. 

For these claims, there is a pre-eminently natural way to try and achieve the response's basic aim of legitimizing the problematic by appeal to the unproblematic. Namely:  show that the problematic discourse's claims {\em `are really a part of'} the unproblematic discourse's claims. Here again: (a) one might  be, in part, an eliminativist---maybe only a favoured subset of the problematic discourse's claims get legitimized; and (b) one might augment the unproblematic discourse with ingredients from other discourses. But the main point is: since claims are expressed in language, {\em `are really a part of'} is here naturally read as {\em `can be deduced from'}.\footnote{Saying this shows that we will take a body of claims, a theory, to be a set of sentences or propositions, rather than a set of models, as in the semantic or structural conception of theories. In Section \ref{defext}, we will briefly defend this: in short, it  will not matter to anything we say.  Note also that (as mentioned above): among the ingredients that augment the unproblematic so as to enable deduction, there might be tools of construction, such as set theory.}

Thus we arrive at the core notion of  {\em reduction}. That is: {\em `Legitimize the problematic by defining (in our liberal sense) its concepts in terms of the unproblematic, in such a way that one can then derive the problematic claims (or at least some favoured subset of them) from within the unproblematic realm (perhaps augmented with other ingredients)'}. 

We shall make this more exact in Section \ref{pleniscarce} onwards. But we end this Section by listing some historically influential examples of proposed reductions (Section \ref{examp}).

\subsubsection{Some proposed reductions}\label{examp}
There are of course many historically influential cases of reduction, including examples (i) to (iv) above; albeit with varying precision and success. For brevity, we will talk only of discourse, not separately of concepts and claims.\\

\indent \indent (i):  Mind and matter. The obvious case is logical behaviourism's project to reduce mental discourse to discourse about behaviour. This looks wrong to most people, since the behaviourist apparently denies that mental states are real.\footnote{The standard reference is Ryle (1949); but note that this over-simplifies Ryle's views (Tanney 2015, especially Sections 8, 9).} But in Section \ref{1111}, we saw how functionalism in the philosophy of mind combined the merits of logical behaviourism with accepting the reality of mental states, viz. as neurophysiological states. \\
\indent \indent (ii):  Ethics and factual description. The obvious case is neo-naturalism in meta-ethics: the project to reduce ethical discourse to factual descriptive discourse. The idea is that Moore's allegation of a `naturalistic fallacy' was wrong: there is a longer and subtler analysis or explication of `good', `right' etc. in terms of facts, for example about what satisfies human desires.\footnote{\label{Hurley}{Cf. Hurley (1989), Jackson (1998), Lewis (1989): all three explicitly invoke functionalism's idea of simultaneous unique  definition, to be  described in Section \ref{homefunc}.  Again, this is part of the Canberra Plan; cf. the Chapters by Colyvan and Robinson in Braddon-Mitchell and Nola (2009).}}    \\
\indent \indent  (iii): Pure mathematics and the empirical. Here, the obvious historical case is the project to reduce mathematics, not to empirical discourse, but to something allegedly even more `secure': to logic, or more plausibly, to set-theory. Thus Frege, Russell and Whitehead claimed to derive from logic, first arithmetic, and then (leaning on previous authors' methods) the rest of pure mathematics, e.g. analysis. Assessing this claim is a large, and still controversial, task---and not for us.\footnote{Among countless references, we recommend  Potter (2020, Chapters 10-13, 31, 37, 42) as a survey of logicism; while Potter (2000) is a monograph focussing on arithmetic, but spanning from Kant to Carnap.}  But even if one objects, as well one should, that these authors' `logic' is really set-theory in disguise: it remains an enormous collective achievement by many authors, from ca. 1880 to 1920, to cast pure mathematics as set-theory, and indeed, to derive much of it in an axiomatized set-theory using a logic as sparse as first-order predicate logic.\\
\indent  \indent (iv):  Unobservable and observable. Let us begin with Bishop Berkeley's phenomenalism. Famously, this takes such a logically weak, i.e. wide, interpretation of `unobservable' as to utterly reverse (i)'s view that the mental is problematic. The idea is: discourse about the material world external to my sensory experience is problematic, while my sensory experience is unproblematic; and so the former should be reduced to the latter.  `Speaking with the vulgar', as the eighteenth-century phenomenalist {\em savant} would put it: material objects like chairs exist in the external world, for example in this room. But such claims as `there is a chair in this room' are to be analysed as very long conjunctions of conditional statements about mental ideas. \\

Although nowadays (iv) seems unbelievable, its influence  within the analytic tradition, on epistemology and philosophy of science, has been enormous: including for our topic of  `theory' vs. `observation' in philosophy of science. Empiricists and positivists, suspicious of concepts and hypotheses about the unobservable, have urged that these should be analysed as compendiously summarizing observable concepts and claims. And some took `observable'  in an avowedly mentalistic, i.e. idealistic, way: {\em not} as ordinary observable properties of material objects,  of `moderate-sized specimens of dry goods', as Austin memorably put it (1962, p. 8).  This was often combined with---indeed inspired by---example (iii). Think of Russell's writings  (e.g. {\em Our Knowledge of the External World}, 1914) propounding a phenomenalist metaphysics, and corresponding foundationalist epistemology, for empirical knowledge, both everyday and scientific. And among the logical empiricists, think of Carnap's {\em Aufbau} of 1928.

This leads to the research programmes which our other papers discuss as spacetime functionalism. For the relationist tradition is that the attribution of geometry to space itself, or of chrono-geometry to spacetime itself, is problematic. It seems to outstrip empirical warrant, since we only experience the material, i.e. ponderable matter and (perhaps) radiation. But it might yet be legitimized by a reduction. Hence research programmes such as the `causal theory of time' and Machian approaches to dynamics.

\subsection{Problems of Faithlessness, Plenitude and Scarcity}\label{pleniscarce}
We already in (2) of Section \ref{1112} introduced these three words as labels for three problems or objections that a programme of reduction is liable to face. We can now develop and illustrate them, using some of Section \ref{problegi}'s examples. Of course, not all problems apply to all versions of the examples. For instance, an objection of Faithlessness (`you are faithless to the meanings of the concept you claim to reduce' ) will apply much better to a version of reduction whose definitions claim to be conceptual analyses, rather than those whose  {\em definiens} makes no claim to synonymy. (Recall from Sections \ref{1112} and \ref{problegi}, our liberal i.e. logically weak usage of `definition', `define', etc. as like `specification' `specify' etc.) 

We will also see that the problems mingle. For example, a problem of Plenitude can underpin one of Faithlessness; and Faithlessness can underpin Scarcity. But despite this variety and mingling, we will  see in the sequel that the labels are worthwhile. For they help us classify the problems or objections that beset reduction---and also functionalism.

 In line with Section \ref{problegi}’s adoption of the word `definition’ as our preferred term, we will present these problems as about definitions. But we again note, in the light of the word’s connotations, that for us, definitions: (i) do not need to be either arbitrary stipulations or faithful to a pre-existing meaning, (ii) can use construction tools like set theory. We also note again that the problems or objections, `More than one!’ or `None!’, can be alleged---not just for the {\em definiens}, to which the word `definition’ naturally attaches, but also---for the {\em definiendum}. To put it in functionalist jargon: there can be a problem of `More than one realizer!’, or `No realizer!’. But these versions of the problems will become prominent only in Sections \ref{rednobstacle} and \ref{homefunc}. So happily, in the rest of this Section, we can construe `definition’ as referring to the {\em definiens}.

\subsubsection{Faithlessness}\label{221}
``The proposed definition is faithless to (the meaning of) the {\em definiendum}. And to make the objection stick: no amendment will work. That is: if you use only the concepts of the unproblematic discourse, you cannot write down a faithful definition of the concept. So you cannot derive the claims of the problematic discourse.'' 

Although this objection is, obviously, more plausible against a reduction whose definitions claim to be conceptual analyses of, or synonymous with, their reduced concepts (their {\em definienda}), it can also apply to a reduction with a  more liberal conception of definition. The objector will say: ``Even by your liberal undemanding standard of what it takes to define, you cannot succeed: the {\em definiendum} has no definition, even in your liberal sense, that uses   only your avowedly unproblematic concepts.''

This objection also takes another form, often as important as the first just stated. It is one of over-shooting, rather than under-shooting. Namely: any definitions using the concepts of the unproblematic discourse will entail claims that are not part of the problematic discourse, and  so are unwelcome. 

We see both forms of the objection in a famous example: Benacerraf's (1965) critique of reductions of arithmetic to set-theory, i.e. of set-theoretic definitions of the natural numbers---definitions that form the core of our example (iii) in  Section \ref{examp}. This example, being about pure mathematics, has the further advantage of being as sharply defined as one could hope a discussion of meanings to be. 

Benacerraf's critique starts from the fact that there are several worked-out reductions of arithmetic to set theory. He takes two examples: the first defines the natural numbers as the Zermelo ordinals, i.e. $0 = \emptyset, 1 = \{ \emptyset \}, 2 = \{\{  \emptyset \}\}, 3 = \{\{ \{  \emptyset \} \}\}$ etc.; and the second defines them as the von Neumann ordinals, i.e. $0 = \emptyset, 1 = \{ \emptyset \}, 2 = \{ \emptyset , \{  \emptyset \}\}, 3 =\{ \emptyset,  \{ \emptyset \}, \{ \emptyset , \{  \emptyset \}\}\}$ etc. That is: the reductions agree that $0 = \emptyset$; but then Zermelo takes each natural number to be the singleton of its successor, while von Neumann takes each natural number to be the set of smaller numbers. 

But both  reductions make claims about numbers that are utterly alien to arithmetic. They agree on some such claims. For example, they both say that 1 is an element of 2: $1 \in 2$. But they also make mutually contradictory claims that are alien to arithmetic: the first says that 1 is not an element of 3, $1 \notin 3$; whereas the second says that 1 is  an element of 3, $1 \in 3$. Thus  with both the agreed, and the mutually contradictory, claims: there is a problem of {\em over-shooting}.

Benacerraf argues that it makes no sense to try to assess such claims. And it hardly matters whether the reductions' claims agree or disagree: even an agreed claim like $1 \in 2$ seems faithless to the meaning of `natural number'. So he concludes, as Mercutio did in {\em Romeo and Juliet}: `a plague on both your houses'. Indeed: he concludes that {\em all} such set-theoretic reductions are wrong, since any of them will imply such claims alien to arithmetic. And he ends by proposing that numbers are not {\em objects} at all: thus supporting a structuralist philosophy of mathematics. 

We will not need to pursue this example, or assess Benacerraf's structuralism about numbers. But these problems or objections about definitions in the foundations of arithmetic  are very similar to those about definitions in the foundations of {\em geometry} that we will discuss, in Sections \ref{implicit} and \ref{beltr} and our other papers.\footnote{{\label{StrucPotter}}{Shapiro  (2000: Chapter 10) is an introduction to the issues. But note that Benacerraf was re-discovering an old theme. Potter (2000, Sections 3.2-3.5) describes how already in 1888, Dedekind articulated Benacerraf's ``over-shooting'' critique, and a version of structuralism about numbers, as immune to it. The similarity of the problems shows up in Potter's Section headings, `Existence' and `Uniqueness' (of natural numbers). For compare how our problem of Plenitude makes existence/uniqueness of the {\em definiendum} easy/difficult, respectively; while our problem of Scarcity makes existence/uniqueness of the {\em definiendum} difficult/easy, respectively.   
 
Besides, there are several other precursors---indeed, luminaries like Frege and Quine; (thanks to Alex Oliver for these cases). (1): Already in the {\em Grundlagen} (1884: paragraph 69, pp. 80-81) Frege himself puts the over-shooting objection to his own official definition of the number of a concept ({\em immediately} after propounding the definition!). He asks: `do we not think of the extensions of concepts as something quite different from numbers?’; and goes on to say that we say `one extension of a concept is wider than another’ (i.e. in modern jargon: is a superset of another), while `certainly we do not say that one number is wider than another’. But after discussion, he concludes that, although `it is not usual to speak of a Number as wider or less wide than the extension of a concept,  . . .  neither is there anything to prevent us speaking this way’. That is, Frege bites the bullet, and allows the definition to mildly revise what we say. To put the point in Carnap’s jargon: Frege says it is enough to give an explication. Dummett (1991, p. 177-179) endorses Frege’s moves. First, he raises the objection of Faithlessness. But then he urges that since nothing Frege will prove, or argue for, turns on the arbitrarily chosen features of his {\em definiens} that go beyond the received sense of the {\em definiendum}, Frege’s choice of {\em definiens} is legitimate. He sums up: `Benacerraf’s problem simply does not arise for Frege’. (2): Quine rehearses the same considerations in {\em Word and Object}’s discussion of the various ways set-theorists and philosophers define an ordered pair. He admits that each proffered definition overshoots in the way we have discussed, but breezily says that he doesn’t care (1960: Section 33, p. 166, Section 53, p. 238). We shall return to this theme---whether to care that one is Faithless---in Section \ref{implicit}.
 
We also note that a cousin of Benacerraf's structuralism, called `structural realism', is prominent in recent philosophy of science. It relates to functionalism through, for example, the use of Ramsey sentences. But it does not bear closely on our main claims; so we postpone a proper discussion to another paper.}}   
For the moment, we just note that Benacerraf's example of Faithlessness turns on the idea of there being two---indeed many---equally {\em bad} definitions of numbers as sets: bad because they over-shoot, viz. by entailing (when taken together with set-theory) claims alien to arithmetic. This idea of many definitions (or reductions) leads to the problem of Plenitude.

\subsubsection{Plenitude}\label{222}
``The unproblematic discourse provides several, even many, definitions of some concept(s) of the problematic discourse. These definitions are equally good, in that they enable deductions of the problematic discourse's claims. So one cannot choose among them in a non-arbitrary way. So they are also equally bad.''

Agreed: we have already seen the  theme here---`many  equally good, and so equally bad': existence of a definition trivialized, and so uniqueness ruled out---in Benacerraf's example. No surprise: as we announced at the start of Section \ref{pleniscarce}, the three problems mingle: a problem of Plenitude can underpin one of Faithlessness. But for us, it is worth articulating Plenitude as a separate problem, for two reasons.

 The first is general, and negative. It is about how equipping the unproblematic discourse with construction tools like set theory can engender a problem of Plenitude. This will lead in to our second reason. Although this reason gives no general answer to the problem, it is positive and is worth stating, since it is specific to our papers’ projects.  The two reasons also differ as regards the realizer-role, or {\em definiendum}-{\em definiens}, contrast. As we noted at the start of Section \ref{pleniscarce}: the problem, `More than one!’ can arise for the realizer, i.e. the {\em definiendum}, just as much as for the role, i.e. the {\em definiens}. And indeed: the first reason will concern a Plenitude of definitions in the sense of {\em definiens}, while the second will concern a Plenitude of the {\em definiendum}---of realizers.  \\

(1): {\em The plethora of mock-ups}:--- The first point is, in short, that Benacerraf's example of two set-theoretic reductions of arithmetic is the proverbial tip of the iceberg. Often, once we see one strategy for defining  concept(s) of the problematic discourse that enables deductions of that discourse's claims, we can instantly see that there are many similar strategies (at least, on a  liberal notion of definition, as adopted in Section \ref{problegi}). ``If you can secure a reduction with this strategy, then here are many similar strategies that work equally well.''  The unproblematic discourse in Benacerraf's example, i.e. set theory, illustrates this point very well. 

Given almost any structure, according to almost any understanding of the word `structure', we can build a set-theoretic mock-up of it.\footnote{We say `almost any understanding' so as not to presuppose sets: we mean, roughly, a plurality of objects with properties and relations among them. And we say `almost any structure', to signal that some vast structures may need proper classes rather than sets: in which case, one could talk of a `class-theoretic mock-up'.}  Namely, we use  iterated curly brackets to represent the internal organization of the structure. But once we do this one way,  we can instantly see how to do it in  many other similar ways, with different conventions e.g. about trivial matters like the order in which items are listed in an $n$-tuple, or how to define an ordered pair (Quine's example in footnote \ref{StrucPotter}). Nor is this just a matter of some half-dozen easily imagined alternative conventions. There are countless ways to define some ghastly forest of curly brackets in which to cast a reduction: and thereby make it incomprehensible, no matter how lucid it was before being consigned to the forest. In short, set-theoretic mock-ups of the given structure are cheap indeed: two a penny, a dime a dozen. 

 Besides, we can build such mock-ups from initial objects, properties or relations (which we place on the lowest or first few ranks of our set-theoretic hierarchy) that are utterly alien to the subject-matter of the problematic discourse. For example, we can build a mock-up of a given structure that involves, say, cats, by using sets with only prime numbers as initial objects (or indeed, with only pure sets). 
  
Nor is this threat specific to set theory. Thus logicians and metaphysicians distinguish set theory from its cousin, mereology. But mereology has sufficiently strong constructive resources to engender the threat. Where we intuitively say there is one cat, Tibbles, on the mat, mereology tells us there is a plethora of objects, differing by the``mental subtraction'' of a single hair, all of them equally good deservers of the name `Tibbles': this is {\em the problem of the many}  (cf.  Lewis 1993).

How to respond to this dismaying plethora? The first thing to say is a reassurance. This plethora does {\em not} trivialize the enterprise of reduction. Agreed, the plethora of (set-theoretic or mereological) objects means that if there is one reduction, i.e. one deduction of claims that are formulated in terms of appropriately defined objects, there is a plethora of other such deductions---almost all of them unthought of, and incomprehensible since cast in some ghastly forest of curly brackets.  But for all that, it is still a substantive enterprise to exhibit just one reduction. Two obvious examples, one from mathematics and one from philosophy, are: (i) Bourbaki’s informal axiomatisation of pure mathematics in set theory, and (ii) Carnap’s attempt in the {\em Aufbau} at a phenomenalist reduction of ordinary talk (example (iv) in Section \ref{examp}). Neither Bourbaki nor Carnap worried about the prospect of these countless complicated alternative reductions. They knew they had enough work to do, to show in detail just one successful reduction; (cf. Carnap 1963, p. 16).

As we see matters, there are two broad strategies for responding to the plethora. One can propose requirements that reduce it (`pruning’). Or one can argue that the Plenitude is okay (`acceptance’). And one can combine these strategies: first prune, and then accept the remainder.
   
As examples of the pruning strategy, one can require any or all of the following:\\
\indent \indent (i) the definition(s) must provide a synonym of the concept(s) of the problematic discourse; and-or that \\
\indent \indent (ii) the reduction must use---the mock-up must be built from---objects, properties and relations in the subject-matter of the problematic discourse, rather than items in some alien subject-matter; and-or that \\
\indent \indent (iii) the deduction of claims, and-or the construction of the mock-up, must be suitably short or natural.\\
 These requirements are undoubtedly tenable, although vague (especially `short' and `natural'). But one must admit that they lead to wider, and mutually related, philosophical controversies about e.g. synonymy, and the distinction between the conventional and the substantive; and thereby to the second strategy, of acceptance.
 
 Some aspects of acceptance are straightforward. One should often just take in one’s stride that there can be different, equally convenient, conventions about e.g. ordering the items in an $n$-tuple, or how to define ordered pairs. For often, nothing important can turn on one’s choice of convention. We expand on this in (2) below; (and again, cf. footnote \ref{StrucPotter} about not caring about being Faithless). But as we just mentioned, there are deeper controversies hereabouts. Recall Putnam’s model-theoretic argument against the objectivity of reference, raised in (4) of this Section’s preamble (and echoed above by our mention of cats and prime numbers). At its simplest, Putnam’s challenge is: `how can you be sure to succeed in imposing the requirement (ii), that a reduction use cats not some alien subject-matter like prime numbers?’ Or more pointedly: `my model-theoretic argument shows that you cannot succeed in imposing such a requirement’.   
 
But obviously, and as we said in (4) at the start of Section \ref{homered}, there are several tenable replies to Putnam’s and similar challenges; and this paper and its sequel will not need to choose among them. It is just that in discussing Plenitude, we are duty-bound to point out this challenging form of it. All the more so, since in Section  \ref{DKL1A} we will advocate the functionalist idea that the functional role of each problematic concept has a unique realizer; thereby providing simultaneous definitions of each of the problematic concepts (recall Section  \ref{1111}).  So the challenge is that these uniqueness claims will stand refuted by a dismaying plethora of realizers: gruesome {\em Doppelgangers}  whose existence seems guaranteed by the constructive power of set theory, and-or of mereology. In Section  \ref{DKL1A} we will return to the question what is the best general response to this challenge. But  whatever it is, we will (as we said in (4)) maintain the objectivity of reference, and so stick to our uniqueness claims, in view of the rich scientific context of the properties, like free mobility, with which we are concerned.\\

(2):  {\em Plenitude is controlled by representation theorems}:--- But the discussion in (1) also reveals a {\em positive} aspect of Plenitude: an aspect which will apply to the examples in our other papers.  Namely: for physical theories, there is often a precise and unproblematic version of the distinction touched on in (1), between the conventional and  the substantive. It is the distinction between: choices of the unit of length, the origin and orientation of spatial coordinate axes etc.; and the coordinate-independent facts (such as the pure-number ratio of two objects' lengths, both expressed in e.g. metres). For formulations of specific physical theories, this distinction can be precise, and so easily treated. 

Indeed, there is a considerable tradition of treating it, in foundational studies in geometry and spacetime theories. Namely, in {\em representation theorems} whose gist can be stated, in the philosophical jargon we have adopted so far, as follows. Given the geometry or spacetime theory, the realizer of each functional role extracted from the theory is {\em not} unique. But this is as it should be. For the non-uniqueness reflects the theory not being committed to any single convention about the unit of length, the origin and orientation of spatial coordinate axes etc. Thus the representation theorem for a geometry or spacetime theory says: \\
\indent \indent (i) For each functional role extracted from the theory, every realizer of that role is related to every other such  by a transformation $T$ that embodies changing one's choices of the unit of length, the origin of axes etc.\\
\indent \indent (ii) Besides, this transformation is to be the same for all the different functional roles, in the natural sense. Namely: if you fix on realizer $r_i$ of role $R_i$ for the various $i$, while for some role $R_j$ I fix on a transformed realizer $T(r_j)$ of $R_j$: then I must also fix on the corresponding  transform  $T(r_i)$ (using the same $T$) as realizer of role $R_i$, for $i \neq j$.

 In mathematical jargon, one says that:\\
\indent \indent (i) the realizer is `unique up to an appropriate transformation $T$ of units and coordinates'; and\\
\indent \indent (ii) the realizers taken collectively e.g. in an $n$-tuple are unique up to a {\em common} transformation $T$.

 In short: although the realizer is non-unique---there is Plenitude---we have complete control and understanding of the variety of realizers, as arising from different conventional choices. The Plenitude is welcome, and right. There should  {\em not} be uniqueness:  it would dictate to us a single convention. As mentioned, we will see this in our examples.\\

\subsubsection{Scarcity}\label{223}
``The unproblematic discourse cannot provide  even one definition of some (at least one) concept of the problematic discourse.'' 

Again, the reasons for the objection vary from case to case. As we mentioned, an objection of Faithlessness can prompt one of Scarcity: ``you cannot write down a faithful definition''. Examples (i) and (ii) in Section \ref{examp} give well-known cases. For (i) (`mind and matter'), some say that the `raw feel' of pain cannot be faithfully defined in material vocabulary. For (ii) (`ethics and facts'), Moore says that `good' cannot be faithfully defined in naturalistic vocabulary (cf. footnote \ref {Moore}). Besides, the objection need not be based on requiring definitions to give conceptual analyses or synonyms. For even on a liberal notion of definition, there can be what we called under-shooting and-or over-shooting.

Sad to say, but here, as in life: scarcity makes for thieving. That is: we connect here with Russell's quip about the advantages of theft over honest toil. As we mentioned at the end of Section \ref{examp}, much of Russell's writing, especially in his phase as a logical atomist, proposed reductions in our sense. He required that the {\em definiendum} must be---not  declared, but---{\em shown} to have, thanks to the {\em definiens}, the properties of the original problematic entity (in his jargon, the `metaphysical entity' (1918) or `supposed entity' (1924)). He is thus opposed to so-called {\em implicit definition}, i.e. to thinking it is enough, for justifying a concept or discourse, to give a set of postulates (`axioms') containing the concept and from which one's claims about it can be deduced. How do you know---he might say---that your deduction sets out from safe ground? (In Section \ref{implicit} we will return to this, in connection with (a) the Frege-Hilbert controversy about implicit definition, and (b) Torretti's work.)  

Hence Russell's famous  {\em bon mot} about `theft over honest toil'. It is in his {\em Introduction to Mathematical Philosophy} (1919), in  his discussion of deducing the truths of arithmetic from logic, or what we today would call `set-theory' (cf. example (iii) in Section \ref{examp}):
\begin{quote}
The method of `postulating' what we want has many advantages; they are the same as the advantages of theft over honest toil. Let us leave them to others and proceed with our honest toil. (1919, p. 71)
\end{quote}
Thus `theft' is here the dogmatic postulation of entities by implicit definitions, viz. as being those things that obey certain axioms or postulates; and `toil' is here the work of finding judicious definitions (including constructions) so that the {\em definienda} can be shown  using logical inference alone to satisfy the claims made about them. And as we said: Scarcity makes for theft.\footnote{\label{toil}{Agreed: advocates of conceptual analysis often do not stress that they must capture all or most of the claims about the {\em analysandum} once it is  interpreted in terms of the {\em analysans}. At least they do not stress this, with `capture' meaning `derive'. That is hardly surprising since, as we have seen: in most cases of philosophical interest, such a derivation is a very tall order. Nevertheless: reduction, as we understand the enterprise, is thus obliged. And as we saw in the quotes above, Russell himself accepted this obligation: as did Carnap (1963, p. 16).

Not that we wish to put Russell or his {\em bon mot} on a pedestal. Oliver and Smiley (2016: 272) call it `one of the shoddiest slogans in philosophy’; and their reasons echo one of our themes for our spacetime examples, viz. that writing down the right functional role, or analysis, of a concept, {\em before} any `construction’ begins, can take considerable `honest toil’.  Thus they point out that : (i) Russell originally aimed it, unfairly, at Dedekind’s treatment on continuity, and (ii) `it assumes [wrongly] that we already know what we want … the examples from Dedekind show just how much honest toil it takes to discover—to formulate precisely---just what it is that we want’.  }} \\  

\subsubsection{Answering these three problems in the sequel}\label{224}
So much by way of stating the problems, or objections, of Faithlessness, Plenitude and Scarcity. We will see illustrations of them, and of how they can be answered, as regards reduction (in Section \ref{rednobstacle}), functionalism (in Sections \ref{homefunc}  and \ref{DKL4}) and spacetime theories (in our other papers). Broadly speaking, the situation will be that:\\

\indent \indent (1):  In general, we will allow some Faithlessness---we do not require a reduction's (nor a functionalist reduction's) definitions to provide synonyms. For recall that in Sections \ref{1112} and  \ref{problegi}, we adopted a liberal i.e. logically weak usage of `definition' etc. as like `specification' etc. We will see in Section \ref{defext} that this fits with the core idea of Nagelian reduction, viz. what logicians call `definitional extension’. For the Nagelian, a bridge law---a proposition that enables the deduction of the reduced theory---can be a definition in this logicians’ liberal usage of the term.  It can even involve constructions: which, in general, logicians’ usage of `definition’ excludes.

But here, we must stress the distinctive features of functionalist reduction: especially its ternary, not binary, contrast of vocabularies (cf. the start of Section \ref{intro}). For definitions extracted from the initial theory tend to be more Faithful than the definitions given by the later or independent theory (which are therefore more naturally called `specifications’). 

Think of Section \ref{1111}'s philosophy of mind example, especially its two-premise inference to the derived bridge law, that pain is C-fibre firing. Here, the initial theory is everyday mental and material-behavioural discourse. The later or independent theory is neurophysiology. The functional definition of `pain' extracted from the former has a much better claim to be Faithful to the meaning of `pain' than does the ``definition''---hence: better called `specification'---given by neurophysiology. We will see that this pattern is typical. By and large: (i) the functional definitions extracted from the initial theory have a good claim to provide synonyms; while (ii) the specifications of the same term given by the later or independent theory do not. In particular,  for the authors and theorems we celebrate in our other papers:\\
\indent \indent (i) the initial theory's functional definitions of `freely mobile', `simultaneous', etc. are certainly explications, and have a strong claim to be synonyms, or conceptual analyses,  of the concepts at issue; while \\
\indent \indent (ii) the specifications of these notions given by the later or independent theory are not analyses or synonyms or explications: they give novel information, for example that free mobility requires a Riemannian metric of constant curvature.\\

\indent \indent (2): As to Plenitude:--- In the literature   within the philosophy of science  on reduction (i.e. disregarding functionalist reduction: as in Section \ref{rednobstacle}), the problem of  a plenitude of definitions, i.e. many {\em definiens}, is hardly discussed. There are two obvious reasons: one creditable, one less so. The creditable reason is that reduction is `uphill work’. It is in general hard to find for each {\em definiendum} in the problematic discourse, even just one {\em definiens} that, taken together with other such, secures a deduction of all the problematic discourse’s claims. In short: one faces a problem of Scarcity, rather than Plenitude. The less creditable reason is that philosophers of science tend to ignore the logical and metaphysical issues that led, in (1) of Section \ref{222}, to the plethora of mock-ups: to the fact that if there is one reduction, there are very many. In effect, philosophers of science assume that some combination of (1)’s pruning and acceptance strategies will control the plethora. (Agreed: sometimes, with good reason: for example, we will see that Nagel proposed that each {\em definiens} should be short and conceptually homogeneous.) 

 But here we again need to recall the realizer-role, or {\em definiendum}-{\em definiens}, contrast. As we noted at the start of Section \ref{pleniscarce}: the objections, `More than one!’ or `None!’, can be alleged for the realizer, for the {\em definiendum}, just as much as for the role, for the {\em definiens}. And the objection `More than one realizer’ goes of course by the name {\em multiple realizability}. This, we of course admit, {\em is} much-discussed in the literature on reduction---and we will address it in Section \ref{rednobstacle}.  So in short: multiple realizability amounts to Plenitude of the {\em definiendum}, but not of the {\em definiens}. Anyway, for us, the more important point will be, as we noted in (2) of Section \ref{222}: the authors and theorems we celebrate in our other papers each give a representation theorem that yields a controlled and welcome Plenitude---of realizers, of the {\em definiendum}. \\

\indent \indent (3):  As to Scarcity:--- In Section \ref{Scarc}, we will discuss this problem in the sense we introduced: namely lack of a definition, a {\em definiens}. It {\em is} emphasised in the literature on reduction: mostly under the heading of {\em multiple realizability}. For although {\em multiple} realizability means there are many ways to realize a predicate of the reduced discourse (theory), each way is sufficient but not necessary; and this means that multiple realizability makes for lack of a definition, i.e. for Scarcity. But we shall argue that  multiple realizability is not really a problem for reduction, but only for reductionism. Scarcity also arises under the heading of {\em circularity}. But this problem will be answered by functionalism's idea of simultaneous unique definition,  already introduced in Section \ref{111}. Thus the problem of circularity will prompt our transition to functionalism, in Section \ref{homefunc}.

\section{Reduction based on definitional extension}\label{rednobstacle}

So much by way of generalities about reduction and about the three problems of Faithlessness etc. that a reduction can face. In this Section, we discuss reduction more precisely, in the jargon of philosophy of science---but without considering functionalism. We begin with the formal notion of {\em definitional extension} (Section \ref{defext}), and then discuss how it has been modified, especially by Nagel (Section \ref{Nagelmodif}). Then we describe how the three problems of Faithlessness etc. play out for this account of reduction. As we have just announced (Section \ref{224}), they do get illustrated, and in part get answered, albeit under different headings---the best known of which are {\em multiple realizability} and {\em circularity}. We shall nevertheless organize our discussion, using our three labels, `Faithlessness' etc., for three Subsections (Sections \ref{Faith}, \ref{Plen} and \ref{Scarc}). This discussion prepares us for the next main Section on    functionalism (Section \ref{homefunc}).

\subsection{Definitional extension}\label{defext}
We now recall the notion of reduction articulated by Nagel (especially 1961, pp. 354-358; 1979, pp. 361-373)); and Hempel (1966, especially pp. 75-77) . In this Section, we begin with its formal core, called {\em definitional extension}. In the next, we will discuss the informal conditions Nagel and Hempel add to it. \\

We take the {\em relata} of reduction to be theories, i.e. bodies of claims. We will also take a theory to be a deductively closed set of sentences. To this, an objection will be made immediately, i.e. quite apart from the topic of reduction. Namely: this {\em syntactic conception} of a theory as a set of sentences is in any case wrong, and should be replaced by the {\em semantic (or structural) conception} of a theory as a set of models: and this will prompt some other treatment of reduction---presumably as something like subset-hood of sets of models. Indeed, Torretti himself would surely make this objection, since he rejects the syntactic conception and endorses the semantic one, in a version similar to that of Sneed and Stegm\"{u}ller (1990, pp. 109-160; 1999, pp. 407-416). We will return to this objection, and to Torretti, in Section \ref{peace}. But here we just set it aside, since {\em pace} Torretti, we believe the recent literature contains convincing replies. Here, we  emphasise: \\
\indent \indent (i): Lutz's detailed arguments for there being no material difference between the conceptions: cf. his (2017), focussed on a three-cornered debate between Halvorson, Glymour and van Fraassen, and his (2017a: especially Section 5.2), focussed on Newman's objection to ``structural realism''. (Hudetz (2019) builds on the former; Halvorson (2019, pp. 107-11, 172-174) introduces the debate.) \\
\indent \indent  (ii): Niebergall’s studies of inter-theoretic reduction (2000, 2002), which argue against a semantic or structural understanding of it. \\

We allow that the language in which a theory is written is natural rather than formal. But we presume that the languages, and so the theories, we are concerned with have rules of deductive inference that make the enterprise of reduction as deduction, announced in Section \ref{problegi}, reasonably precise. Then, taking a theory as a deductively closed set of sentences, the formal core of reduction is that one theory $T_t$ (`{\em t}' for `top', or `tainted') {\em is reduced to} another theory $T_b$ (`{\em b}' for `bottom' or `better') iff:
 \begin{quote} 
 by adding to $T_b$, a set $D$ of definitions, one for each vocabulary item in the language of $T_t$, one can, within this augmented $T_b + D$ (i.e. using its underlying logic, and any set of its sentences, including the definitions in $D$), deduce every sentence of $T_t$.
  \end{quote} 
In such a case: we say that $T_t$  {\em is a definitional extension} of $T_b$. 

This is a standard idea in formal logic. Here of course, the theories are cast in a formal language, almost invariably a predicate logic with the vocabulary items being predicates, and maybe also functional expressions and singular terms. So for predicates, the formal proposal is that each $n$-place predicate $F$ of the language of $T_t$ gets as a definition a universally quantified biconditional stating $F$ to be co-extensive with some open formula $\Phi$ within the language of $T_b$ that has $n$ free variables.  Thus the definition, which is to be an element of $D$, looks like:
 \begin{quote} 
  $(\forall x_1)...(\forall x_n)[F(x_1,...,x_n) \equiv \Phi(x_1,...,x_n)]$.
  \end{quote}
Recall the start of Section \ref{problegi} with our liberal, i.e. logically weak usage of `definition' and cognate words: we do not require that the definition be faithful to a pre-existing meaning, or that $F$ and $\Phi$ are co-extensive in domains other than the given one. 

But we emphasise that despite this liberality or logical weakness, providing the set of definitions $D$ is not a matter of `theft’: it is `honest toil’. To obtain a deduction of all of $T_t$, the definitions in $D$ will have to be {\em judiciously chosen}; and it is easy to write down simple examples of $T_b$ and $T_t$ for which it is impossible to formulate such definitions.

In philosophy of science, the jargon is of course {\em bridge law}, rather than `definition'. But we shall mostly keep to `definition' (or `specification'), for two reasons (cf. (1) in Section \ref{224}):\\
\indent \indent  (i) `bridge law', like `correspondence rule', connotes various controversies from mid-twentieth century philosophy of science, about whether they are contingent hypotheses (even laws of nature?) or stipulations (always or sometimes?): controversies which we will be able to avoid; \\
\indent \indent (ii) our advocacy of functionalist reduction (Section \ref{DKL4})  will be clearer if we reserve `bridge law' for its {\em derived bridge laws}.
    
So much for the form of definitions of $T_t$'s predicates. Similar proposals are made for its functional expressions $f, g, ...$ and individual constants $a, b, ...$: (though these usually involve admissibility conditions requiring that a function be single-valued and that an individual constant have a bearer (Hodges 1997, p. 52)). But in philosophical logic (and so quite independently of one’s account of reduction), it is standard practice to simplify matters by eliminating these expressions in favour of predicates (i.e. within $T_t$---before reduction). Thus a  functional expression with $n$ arguments is eliminated  in favour of a $(n+1)$-place predicate, and a singular term is eliminated in favour of a $1$-place predicate, in the spirit of Russell's theory of descriptions; (e.g. Quine (1960, Sections 37, 38); of course, Russell himself saw that theory as a good example of `logical construction': cf. Section \ref{examp}).\\

Of course, this is just the beginning of a large topic in logic. Among the questions to be addressed are:\\
\indent \indent  (i) What about  many-sorted languages? \\
\indent \indent (ii) What about the choice of logic, for example allowing higher-order not just first-order quantification? \\
\indent \indent (iii) How does the notion of definition just sketched, called {\em explicit definition} by logicians, relate to what they call {\em implicit definition}---which is the logicians' precise version of philosophers' notion of {\em supervenience} or {\em determination}? \\
\indent \indent (iv) And when one notices that the sketch just given will not provide new objects---for the quantifiers $(\forall x_1)$ etc. range merely over $T_b$'s given domain of quantification---one naturally asks: What about  endowing $T_b$ (or its language) with `construction tools', so that the reduction can indeed build new objects? The obvious candidates for such tools are of course, set theory and mereology: cf. the mock-ups discussed in Section \ref{222}. 

However, we do not need to develop answers to these questions, or even to taxonomise the possible rival answers and choices. It will be enough for this paper (and its companions) that we here raise the questions; so that we can later see how the various proposals we discuss illustrate various answers and choices.\footnote{\label{rigorlogic}{For rigorous details about the notion of definitional extension, and about answers and choices for these questions, cf. for example, Boolos and Jeffrey (1980, pp. 245-249), Button and Walsh (2018, Sections 5.1-5.5) and Hodges (1997, Sections 2.3, 2.6.2, and 5.5). Note that the idea of `implicit definition' in (iii) above, originally due to Padoa (in 1900),  is {\em not} the idea usually understood by this phrase, that was advocated by Hilbert, and denounced by Frege. We will return to clarify both ideas in  Section \ref{precurs}. And in Section \ref{Plen} we will briefly relate definitional extension to the question when two theories count as equivalent.}} 

What we of course must do is assess definitional extension being adopted as the formal core of the conception of reduction. This we undertake in the following Sections: the focus will of course be on the {\em pros} and {\em cons} of the informal conditions that might get added to definitional extension. 

\subsection{Nagel's modification of definitional extension}\label{Nagelmodif}
Agreed: it was shown long ago (in the 1960s) that definitional extension was  wrong---extensionally wrong, so to speak---as a description of reduction between scientific theories. Examples were given showing that definitional extension was too weak, i.e. some examples of definitional extension are not reductions. And other examples showed that it was too strong,  i.e. some examples of reductions are not definitional extensions. 

For the most part, these examples were offered as criticisms of Nagel's (1961) account of reduction. Our answer is, in short: `Unfair to Nagel!' There are two concerns here: (1) about what Nagel actually said about reduction; and (2) about whether he was right. We shall discuss them in turn: though only briefly, since more details---broadly, {\em pro} Nagel---are elsewhere.\footnote{Cf. Butterfield (2011 Sections 3.1, 3.2; 2014, Sections 1.2, 4); and the papers by Dizadji-Bhamani et al. and Schaffner cited below.}  \\

(1): {\em Nagel’s account}:--- One must distinguish definitional extension from Nagel's own account. Nagel did not say that reduction is just definitional extension, for two reasons. 

First, there is a topic that this Section has so far set aside. Namely: do the vocabularies of the two theories $T_t$ and $T_b$ {\em overlap}? That is, do they have any (non-logical) vocabulary in common?  In some examples, both in general philosophy  and in philosophy of science---including examples we have mentioned---the answer is `No'.  Thus elementary arithmetic makes no mention of set theory; the wave theory of light  makes no mention of electromagnetism.\footnote{Or rather: that is so, in a suitably historically sensitive sense of `wave theory of light', e.g. up till 1870; since the success of Maxwell's theory, most expositions of the wave theory of light of course emphasise that light is electromagnetic waves. We will return to the issue of  meanings shifting over time.} But in many cases, the answer is `Yes'. And in the vaguer and more complicated cases of discourses, rather than scientific theories, which we treated in Section \ref {problegi}, it is likely to be very difficult to sift out two sets of claims, each cast in one vocabulary, with a view to deducing the one from the other, once it is augmented with suitable definitions (in our liberal sense) of the one's terms. Thus consider examples (i) and (ii) in Sections \ref{problegi} and \ref{examp}: i.e. mind and matter; and the ethical and the factual: or within the philosophy of science, theory and observation ((iv) of Section \ref{examp}), and the case of interest to spacetime functionalism---a theory of chrono-geometry and a theory of matter-and-radiation. 

In any case, Nagel allows the  non-logical vocabularies to overlap, or even be identical; (in which case he speaks of `homogeneous reduction'; otherwise, of `heterogeneous reduction'). And whether or not they overlap, he is not committed to the form of the definitions, i.e. the bridge laws, always being as in Section \ref{defext}, i.e. to their being understood as in a logic book. It is enough that the bridge laws state connections---make assertions using both vocabularies---in such a way that, once they are added to $T_b$, $T_t$ can be deduced (Dizadji-Bahmani et al. (2010: p.398); Schaffner (2012: p. 538)).    

Second, even for cases where the definitions, the bridge laws, take the form in Section \ref{defext}, Nagel added to definitional extension, further informal conditions---and in advance of much of the 1960s criticism, to boot (1961, pp. 358-363). These conditions were motivated by the idea that the reducing theory $T_b$ should {\em explain} the reduced theory $T_t$; and following Hempel, he conceived explanation in deductive-nomological terms (cf. Schaffner 2006, pp. 380-382; 2012, p. 536). Thus he says, in effect, that $T_b$ reduces $T_t$ iff:\\
\indent \indent (i): $T_t$ is a definitional extension of $T_b$; and \\
\indent \indent (ii): in each of the definitions of $T_t$'s terms, the {\em definiens} in the language of $T_b$ must {\em play an explanatory role} in $T_b$. This is a matter of it being reasonably short, and conceptually unified; so it cannot be, for example, a long and heterogeneous disjunction.

Agreed, condition (ii) with its phrases, `playing a role' and `being reasonably short, and conceptually unified', is vague. And even if one made it precise,  many  reject Nagel's (and Hempel's)  deductive-nomological account of explanation that motivates it, while  still wanting reduction to include explanation. So a consensus about the account of reduction will require a consensus about the much-contested concept of explanation.  Obviously, we cannot settle such controversies here. Suffice it to say that if there are scientific reductions exemplifying the Nagelian account, that makes a good case that Nagel's added condition (ii) answers the first part of the above  criticism, i.e. the allegation that definitional extension is too weak. That is, Nagel can reply: `Yes it is too weak, but condition (ii) disposes of the objection'. 

Besides, Nagel replied to the second part of the above criticism, i.e. the allegation that definitional extension is too strong.  The idea of the criticism, is that in many cases where $T_b$ reduces $T_t$, $T_b$ corrects, rather than implies, $T_t$. One standard case is Newtonian gravitation theory ($T_b$) and Galileo's law of free fall ($T_t$). This $T_t$ says that bodies near the earth fall with constant acceleration. This $T_b$ says that as they fall, their acceleration increases, albeit by a tiny amount. But surely $T_b$ reduces $T_t$. To which, Nagel's reply is `I agree': a case in which $T_t$'s laws are a close approximation to what strictly follows from $T_b$ {\em should} count as reduction. (Nagel called this `approximative reduction' (1979, pp. 361-363, 371-373); cf. also Hempel (1965, p. 344-346; 1966, pp. 75-77).) Besides, cases where $T_b$ corrects  $T_t$ do not always involve merely quantitative approximation, with no change of the concepts involved.  Thus, Schaffner's proposed modification of Nagel's account requires only that there be a strong analogy between $T_t$ and its corrected version, i.e. the theory that strictly follows from $T_b$ (1967, p. 144; 1976, p. 618). For a recent study of approximation and analogy, as they apply here, cf. Fletcher (2019). \\

(2): {\em Right?}:--- Is Nagel’s account of reduction (as modified, e.g. to allow approximative reduction or an analogue of $T_t$) right? This of course raises raises controversies, e.g. about how reduction relates to explanation, that we cannot hope to resolve. (Arguably, in the present state of knowledge, no one could.) But our advocacy of functionalist reduction, and the examples in our other papers, do not need a general defence of Nagel’s account. After all, what matters scientifically and conceptually is---not the best conceptual analysis or explication of the word `reduction', but---to understand the various relations between scientific theories. So it will be enough for us that, as modified by functionalism, Nagel’s account fits the examples in our other papers. Nevertheless, we submit that over the years, it has stood up well. Here we commend: Dizadji-Bahmani et al. (2010) who defend it, as modified by Schaffner, against a battery of objections (p. 400f.);  and Schaffner (2012), who gives a historical survey of its reception including modifications by himself (pp. 539-549), a partial defence and a detailed example from optics (pp. 551-559).  Cf. also Niebergall (2000, 2002).

But our advocacy of functionalist reduction will need some details of how the three problems we have labelled Faithlessness etc. play out for Nagel’s account of reduction: details which will match some of what these authors say. So to this, we now turn.

\subsection{Faithlessness as a problem for Nagelian reduction}\label{Faith}
Clearly, a reduction based on definitional extension is liable to face a problem, or objection, of Faithlessness,  simply because a `definition’, as we (and the logic books) use the term, need not be Faithful to the pre-existing meaning of a term in $T_t$. Nor is this liability lessened by Nagel’s modifications of definitional extension. For as to his informal condition (ii) in Section \ref{Nagelmodif}:   a short and conceptually unified {\em definiens} can be Faithless to the {\em definiendum}'s pre-existing meaning. And allowing approximative reduction and-or analogy does not secure that the {\em definiens} used is Faithful.

Clearly, how severe this problem is---how convincing the objection is---will vary from case to case. Broadly speaking, it will be more of a problem for reductions as definitional extensions   in general philosophy (i.e. metaphysics, philosophy of mind and ethics), where there is a strong tradition of requiring the definitions to be conceptual analyses or explications (cf. Section \ref{examp}), than for  scientific reductions as described by philosophy of science, which of course makes no such requirement on the definitions, i.e. bridge laws. (Obviously, this contrast holds good whether one takes reductions just as definitional extensions, {\em a la} Section \ref{defext}, or follows Nagel in adding conditions like (ii): it turns just on whether definitions must be  Faithful.) 

But we should also notice that (as we announced in Section \ref{224}) the ternary contrast involved in functionalist reduction will make a difference. That is: we will see that the functional definitions extracted from the initial theory can claim to be Faithful to the meaning, even in scientific reductions like our examples of spacetime functionalism; (they will also be short and conceptually unified, as Nagel requires). On the other hand, the definitions---better called `specifications'---given by the later or independent theory are not, nor aim to be, Faithful to a pre-existing meaning.

In any case, the strategy for replying to the problem must be the same for reductions in (1) general philosophy and in (2) science: viz. as we put it above, `to allow a little Faithlessness'. We briefly discuss the cases (1) and (2).  \\

(1): {\em In philosophy}:--- The concepts of $T_t$, the concepts of the initially problematic discourse, are usually vague. And even setting aside vagueness, criteria for the identity of concepts (properties etc.: cf. footnote \ref{fnstateconcept}) are much disputed in general philosophy. Should we judge identity by some notion of logical equivalence, or more finely (i.e. hyper-intensionally), or more coarsely e.g.  up to some nomological equivalence? (Cf. e.g. Oliver (1996: pp. 16, 20-25, 44).) So it is hardly surprising that there is rarely a consensus about whether a reduction has been faithful to $T_t$'s concepts. (Or `faithful enough': recall Section \ref{problegi}'s allowance that reduction capture only part of the problematic discourse's concepts and claims.)  In philosophy as a whole, one of the most discussed examples is {\em qualia}, also known as ``raw feels''---cf. example (i) in Sections \ref{examp} and \ref{223}. Here, the standard example is pain. Whether a reduction proceeds by conceptual analysis or explication (Section \ref{problegi}) or by definitional extension as in Section \ref{defext}, the reduction will necessarily express the {\em definiens} of `pain' {\em in words} (perhaps technical, as well as everyday). And some people deny that any amount, or any type, of discursive information, necessarily expressed in words, can capture  the `quale', the `raw feel', of pain. But of course, this is not the place to address this denial.\footnote{As we mentioned at the end of Section \ref{1112}: we in fact reject the denial, and associated views like epiphenomenalism, since we endorse a Lewis-Armstrong-style functionalism about mind. But nothing in this or our other papers will turn on this.} Here, we just note the obvious strategy for replying to the problem: allow a little Faithlessness. \\

(2):  {\em In science}:--- In philosophy of science, the problem of Faithlessness is discussed under the heading of {\em meaning variance}, especially in discussing the nature of scientific progress. Recall that `{\em t}' in $T_t$ could stand for `tainted' as well as `top'; and `{\em b}'  in $T_b$ for `better' as well as `bottom'.  When one theory is succeeded by another, it is natural, on a cumulative view of the transition, to think the successor theory gives a reduction of the predecessor. But usually, the successor will not imply exactly the predecessor. And as we said in  Section \ref{Nagelmodif}, this is not always a matter of just quantitative approximation (cf. Newtonian gravitation theory succeeding Galileo's law of free fall). In some cases, the {\em definiens} in the language of $T_b$, of a vocabulary item in $T_t$, is not completely Faithful to the item's meaning. And this point is not just a matter of being content, as a matter of one's philosophical method, to be revisionary, or to admit Carnapian explications. Undoubtedly,  transitions from one scientific theory to another often involve conceptual change that is significant enough to make the {\em definiens}  Faithless. But as we said above:  in our spacetime examples, the {\em functional} definitions will not face this problem of Faithlessness.

\subsection{Plenitude as a problem for Nagelian reduction}\label{Plen}
As we said in (2) in Section \ref{224}:  the problem of Plenitude, of there being too many definitions of a {\em definiendum}, is hardly discussed in the philosophy of science.\footnote{\label{Newman}{The problem should be distinguished from what has come to be called `Newman's objection': which was originally made by Newman against Russell's (1927), but is now seen as a problem for various forms of  `structural realism'. For while Plenitude is a matter of there being too many definitions ,  Newman's objection is a matter of it being {\em logically guaranteed} that there is a  {\em realizer}, a  {\em definiendum}.  The idea then is that this guarantee is a problem since structural realism wants the existence of the {\em definiens} to be its main substantial assertion.  As mentioned in footnote \ref{StrucPotter}, we postpone discussion of structural realism to another paper.}} The focus is instead on Scarcity; cf. the next Section. But the problem is worth stating, as it affects definitional extension or any notion of reduction built on it, such as Nagel's notion. For although our statement of the problem, in (1) of Section \ref{222}, made no mention of definitional extension (which we had not yet introduced), it is clear that invoking definitional extension does nothing to avoid the threatened plethora of equally good---and so equally bad---definitions. Rather, invoking definitional extension just makes the threat more precise. Similarly, invoking definitional extension does not mitigate how the the strategy, `allow a little Faithlessness', that we endorsed in our liberal, i.e logically weak, construal of `definition' and in Section \ref{Faith} aggravates the problem of Plenitude.    

Besides, it hardly helps to appeal to reduction having requirements additional to definitional extension, such as Nagel's condition (ii), that a {\em definiens} be short and unified. For  some constructions using set-theory and-or mereology, as described  in (1) of Section \ref{222}, will yield, when applied to a short and unified {\em definiens}, a rival---since equally short and unified---{\em definiens}. That is: there could be a plethora of definitions that each satisfy the additional  requirements. We made essentially this point in different words, i.e. independently of definitional extensions, at the start of the response `acceptance' in (1) of Section \ref{222}---emphasising that one can often take it in one's stride.

But here, we should also note {\em another kind of plenitude}. For it yields a moral about meanings, that will return in Section \ref{implicit}. It is not a plenitude of ways to make a given $T_t$ a definitional extension of a given $T_b$; but a plenitude of surprising cases of  definitional extension.   This plenitude is well recognised by logicians and model-theorists who work on formal relations between theories such as definitional extension (and its cousins, studied in the theory of definability: cf. Button and Walsh (2018: Chapter 5)). That is: they are aware that  often, one theory turns out to be a definitional extension of another, even though we usually think of them as being about very different topics, or indeed as contradicting one another about a given topic. 

Agreed: on reflection, it is not surprising that this should happen, despite the theories' different intended topics or claims. For the definitions that yield a  definitional extension are {\em not} required to respect any pre-existing meanings of the terms in either theory. So this allows what one might call definitional extension `thanks to Faithlessness': or  `thanks to typographic accident'. Besides, over the course of time, the mathematical  community's knowing of a definitional extension can shift the meanings even of central mathematical words. For example, think of the nineteenth-century rigorization of analysis: at the end of that process, but not at the beginning, mathematicians could understand `real number' as `an equivalence class of Cauchy sequences of rational numbers'. This of course echoes the theme of meaning variance, between scientific theories, as discussed in (2) at the end of Section \ref{Faith}.

But this plenitude of definitional extensions is worth emphasising to philosophers of science. It should give them pause in their discussions of formal notions of theoretical equivalence.\footnote{Recent discussions include Butterfield (2018: Section 5), De Haro (2020), Dewar (2019), Halvorson (2019),  Hudetz (2019, 2019a) and Weatherall (2018, 2018a). For general arguments against formal analyses of theoretical equivalence, cf. Sklar (1982) and Coffey (2014).} For in some cases, the surprising relation of definitional extension is symmetric. That is, each of two theories, that we think of as about very different topics or as making contradictory claims, is a definitional extension of the other. This is called being `definitionally equivalent'. Hudetz (2019a, pp. 60-62, especially Prop. 4) gives a telling example. One can formalize Minkowski and Euclidean geometry for $\mathR^4$---which we usually think of as inequivalent, indeed contradictory---as definitional extensions of each other. For each of them can be formulated as a definitional extension of the theory of the real line; and then each can `recover' the other, by `building'  from the real line. So each `contains' the other. Besides, this situation cannot be readily overcome by appealing instead to other model-theoretic notions of equivalence for theories. For definitional equivalence is stronger than several other natural notions of equivalence studied in model theory; (Button and Walsh 2018: Proposition 5.10, p. 117; and cf. the discussion of Feferman's theorem in ibid. pp. 118-120, and in Niebergall (2000, pp. 44, 52)). The overall moral here---that one cannot expect formal structures to completely capture meanings---will return in Section \ref{implicit}.

So much by way of stating the problem of Plenitude, in relation to definitional extension. Fortunately, as we noted in (2) of Section \ref{222}: the problem will not affect the authors and theorems we celebrate in our other papers. Their representation theorems will give a controlled and welcome Plenitude.  
    
\subsection{Scarcity as a problem for Nagelian reduction: multiple realizability and circularity}\label{Scarc}
For Nagelian reduction, the problem or objection of Scarcity is: `there is not even one definition of a certain {\em definiendum} of $T_t$, that enables (along with other definitions) the deduction of all $T_t$'s claims'. As we see matters, there are two main ways this problem arises, which we will discuss in turn: (1) multiple realizability and (2) circularity.\footnote{\label{MRscarce2}{For why multiple realizability threatens Scarcity of definitions, despite being a plenitude of realizations at the level of $T_b$, cf. (3) at the end of Section \ref{224}.}}  Or rather, the problem of Scarcity {\em seems to} arise. For we will argue: (i) following Sober (1999): that multiple realizability is not really a problem for  Nagelian reduction; and (ii) as presaged in Section \ref{111}: that the problem of circularity is solved by  functionalism's idea of {\em simultaneous unique definitions}.

Note that (1) and (2) relate differently to Section \ref{intro}’s distinction between binary and ternary contrasts of vocabulary. (1) relates to the usual binary contrast invoked in discussions of reduction. For (1) does not consider functional definitions in $T_t$: the focus is just on instances of a predicate in $T_t$’s vocabulary satisfying disparate predicates in $T_b$’s vocabulary. But on the other hand, (2) involves the ternary contrast: for it is about functional definitions within $T_t$.\\

(1): {\em Multiple realizability: no worries}:--- Very often, the instances of a concept (predicate) $F$ in one theory  $T_t$ vary greatly in how they are described by another theory  $T_b$. Indeed, they vary greatly even in respects that are candidates to occur in the {\em definiens} of $F$ in a putative Nagelian reduction of $T_t$ to $T_b$. This is  {\em multiple realizability}. Recall from the end of Section \ref{1111}, the philosophy of mind's standard example: pain might be one brain state in humans, another in molluscs (Lewis 1969, p. 25).\footnote{ Some say that there might even be {\em infinitely} many ways, according to the taxonomy provided by  the vocabulary (the concepts) of $T_b$, to be an instance of (to realize) $T_t$'s predicate $F$. This is the idea of {\em supervenience} or {\em determination}: that the  $T_t$-facts merely supervene on, are determined by, the  $T_b$-facts, and a {\em definiens} would have to be an infinite disjunction. Cf (iii) in Section \ref{defext}. But we  doubt there are such truly infinite cases; and if they occur, we doubt their scientific importance (cf. Butterfield (2011, Sections 4.1, 5.1, pp. 940-944, 948-951; 2011a, Sections 4.2.3, 5.2.3 and 6.3.4, at pp. 1070, 1089, 1100, 1127). Anyway, such cases will not occur in our other papers' examples. So we will only consider finite disjunctions.}   

Such cases are often pressed as objections to the reductionist picture of a hierarchy of levels (of scale or of description), with reductions between successive levels.  As we said in (2) of the preamble to Section \ref{homered}: we have no brief to endorse the reductionist picture. And we agree that multiple realizability undoubtedly makes higher-level theories autonomous from lower-level theories; (in another jargon: theories in the  special sciences  autonomous from theories in the basic sciences). Here, {\em autonomy} means, so to speak, never having to care: to develop a theory of capital growth, you never need to consult chemical theories; and for a theory of genetics, you never need to consult nuclear physics. 

But on the other hand, we do {\em not} agree that multiple realizability, and it being so widespread, is an objection to a broadly Nagelian conception of reduction. Agreed: it certainly implies that either:\\
\indent \indent (a) the  {\em definiens} (of $T_t$'s multiply realized concept) that is given by $T_b$ is very disjunctive; or \\
\indent \indent  (b) using a non-disjunctive {\em definiens}, the reduction is in an obvious sense {\em local}. \\
We also agree that either (a) or (b) makes for $T_t$ being autonomous in the above sense. For (a) implies that to do science in terms of the vocabulary (the concepts) of $T_t$, while avoiding hopeless complexity and confusion,  we will not care about  the many varied disjuncts in the  {\em definiens}. And similarly, (b) implies that to do science with $T_t$---to describe and further investigate the patterns of co-occurrence of the vocabulary, the concepts, of $T_t$---we will not care about (b)'s single local reduction: for we are focussed on the $T_t$-patterns, that we can see to be realized in many ways other than {\em via} this local reduction. 

But we follow Sober (1999) in maintaining that (a) and-or (b) do {\em not} make trouble for Nagelian reduction. To explain this, it will be clearest to distinguish the contentions: (i) that multiple realisability provides an argument against reduction; and (ii) that it provides an argument against Nagel's account of reduction. The idea of (i) is that the {\em definiens} of a multiply realisable concept (predicate) will have to be so disjunctive that it cannot enter into scientific explanations, and-or cannot enter into laws. The idea of (ii) is that, as we reported in Section \ref{Nagelmodif}, Nagel himself required that the {\em definiens} play a role in the reducing theory $T_b$; and in particular, it cannot be a very heterogeneous disjunction.

Our reply is that (i) is wrong; and that while (ii) is right about Nagel, it hardly matters.  (For more details, cf.  Butterfield (2011: Section 3.1.1, (5), p. 933, Section 4.1.1, p. 941): Dizadji-Bahmani et al. (2010: p. 406) and Schaffner (2012: pp. 543-544) are concordant replies.) As to (i), we endorse a persuasive reply by Sober (1999). Sober says: a disjunctive {\em definiens} in the language of $T_b$ for a concept (predicate) $F$ that occurs in $T_t$ is no bar to a deduction of a law of $T_t$ involving $F$ and other concepts in $T_t$ (perhaps each also with disjunctive {\em definiens} within $T_b$). Nor is it a bar to this deduction being an explanation of the law. Sober sums up this reply as a rhetorical question (1999, p. 552): `Are we really prepared to say that the truth and lawfulness of the higher-level generalization is inexplicable, just because the ... derivation is peppered with the word `or'?' We agree with Sober: of course not!

This reply also provides our reply to (ii). Agreed, and as we reported: Nagel himself vetoed very heterogeneously disjunctive {\em definiens}. Following Sober, we think this was unnecessary. But agreed: if you veto them, one has {\em local} reductions. But we  need have no quarrel with such a veto. As we said in (2) of Section \ref{Nagelmodif}, there is surely no single best sense of `reduction'. And a stronger sense along these lines, i.e. requiring non-disjunctiveness, will unquestionably make for reductions narrower in scope. What really matters, scientifically and philosophically, is to assess, in any given scientific field, just which such reductions hold good, and how narrow they in fact turn out to be. Cf. also Sober (1999, pp. 558-559).  
 
So much by way of discussing multiple realizability, as a  form of Scarcity that threatens reductions based on definitional extensions. Fortunately, as we announced in (3) of Section \ref{224}: the problem will not confront our examples about space and time. In each example, the {\em definiendum} predicate in $T_t$  is {\em not} multiply realizable in the relevant sense. Its instances do {\em not} vary greatly in respects that are mentioned in the {\em definiens} provided in the vocabulary of $T_b$. For example, in the first case-study: although the instances of the {\em definiendum} predicate `... is a freely mobile rigid body' of course differ from one another in their properties, they do {\em not}  differ from one another in the geometric properties that enter in to the {\em definiens} for this predicate.\\
     
(2): {\em Circularity avoided}:--- Finally, we note how Nagelian reduction (more generally: reduction based on definitional extensions) runs up against an apparent problem of logical circularity. This counts as a problem of Scarcity, since if this circularity were vicious, the enterprise would fail---there would be no definitions fit for their purpose. But we will urge that happily, the problem is by and large, {\em only} apparent---thanks to functionalism. 

The problem is simply stated.  When we try to formulate definitions for {\em all} all the non-logical vocabulary (say: predicates) of $T_t$ in terms of that of $T_b$, it seems that the {\em definiens} for a predicate (concept) $F$ of $T_t$ may well need to also use another predicate (concept)  $G$ of $T_t$; while also the {\em definiens} for $G$ needs to use $F$---a logical circle. We saw this illustrated, already in Section \ref{1111}, for logical behaviourism (without the $T_t$/$T_b$ distinction): i.e. for the reduction of mental discourse to material and behavioural discourse, with reduction taken as conceptual analysis, so that definitions must be Faithful to the pre-existing meanings of mental terms.  The problem seems likely to beset the enterprise of finding definitional extensions, whenever the reduced discourse or theory $T_t$ includes a reasonably large or rich set of propositions mixing its predicates (concepts), $F, G$ etc. And all the more likely, if the definitions are required to be Faithful to the given meanings of $F, G$ etc. 

In reply, the first thing to say, of course, is just to admit that `yes, there can be such obstacles to definitional extension’. After all, the advocate of reduction never claimed that definitions of each of $T_t$’s non-logical vocabulary (predicates), in terms of $T_b$’s non-logical vocabulary, can in all cases be constructed so as to derive $T_t$ from $T_b$. (Recall Section \ref{defext}’s definition of `definitional extension’.)   On the contrary, the advocate of reduction recognises that the enterprise of reduction is a risky business---it can fail. Indeed, it can fail even if one allows the definitions not to be Faithful; and even if---in line with the allowances in  the italic schema of Section \ref{problegi}---one seeks only to derive a certain subset of $T_t$, and-or one allows an augmented $T_b$.\footnote{\label{circularimplicit}{This admission is like the point often made by mathematicians and logicians about axiom systems that are said to `implicitly define’ the (non-logical) words within them (or the concepts referred to be those words). Namely, that `implicit definition’ is a misnomer, since in general, one cannot extract genuine definitions of each term from the axiom system. As it is often put: the elementary analogy with solving $n$ simultaneous linear equations for $n$ unknowns is misleading. We shall return to this, especially  in Section \ref{implicit}.}}

Besides, the problem was historically influential: advocates of reduction recognised it. Think of how, in philosophy of mind, the threat of circularity was lodged as an objection to logical behaviourism. And in philosophy of science, the logical empiricists’ efforts to reduce theory to observation (example (iv) in Section \ref{examp}) were beset by it. To take one famous example: think of how in his `Testability and meaning’ Carnap lowered his sights from defining the (putatively theoretical) predicate `is soluble’ in terms of observational predicates such as ‘is in water’ and `dissolves’, and proposed only what he called `reduction sentences' such as, in logical notation:  $(x)[InWater(x) \supset [Soluble(x) \equiv Dissolves(x)]]$ (1936: pp. 439-444) (Cf. also Braithwaite (1953: 66-68, 76-79).)    

But as we announced in Section \ref{111}: in fact, {\em all is not lost}. That is:  in many cases---scientifically important cases, and philosophically important cases---the threat of circularity is only apparent.  This is the distinctive insight of functionalism: each concept of the problematic discourse or theory is vindicated (and so claims involving them can be vindicated) by displaying a {\em simultaneous definition} of each of them, in terms of the  unproblematic concepts. 

But beware: this insight needs to be stated carefully, so as to respect the distinction between the binary and ternary contrasts of vocabularies, introduced already in Section \ref{intro}; i.e. the distinction between the first and second steps of the Canberra Plan (cf. (3) in Section \ref{1112}). That is: the functionalist idea of simultaneous definitions of each of many terms applies in a single given theory (which we called the `initial theory’); whose vocabulary is divided into:\\
\indent \indent (i) terms that are taken as unproblematic (understood); and \\
\indent \indent (ii) others that are each defined (simultaneously) as the unique occupant of a certain role spelt out using the terms in (i). \\
This is the binary contrast. On the other hand, there can be {\em another} theory, accepted later than or independently of the first: a theory that also specifies these occupants---in terms different from both the classes (i) and (ii). In this situation, i.e. if there is such a theory, we have a ternary contrast of vocabularies---and the possibility of derived bridge laws (`definitions’ in our weak sense) that imply a reduction.  

So the care is needed because: \\
\indent \indent (a) we philosophers all first learn about functional roles, and simultaneous functional definitions etc., in the context of a single theory, with examples like logical behaviourism and the empiricist attempt to define theoretical terms in terms of observational ones---all a matter of the binary contrast, and the first step of the Canberra Plan: while on the other hand, \\
\indent \indent (b) the jargon of reduction, and our mnemonic notations $T_t$ and $T_b$ (`t’ for `top’, `b’ for `bottom’), and the usual examples like thermodynamics being reduced to statistical mechanics (which was Nagel’s main example), all fit the situation where another theory is added, later or independently---all of which means one is considering a ternary contrast, and the second step of the Canberra Plan.

So much by way of a warning. We now spell out, first, the functionalist insight about simultaneous definitions, with its binary contrast (Section \ref{homefunc});  and then, functionalist reduction, with its ternary contrast (Section \ref{DKL4}).

\section{Functional roles and simultaneous definitions}\label{homefunc}

 In Section \ref{111}, we introduced functionalism by briefly explaining, in turn, the ideas of:\\
 \indent \indent (a) a functional role, e.g. of a mental state or concept, and its realizer or occupant;\\
 \indent \indent  (b) simultaneous unique definitions of those occupants;\\
 \indent \indent (c) a theoretical identification, e.g. of a mental state with a brain state, being compulsory, i.e. being the conclusion of a valid argument with true premises (premises that describe the unique occupant of a functional role in two ways); rather than just recommended as ontologically parsimonious---so that functionalism gives reductions.

In this Section and the next, we give more details: (a) in Sections \ref{DKL1} and \ref{DKL1A}, (b) in Section  \ref{DKL2}, and (c) in Section \ref{DKL4}. So as we warned at the end of Section \ref{rednobstacle}: (a) and (b), and so all of Section \ref{homefunc}, will be about a  binary contrast of vocabularies within a single theory. But Section \ref{DKL4} will return us to the scenario of two theories, the second accepted later than or independently of the first: the usual scenario of reduction, though now involving a ternary contrast of vocabularies

\subsection{Functional roles}\label{DKL1}

  Undoubtedly, many concepts are {\em functional}, in philosophers' sense of that term. That is: {\em the concept is individuated by---i.e. uniquely specifiable by---its pattern of relations to other concepts}.  This pattern is then called the concept's {\em functional role}; and the concept is the {\em occupant} or {\em realizer} of the role. Following our usage since Section \ref{problegi}, we shall often say, instead of `specification': `definition'---without connotations of synonymy.
    
  As we have discussed, pain is the standard example from the philosophy of mind. A sketch of the functional role is: to be in pain {\em is} to be in a state that is (a) typically caused by tissue damage, (b) typically causes aversive behaviour, i.e. avoidance of the damage's cause, and (c) is related in such-and-such ways to other mental states, for example, implying the emotion of distress, and typically causing both belief that one is in pain and an intention to avoid the damage's cause. 
  
Of course, details vary as regards the four main words: (1) `concept', (2) `relations', (3) `definition' (`specification') and (4) `(unique) occupant'; (though the ensuing variety in versions of functionalism need not spell incompatibility). We will treat these in order, with `(unique) occupant'  getting a Section of its own (Section \ref{DKL1A}).\\

(1): {\em The concept}:  The concept at issue can be a property, or what many would rather call a way of thinking, a mode of presentation, of a property; (often called `a concept of the property'). In some examples  (including that of pain), what is specified is more naturally called a `state', or `state of affairs'. As we said in footnote \ref{fnstateconcept} (and also touched on later, e.g. (1) of Section \ref{Faith}), this variety is not worrisome. For  `concept , `property' and `state' are philosophical terms of art, and we will not need to decide on an exact usage:  nor on exact criteria of individuation, nor on an exact range of possibilities across which the concept or what-not is claimed to be uniquely specified. What is important for us, both in this paper and our others, is just that the concept or property specified:\\
\indent \indent (i)  can be a physical quantity: our other papers will have examples like being freely mobile (a property of bodies), and being simultaneous (a relation between events);\\
\indent \indent (ii) can have  properties and relations (maybe many such) beyond those by which one specifies it: this of course  opens the door to it being specified in a different, independent, way---and thus to  functionalist reduction (cf. Section \ref{DKL4}).\\
We emphasise that there is nothing problematic about (ii). Just think of an attributive definite description such as `the tallest Swede now alive'. Whoever that person is, he or she has countless properties not encoded in that description: properties which---if only we knew them---we could use to refer the person. Such examples make vivid how  the idea of specifying something by its pattern of relations to other things applies equally well to concrete objects, like people, as to concepts and properties. Think of drama: Ian McKellen is the occupant of the role, Macbeth; and Judi Dench the occupant of the role, Lady Macbeth (both in a famous production in the 1970s).  Similarly, Lewis' own exposition sketches a detective story: the detective uniquely specifies the culprit or culprits by their actions and their relations to other people and things (1972, p. 250-251: cf. Section \ref{DKL2parable} below).\\

(2): {\em The relations}: The relations to other concepts can be a matter of: either \\
\indent \indent  (a) causation, whether deterministic or probabilistic (as suggested by the `typically caused' in our sketched functional role for pain); or \\
\indent \indent (b) law-like association, i.e. Hume's `constant conjunction' of properties without the interpretation as causation (as in many quantitative non-causal putative laws, e.g. Ohm's law); or \\
\indent \indent (c) logical relations, such as implication (e.g. being in pain implies being in distress---as in the sketch above). \\
Of these three types of relation, philosophers focus on (a) and (b); and accordingly, often say `causal-nomological role' instead of `functional role'. But we shall keep to the phrase, `functional role': not least because in the roles figuring in our other papers' spacetime examples, logical relations like implication will be {\em much} more prominent than causal relations. Note also that when a functional role invokes a relation of any of these types, (a), (b) and (c), it may well be qualified by a `typically' or `normally'. Thus our sketch for pain said `typically caused by tissue damage'. (Besides, these words `typically' and `normally', and especially the latter, might mean more than a simple statistical majority: `normal' connotes `norm' and thus `goal' or `purpose'---for example in biology, the proper functioning of the organism.  In this way, the `function'  in `functionalism' has a slight connotation of purpose. But though this is a theme in philosophy of biology, it will not be an issue for us.) \\ 

(3): {\em The definition}:  In our logically weak usage (since Sections \ref{1112} and \ref{problegi}), `definition' need not be either (a) stipulatively defining a new word, or (b) conceptual analysis: where (a) and (b) deploy some agreed notion of logical or metaphysical equivalence. A definition can merely state a reference (extension) for the word, whether  pre-existing or new, that is unique, i.e. unambiguous, for the context and purposes at hand. Thus the context and purposes may require much less than faithfulness to a pre-existing use, and-or much less than a statement of the reference in all ``possible worlds'', i.e. much less than uniqueness on some agreed notion of logical or metaphysical equivalence. Cf. Section \ref{Faith} on allowing a little Faithlessness. What matters is only that the definition coheres with the claims using the word that we hold true in the context and purposes at hand.\footnote{Agreed: if we are undertaking a reduction with a definitional extension, we ask for more: the definition, taken together with others and with the claims of the reducing theory, must imply the claims of the reduced theory. Cf. Section \ref{defext}.}  

This logically weak usage may suggest that such definitions are not much of a topic: small beer, as we say in England. But not so. And not just because of the idea (coming up in Section \ref{DKL2}) of simultaneous definitions of several words. For even when aiming---within a limited context, and allowing a little Faithlessness---for just one definition, it can be a very considerable achievement to give a definition of a word using the vocabulary prescribed by the context, that coheres with the claims  we hold true. 

We already touched on this in footnote \ref{toil}, when endorsing Oliver and Smiley's point, {\em a propos} of Dedekind, about `just how much honest toil it takes to discover—to formulate precisely---just what it is that we want’ (2016, p. 272). Their example is Dedekind's definition of natural numbers in terms of the successor relation, by what are now called `Peano's axioms'. For details, including a comparison of Dedekind, Peano and Frege, and also Dedekind's categoricity theorem (i.e. that all models of the axioms are isomorphic), cf. Potter (2000, pp. 81-89). 

Furthermore, our other papers will give examples in geometry and spacetime physics. For example: the definition of `freely mobile' for rigid bodies, that was articulated by Helmholtz, Lie and their successors, and the definition of `simultaneity'  for special relativity in terms of causation, that was articulated by Malament (following work by Robb and others), are both considerable achievements. Although the context is limited (viz. rigid bodies and special relativity, respectively), and only one word or phrase is defined, it is a considerable achievement   to {\em prove} uniqueness of reference---analogous to Dedekind's categoricity theorem. This leads in to Section \ref{DKL1A}.

\subsection{The unique occupant}\label{DKL1A}
The claim that there is a unique occupant (realizer) of a given functional role can be challenged in various ways: and of course, most of the variety will arise from the specific details of the role we are considering. Much that one might say by way of challenge, and much one might say by way of response, is straightforward, or even common sense: especially when the functional role is of a concrete object like a person. 

Challenge: `There are many players of the role `Macbeth'. Besides, one might say there is no single role, since interpretations of the part vary'. Response: `But I only claim uniqueness relative to the context of a specific production, with the role's interpretation agreed by the director and players'. Challenge: `But what about understudies?' Response: `OK: I only claim uniqueness for a specific production, on a specific night'. 

Similarly, about philosophically interesting concepts or properties like pain.   Challenge: `Pain is multiply realized: there are many players of the pain-role. Besides,  one might say there is no single pain-role, since there is flexibility and vagueness about what to include in it.\footnote{Recall the traditional objection to logical behaviourism, that it ruled out what seems possible:  for example, the conjunct in our sketched pain-role, `typically causes aversive behaviour’, seems to rule out perfect-actor ``super-Spartans'' who never flinch when in pain.} Response: `But I only claim uniqueness relative to a kind, which may well be narrower than a species. As Lewis says:  `[pain] might even be one brain state in the case of Putnam, another in the case of Lewis. No mystery: that is just like saying that the winning number is 17 in the case of this week's lottery, 137 in the case of last week's' (1969, p. 25): cf. (1) in Section \ref{Scarc}.

There is also the Challenge of no realizers, rather than many. Consider eliminativism about folk-psychological propositional attitudes like belief, or about the theoretical entities (whether properties, relations or objects)  of yesteryear's theories. Challenge: `Nothing fits the role of belief, or desire; or the role of phlogiston, or caloric'. Again, there are two straightforward Responses, each of which is surely right in many cases. First, one just agrees with eliminativism: the role in question is unrealized, and the theory or discourse it comes from, should be abandoned (cf. the `cognitive rubbish' response in Section \ref{problegi}).  Second, one says that the role has near-realizers, and one of the nearest is near enough to deserve the name in question (`belief’ or `phlogiston’, as the case may be). And of course, one must admit that there can be ambiguity and vagueness: perhaps none of the near-realizers counts unambiguously as nearer than all the others. 

Agreed; both these Responses look like qualifications of functionalism’s leading idea, viz. unique realizers of functional roles. But no worries. For of course, no functionalist claims that any role, any pattern of relations, that you can extract from (think of, or write down, within) any theory or discourse, must have a unique realizer.  The claim is instead (as we shall see in detail in Section \ref{DKL2}) that the advocate of a theory claims that it is sufficiently informative (logically strong) that each of the entities (whether properties, relations or objects) that it newly introduces is the unique realizer of a pattern of properties and relations that can be extracted from the theory. Besides, the pattern can be extracted systematically, in the same way for all the new entities. To put the claim in terms of language, not entities: the functionalist claims that each of the terms the theory newly introduces can be defined by the term's pattern of occurrence in all the assertions of the theory. Besides, these definitions can be extracted systematically from the theory, and presented simultaneously.

For our purposes, especially in our other papers, what matters is that the claim of a unique occupant (realizer) holds good: indeed, {\em provably so}. To put it in terms of the Challenges, the straightforward Responses apply: for instance, one responds to multiple realization by limiting the context enough to prove the uniqueness. 

Let us illustrate with the example of simultaneity, mentioned at the end of Section \ref{DKL1}. In the limited context of special relativity, there is a theory about causal connectability of spacetime points that is sufficiently informative that one can prove there to be a unique equivalence relation satisfying certain conditions that, most would agree, are part of the meaning of the term `simultaneity'. To put it in the jargon of functionalism: one proves that there is a unique occupant of the simultaneity-role; and the term `simultaneity' can be thereby defined; besides, the definition is Faithful to the pre-existing meaning of the term. And for our papers’ other examples, the situation will be similar: one proves unique occupancy of the relevant role. (Or in some cases: uniqueness {\em modulo} a choice of units and coordinate system---a controlled and welcome Plenitude; cf. (2) in Section \ref{222}.)

But agreed: not all that one might say hereabouts, by way of challenge to the  functionalist claim of unique occupancy, or by way of response, is straightforward. Already in (4) at the start of Section \ref{homered}, and at the end of (1) in Section \ref{222}, we acknowledged that Putnam’s model-theoretic argument, and Newman’s objection to Russell’s structural realism, make a deep and general challenge about reference: roughly, that for realists like us (or Lewis), reference and truth are all too easy to attain. (In (1) of Section \ref{222}, the specific challenge was a problem of Plenitude: the plethora of mock-ups.\footnote{As we noted in (3) in Section \ref{224} and footnote \ref {MRscarce2}: the labels `Plenitude' and `Scarcity' can be confusing in the present context. For they were introduced in Sections \ref{222} and \ref{223} as about, respectively, having too many, or no, definitions. But in this Section, the focus has been on having too many, or no, occupants (realizers) of a definition: a different topic.})  In reply, we emphasised that there are several cogent responses to these challenges, between which we (and our other papers) do not have to choose; (similarly, cf. footnotes \ref{StrucPotter} and \ref{Newman} about postponing structural realism). But here, in the more specific context of functionalism, we should give more detail about our views on reference. For in the next Section we will report Lewis’ (1970, 1972) expositions, especially of functional definition. And since it assumes that the $O$-terms used by a functional definition---the {\em old} or {\em original} terms: what Section \ref{problegi} called `unproblematic terms’---have a reference, we owe some discussion of how {\em they} get a reference. 

In short, our answer is twofold: partly Lewisian and partly not. The Lewisian part is: {\em causal descriptivism}, so labelled (and endorsed) by him in his reply to Putnam’s argument (1984: pp. 226-227). That is: a descriptivist account of reference that: \\
\indent \indent  (i) takes to heart various points engendered by the criticisms launched by the causal theory of reference (1984: p. 223); and \\
\indent \indent  (ii) couches its descriptions in largely causal terms---and thereby, urges Lewis, often accounts for puzzle cases at least as well as a causal theory.

But care is needed. {\em Global descriptivism} is the view that for {\em all} the terms of our language, their reference is determined solely by requiring any assignment of candidate referents to render true, {\em according to that assignment}, whatever we assert. Evidently, this view will make truth all too easy to attain. And indeed: Lewis argues that the lesson of Putnam’s model-theoretic argument is precisely that global descriptivism is {\em false} (1984: 224, 226). Again, we agree: we read Putnam the same way.  Besides, we agree with Lewis that this conclusion is {\em not} avoided by making one's global descriptivism also causal, i.e. by also adopting causal descriptivism's (ii) above. That would be of no avail; since for global descriptivism, this tactic amounts to invoking `just more description, just more theory'. So an advocate of causal descriptivism, such as Lewis or us, must keep their descriptivism {\em non}-global; and so, in order to answer Putnam, they  `must seek elsewhere [than causation] for the saving constraint'  (p. 227).   

This is the point at which Lewis turns to his theory of natural properties (expounded in his 1983)---and at which we part company with him. For while we admire the way this theory fulfils several needs in his overall metaphysical system, it uses a notion of similarity this is given once for all, across all of modal reality. In particular, it is not relativized to either a theory or a possible world. This we admit that we cannot believe. And so as we said in (4) of Section \ref{homered}, we must seek reference's `saving constraint' in some other direction.

The upshot is that like Lewis, we endorse causal descriptivism and deny global descriptivism (even using causal descriptions); but as a reply to Putnam’s model-theoretic argument, we would opt for a less gung-ho realism than Lewis'.\footnote{ A wrinkle about terminology. Nowadays, some (e.g. Janssen-Lauret and McBride (2020)) use `global descriptivism’ for the weaker doctrine that the reference of all terms is settled {\em en bloc} by  total theory. This is weaker than our definition above, since it makes no claim that the only constraint on reference-assignment is making true whatever we assert. So it can be combined with the sort of `reference magnetism’ Lewis espoused; or with some other constraint that is not `just more theory’. So there is no disagreement here. } Obviously, this is not the place to develop this position. We have no space; and anyway, our position is not unusual. Causal descriptivism has been defended both in philosophy of language (e.g. Kroon 1987), and in philosophy of science, specifically as part of defending scientific realism (Psillos 1999, 2012). (Our endorsement of scientific realism will return in Section \ref{roberto}.)

\subsection{Simultaneous unique definitions}\label{DKL2}
 In Sections \ref{111} and \ref{Scarc}, we saw how all the concepts in some relevant set being definable or specifiable by some of their relations to each other and to other concepts faces a threat of logical circularity. (For the choice between `definition' and  `specification', cf. (1) in Section \ref{1112}, the start of Section \ref{problegi} and (3) in Section \ref{DKL1}.) For if each of two concepts is to be defined, in part, by its relations to the other, we apparently cannot define either of the concepts without a circularity. Similarly for more than two concepts: there could be a sequence of putative definitions of concepts, where the last definition invokes the concept  defined by the first---a logical circle. Besides, in branches of philosophy where functionalism has been attractive, such logical circles seem all too likely. Recall the example of how a logical behaviourist might try to define belief that it is raining.

But in fact, this threat can be answered. We {\em can} make perfectly good sense of simultaneously defining, with no logical error, each concept in a set. We only need all the definitions to be implied by a sufficiently rich body of information, which spells out the functional roles of each concept in the set. And this requires simply that the body of information be true only if for each such role, there is a unique occupant or realizer of that role.  

As so often, it is in Lewis' work that the idea is stated most clearly. So this Section summarizes his exposition. We begin with a parable from his (1972): Section \ref{DKL2parable}. This leads to the Ramsey and Carnap sentences of a theory, and how to modify them: (Section \ref{DKLsentences}; also drawing on his (1970)). This stage-setting prepares us for the {\em denouement}: explicit simultaneous definitions of many terms---partly in terms of each other: Section \ref{DKLdefns}.  

\subsubsection{A parable}\label{DKL2parable}
Lewis (1972) begins with an example of simultaneous unique specifications of, not concepts, but people---in a country-house detective story.
\begin{quote}
We are assembled in the drawing room of the country house; the detective reconstructs the crime. That is, he proposes a theory designed to be the best explanation of phenomena we have observed: the death of Mr.~Body, the blood on the wallpaper, the silence of the dog in the night, the clock seventeen minutes fast, and so on. He launches into his story: 
\begin{quote}
$X, Y$ and $Z$ conspired to murder Mr. Body. Seventeen years ago, in the gold 
fields of Uganda, $X$ was Body's partner \dots Last week, $Y$ and $Z$ conferred in 
a bar in Reading \dots Tuesday night at 11:17, $Y$ went to the attic and set a 
time bomb \dots Seventeen minutes later, $X$ met $Z$ in the billiard room and 
gave him the lead pipe \dots 
\end{quote}
And so it goes: a long story. Let us pretend that it is a single long conjunctive
sentence  \dots

Suppose that after we have heard the detective's story, we learn that it is
true of a certain three people: Plum, Peacock and Mustard [respectively] \dots
 
We will say that Plum, Peacock and Mustard together {\em realize}
(or are a {\em realization} of) the detective's theory \dots We may also find out that the story is not true of any other triple \dots

In telling his story, the detective set forth three roles and said that they were
occupied by $X, Y$ and $Z$. He must have specified the meanings of the three terms `$X$', `$Y$' and `$Z$' thereby ... They
were introduced by an implicit functional definition, being reserved to name
the occupants of the three roles. When we find out who are the occupants
of the three roles, we find out who are $X, Y$ and $Z$. Here is our theoretical
identification. (1972, p.~250-251)
\end{quote}
The point is crystal-clear, especially from the last paragraph. In short: we interpret the detective's story as truly indicting Plum, Peacock and Mustard iff they as a triple are the unique realization of it; and it is true, i.e. true of some or other trio of culprits, iff some triple of people are the unique realization of it.\footnote{\label{nearrealize}{Here, `true' means `completely true'.  This logical strength prompts a clarification. Shortly after the quoted passage, Lewis writes:
\begin{quote}
 A complication: what if the theorizing detective has made one little
mistake? He should have said that $Y$ went to the attic at 11:37, not 11:17.
The story as told is unrealized, true of no one. But another story is realized,
indeed uniquely realized: the story we get by deleting or correcting the little
mistake. We can say that the story as told is {\em nearly realized}, has a unique
{\em near-realization} ... In this case the T-terms ought to name the components
of the near-realization ... But let us set aside this complication for the sake of simplicity, though we know well that scientific theories are often nearly realized but rarely realized, and that theoretical reduction is usually blended with revision of the reduced theory.
\end{quote}
Well said. Indeed in (1) of Section \ref{Nagelmodif}, we saw Nagel, Schaffner and Fletcher also say what `we know well'.
}}

\subsubsection{Ramsey sentences and Carnap sentences---modified}\label{DKLsentences}

With a little notation, we can compendiously state both Lewis' general idea and the battery of explicit definitions.\footnote{In this Section and the next, we are very indebted to Adam Caulton: whose insightful  comparison of Lewis' and Carnap's views we have regretfully suppressed, for the sake of space.}  As in Lewis' parable (and his 1970), we adopt the notation: $T$-terms and $O$-terms. But  as Lewis says: `$T$-term' need not mean `theoretical term', and `$O$-term' need not mean `observational term'. In the parable, he wrote: `$O$ does not stand for `observational'. Not all the $O$-terms are observational terms, whatever those may be. They are just any old terms' (1972, p. 250). Similarly in his (1970), he writes:
\begin{quote}
I do not understand what it is just to be a theoretical term, not of any theory in particular, as opposed to being an observational term (or a logical or mathematical term).[A footnote endorses Putnam's article `What Theories Are Not'.]  I believe I do understand what it is to be a $T$-term: that is, a theoretical term introduced by a given theory $T$ at a given stage in the history of science. If so, then I also understand what it is to be an $O$-term: that is, any {\em other} term, one of our {\em original} terms, an {\em old} term we already understood before the new theory $T$ with its new $T$-terms was proposed. An $O$-term can have any epistemic origin and priority you please. It can belong to any semantic or syntactic category you please. Any old term can be an $O$-term, provided we have somehow come to understand it. And by understand I mean ``understand''---not
``know how to analyze.'' (1970, p. 428)
\end{quote}
So despite the letters `T' and `O', Lewis' proposals are not only about the theory-observation distinction. Indeed, these papers' influence in the years since 1970 has led to the framework of functional definition being applied to many of philosophy's contrasts between what Section \ref{problegi} called `the problematic' vs. `the unproblematic': for instance, the ethical vs. factual contrast, as in footnote \ref{Hurley}. In short, it has led to the Canberra Plan.  

Happily, Button and Walsh (2018: p. 55)  suggest a helpful mnemonic to replace the overly restrictive, and misleading, `theory' and `observation': namely, `T' stands for {\em troublesome}, and `O' stands for {\em okay}. We will from now on use this mnemonic. But the main point is as in the quote: the $O$-terms are understood, their reference is settled ({\em modulo} the deep issues set aside at the end of Section \ref{DKL1A}!);  while the $T$-terms are  yet to be understood, either because they are new or because they are problematic.  

So we begin by assuming we have a theory $T$, for which there is some distinction between troublesome and okay terms: $T$-terms and $O$-terms. We take $T$ as a long conjunction of claims, which we call the {\em postulate}. The leading idea will be to use the patterns of relations to the $O$-terms, that the $T$-terms enjoy, to fix the meanings of the $T$-terms. And this will involve commitment  to unique realizations, which will yield explicit definitions, one for each $T$-term.\footnote{So in our other papers, we will use this notation in examples of defining a spacetime structure as the unique realizer of some role. Broadly speaking: it will be vocabulary about the physics of matter and radiation that are the $O$-terms, and vocabulary about chrono-geometry that are the $T$-terms.} 

In the theory $T$, let us make all the theoretical terms $t_1, \ \ldots\ t_n$ used in $T$ explicit by writing  $T(t_1, \ldots t_n)$. Here, we treat all theoretical terms as first-order, i.e. as names. This is as in Lewis, who says that this choice
\begin{quote}
 `is of no importance. It is a popular exercise to recast a language so that its non-logical vocabulary consists entirely
of predicates; but it is just as easy to recast a language so that its non-logical
vocabulary consists entirely of names (provided that the logical vocabulary includes a copula). These names, of course, may purport to name individuals, sets, attributes, species, states, functions, relations, magnitudes,
phenomena or what have you; but they are still names. Assume
this done, so that we may replace all $T$-terms by variables of the same sort' (1972, p. 253).\footnote{\label{nonchalant}{This may seem nonchalant: Lewis (1970, p. 429) gives a longer defence of this choice. But agreed: questions remain, both in general (e.g. `How does this bear on the usual construal of Newman's objection as depending on the Ramsey sentence using second-order quantifiers?') and for our own advocacy of spacetime functionalism (e.g. `In our examples, can the defined spacetime structures, e.g. a metric, be named?'). We address these questions in our other papers.}}
\end{quote}

Now form the \emph{realization formula} of $T$ by replacing each $T$-term in all its occurrences with the same variable, $x_1, \ldots x_n$ respectively (where we of course assume that none of $x_1, \ldots x_n$ already occur in $T$, even as bound variables). Call this formula: $T(x_1, \ldots x_n)$. Any $n$-tuple of entities that satisfies the realization formula is a \emph{realization} of $T$.  The entities \emph{realize} $T$. 

The \emph{Ramsey sentence} of $T$ is just the claim that $T$ is realized. That is to say, it is: 
\be\label{RamSent1}
\exists x_1, \ldots , x_n T(x_1, \ldots , x_n).
\ee
 The Ramsey sentence has exactly the same purely okay consequences as $T$; i.e.~it has as consequences exactly the same sentences containing \emph{only} okay terms as does $T$.  But the Ramsey sentence has no purely troublesome consequences---\emph{a fortiori} no purely troublesome consequences entailed by $T$---since it contains no troublesome vocabulary.\footnote{To be precise: its purely troublesome consequences are merely all the logical truths (logically valid formulas) containing only $T$-terms.}  So the Ramsey sentence serves as a completely adequate okay surrogate for the original theory $T$. 

Thus the residual content of $T$ lies in what $T$ says by way of ``implicit definitions'' of the $T$-terms.  That seems to be encapsulated in what is called the \emph{Carnap sentence} of $T$. This says that if $T$ is realized, then it is realized by $t_1, \ldots , t_n$.  It says: if anything realizes $T$, then  $t_1, \ldots , t_n$ does. Thus the \emph{Carnap sentence} of $T$ is:
\be\label{CarSent1}
\left[(\exists x_1, \ldots , x_n T(x_1, \ldots , x_n)) \supset T(t_1, \ldots , t_n) \right].
\ee
Conjoining the Carnap sentence with the Ramsey sentence entails $T$. But the Carnap sentence entails no purely okay sentences except logical truths (logically valid formulas). So its role is apparently to interpret the troublesome terms. Thus Lewis writes: `it does seem to do as much toward interpreting them as the postulate itself does. And the Ramsey and Carnap sentences between them do exactly what the postulate does' (1970, p.~431). But Lewis will shortly argue that these sentences need to be {\em modified}. 

First, let us press the question: what do the variables $x_i$ range over?  As we know, Lewis  is a realist. He sees no ontological distinction between ``observable" and ``theoretical" {\em entities}: where `a theoretical entity is something we believe in only because its existence, occurrence, etc. is posited by some theory---especially some recent, esoteric, not-yet-well-established scientific theory' (1970, p. 428).\footnote{Note that although `entity' connotes `object', and Lewis mentions some objects as his examples of theoretical entities, viz. living creatures too small to see and the dark companions of  stars,  Lewis' taking all $T$-terms as names means that properties and relations, like physical magnitudes or a spacetime metric, will count as theoretical entities. Cf. the end of footnote \ref{fnstateconcept} and footnote \ref{nonchalant}.} He admits only a distinction between troublesome (i.e. yet-to-be-understood, e.g. because newly introduced), and okay (i.e. already understood), {\em terms}. Thus he writes
\begin{quote}
I am also not planning to ``dispense with theoretical entities."
Quite the opposite. The defining of theoretical terms serves the cause of scientific realism. A term correctly defined by means of other terms that admittedly have sense and denotation can scarcely be regarded as a mere bead on a formal abacus. ...  Theoretical entities are not entities of a special category, but entities we know of (at present) in a special way. (1970, p.~428)
\end{quote}
So pinning down newly-introduced entities using a theory is like the detective in the parable identifying a suspect. We should think of the entities corresponding to the $T$-terms as \emph{out there, somewhere}, just like the people $X, Y$ and $Z$: we just need sufficient detail in our theory to pick them out. 

In this endeavour, Lewis distinguishes three cases: {\em unique, failed, and multiple, realization} (1970, pp. 431-433). In presenting Lewis' parable, we emphasised the first of these. But the situation will also be clear enough in the other cases.
\begin{enumerate}
\item \emph{$T$ is uniquely realized.} Here, the Ramsey sentence is true; and the Carnap sentence `clearly gives the right specification' (p. 431), i.e. determines the correct unique referents for $t_1, \ \ldots , \ t_n$.

\item \emph{$T$ is not realized.} Then `the Carnap sentence says nothing about the referents of the $T$-terms' (p. 432). But here, we should distinguish two sub-cases:
\begin{enumerate}
\item $T$ is nowhere near being realized.  Lewis gives a familiar example: we know what phlogiston is supposed to be, or what it would be were it to exist, even while we also know that it doesn't exist.  So in this sub-case, the Carnap sentence's silence seems wrong, i.e. overly modest. Where it is silent, one should be opinionated: one should say that the $T$-terms do not refer.   

\item  $T$ is \emph{nearly} realized. That is: We want to say that a certain $n$-tuple comes near enough to realizing $T$. Then as in the  quotation from Lewis in footnote \ref{nearrealize}, the $T$-terms surely name the components of the near-realization. In such a case, the Carnap sentence can probably be regarded as correctly specifying the referents, viz. by taking the term-introducing theory to be some theory $T'$ that is similar to $T$ (maybe logically weaker, i.e. implied by $T$) and that \emph{is} realized by $T$'s near-realization.  
\end{enumerate}

\item \emph{$T$ is multiply realized.} Here again, the Carnap sentence seems  wrong to Lewis, because unduly instrumentalist. He writes:
\begin{quote}
In this case, the Carnap sentence tells us that the $T$-terms name the components of some realization or other. But it does not tell us which; and there seems to be no non-arbitrary way to choose one of the realizations. So either the $T$-terms do not name anything, or they name the components of an arbitrarily chosen one of the realizations of $T$. Either of these alternatives concedes too much to the instrumentalist view of a theory as a mere formal abacus. Neither does justice to our naive impression that we understand the theoretical terms of a true theory, and without making any arbitrary choice among realizations. We should not accept Carnap's treatment in this case if we can help it. Can we? (1970 p.~432).
\end{quote}
\end{enumerate}

In short, Lewis' idea here is that unique realization is preferable to multiple realization, and we ought to say that multiply realized theories have denotation-less $T$-terms. He also urges two reasons why in scientific theorizing, unique realization is a reasonable, not extravagant, hope (1970 pp.~433):\\
\indent \indent (i) `I am not claiming that there is only one way in which a given theory \emph{could} be realized; just that we can reasonably hope that there is only one way in which it \emph{is} realized.'  So in the jargon of possible worlds: multiple realization is still allowed \emph{across possible worlds}, just not within a world. (This leads to some subtleties we need not report---since they are not defects, though they have influenced philosophers' usage of the word `functionalism'; cf. pp. 436-437, Lewis (1994, p. 419-421) and our footnote \ref{Lewis94}.) In fact, the $T$-terms are probably \emph{not}  rigid designators (`logically determinate names'; cf. also p. 435-436).\\
\indent \indent  (ii) The $O$-terms, whose interpretations are fixed, are a large and miscellaneous set; specifically, they are not confined to being observational. (This point yields an answer to a theorem of Winnie.) \\
Here, we would add a third reason to be hopeful. Namely: unique realization does not require $T$ to establish \emph{every} fact about the $T$-term's referents: just enough facts to secure uniqueness.  There can be plenty of room for us to discover more about these referents. This of course leads to the {\em ternary} contrast of vocabularies, and the {\em functionalist reduction} that we advertised in Section \ref{intro}: which we develop in Section \ref{DKL4}.\\

We can sum up this discussion---specifically, the proposal that with no or multiple realization, the $T$-terms do not refer---as proposed modifications of the Ramsey and Carnap sentences. Thus the Ramsey sentence eq. \ref{RamSent1}  is to be replaced by the {\em modified Ramsey sentence}, which says that $T$ is \emph{uniquely} realized:
\be\label{ModRamSent}
\exists y_1, \ldots , y_n\forall x_1, \ldots , x_n \left[T(x_1, \ldots , x_n) \equiv (y_1 = x_1\ \&\ \ldots\ y_n = x_n)\right].
\ee
Thus Lewis writes:
\begin{quote}
The Ramsey sentence has exactly the same $O$-content as the postulate of $T$;
any sentence free of $T$-terms follows logically from one if and only if it follows from the other.[footnote suppressed] The modified Ramsey sentence has slightly more $O$-content. I claim that this surplus $O$-content does belong to the theory
$T$---there are more theorems of $T$ than follow logically from the postulate alone. For in presenting the postulate as if the $T$-terms have been well-defined thereby, the theorist has implicitly asserted that $T$ is uniquely realized. (1972, pp. 253-254)
\end{quote}

The Carnap sentence is to be replaced by three postulates, as follows: which together say that  $T$ is uniquely realized iff it is realized by $t_1, \ldots , t_n$.  (Here, it is important to adopt a system of logic that allows bearerless terms, so that a formula such as $(\exists x)(x=t_1)$ is not a logical truth. Lewis adopts a system by Scott (1970, p. 430).)
\begin{enumerate}
\item If $T$ is uniquely realized, then it is uniquely realized by $t_1, \ \ldots ,\ t_n$:
\be\label{ModCarSent}
\left(\exists y_1, \ldots , y_n\forall x_1, \ldots , x_n \left[T(x_1, \ldots , x_n) \equiv (y_1 = x_1\ \&\ \ldots\ y_n = x_n)\right]) \supset T(t_1, \ldots t_n)\right).
\ee
Lewis calls this the {\em modified Carnap sentence} (1972, p. 254); it is logically implied by the Carnap sentence.

\item If $T$ is not realized at all, then $t_1, \ \ldots\ t_n$ don't refer:\\
$\left(\neg\exists x_1, \ldots , x_n T(x_1, \ldots , x_n) \supset \left(\neg\exists x\ x = t_1\ \&\ \ldots\ \neg\exists x\ x = t_n\right)\right)$

\item If $T$ is multiply realized, then $t_1, \ \ldots\ t_n$ don't refer:\\
$(\exists x_1, \ldots , x_n T(x_1, \ldots , x_n) \ \&$\\
$ \neg\exists y_1, \ldots  y_n\forall x_1, \ldots  x_n \left(T(x_1, \ldots x_n) \equiv (y_1 = x_1\ \& \ldots\ y_n = x_n)\right)) \supset \left(\neg\exists x\ x = t_1\ \& \ldots\ \neg\exists x\ x = t_n\right)$.
\end{enumerate}
Lewis (1972) introduces a helpful concise notation. He uses {\bf boldface} names and variables to denote $n$-tuples. So our original statement of the theory that displayed the $T$-terms, $T(t_1, \ldots , t_n)$, is now written as  $T[{\bf t}]$; the   realization formula $T(x_1, \ldots , x_n)$ is written as  $T[{\bf x}]$; and the Ramsey sentence is written as $\exists {\bf x} T[{\bf x}]$. So Lewis' modified Ramsey sentence, eq. \ref{ModRamSent}, is now written as:
\be\label{ModRamSentBold}
\exists_1 {\bf x} T[{\bf x}]; \;\;  \mbox{that is:} \; \exists {\bf y} \forall  {\bf x} (T[{\bf x}] \equiv {\bf y} = {\bf x});
\ee
and similarly, the modified Carnap sentence, eq. \ref{ModCarSent}, is now written as:
\be\label{ModCarSentBold}
\exists_1 {\bf x} T[{\bf x}] \supset T[{\bf t}]; \;\;  \mbox{that is:} \; (\exists {\bf y} \forall  {\bf x} (T[{\bf x}] \equiv {\bf y} = {\bf x})) \supset T[{\bf t}].
\ee

Lewis also points out (1972, p. 254) that using this notation, we can state conditions 2 and 3 just above, about failed and multiple realizations (respectively), as a conditional. We just need to adopt the neat device of a necessarily denotationless term, say $*$. (For more reasons why this is neat, cf. Oliver and Smiley (2013).) Thus consider
\be\label{ModCarSentBoldNoReference}
\neg \; \exists_1 {\bf x} T[{\bf x}] \; \supset \; ({\bf t} = {\bf *}). 
\ee
Here, $ {\bf t} = {\bf *}$ means that each $t_i$ is denotationless. For if $*$ is some chosen necessarily denotationless name, then ${\bf *}$ is $\langle * , \ldots * \rangle$ and ${\bf t} = {\bf *}$ is equivalent to the conjunction of all the identities $t_i = *$.

\subsubsection{Simultaneous explicit definitions}\label{DKLdefns}
Lewis' proposal in Section \ref{DKLsentences}, that the $T$ terms refer iff the realization formula $T(x_1, \dots x_n)$ (also written $T[{\bf x}]$)  is uniquely realized, immediately yields  {\em explicit definitions} for each of the $T$-terms. 

For if as usual we write  $\iota$ for the description symbol, i.e. we read $\iota x F(x)$ as `the $F$', then the conjunction of:\\
\indent \indent (i) the modified Carnap sentence, eq. \ref{ModCarSent} or equivalently eq. \ref{ModCarSentBold}, and \\
\indent \indent (ii) our `veto' on denotation for the cases of no or multiple realizations, i.e. eq. \ref{ModCarSentBoldNoReference}, \\
is logically equivalent to the battery of definitions:   \\ 
\indent \indent \indent  \indent $t_1 =_{df} \iota y_1 \exists y_2, \ldots , y_n\forall x_1, \ldots , x_n \left(T(x_1, \ldots , x_n) \equiv (y_1 = x_1\ \&\ \ldots\ y_n = x_n)\right)$,\\
\indent \indent \indent \indent $\ldots$\\
\indent \indent \indent \indent  $t_n =_{df} \iota y_n \exists y_1, \ldots , y_{n-1}\forall x_1, \ldots , x_n \left(T(x_1, \ldots , x_n) \equiv (y_1 = x_1\ \&\ \ldots\ y_n = x_n)\right)$.\\
Or more compactly, in the boldface notation: the conjunction of (i) and (ii) is logically equivalent to
\be\label{explicdefneqn}
{\bf t} = \iota {\bf x} T[{\bf x}].\footnote{A minor clarification. The equivalence requires an appropriate stipulation about the truth-values of identity statements using descriptions that are improper, i.e. not realized or multiply realized. But no worries: as Lewis says (1970,  pp. 430, 438; 1972, footnote 11, p. 254), the stipulations in Scott's system are appropriate.}
\ee

Immediately after exhibiting these explicit definitions, Lewis  sums up.
\begin{quote}
This is what I have called functional definition. The $T$-terms have been
defined as the occupants of the causal roles specified by the theory $T$; as
the entities, whatever those may be, that bear certain causal relations to one
another and to the referents of the $O$-terms.

If I am right, $T$-terms are eliminable---we can always replace them by
their definientia. Of course, this is not to say that theories are fictions, or
that theories are uninterpreted formal abacuses, or that theoretical entities
are unreal. Quite the opposite! Because we understand the $O$-terms, and
we can define the $T$-terms from them, theories are fully meaningful; we have
reason to think a good theory true; and if a theory is true, then whatever
exists according to the theory really does exist.

I said that there are more theorems of $T$ than follow logically from the
postulate alone. [Cf. the last displayed quote in Section \ref{DKLsentences}, just after eq. \ref{ModRamSent}; and cf. (1970, pp. 438-440)] More precisely: the theorems of $T$ are just those sentences
which follow from the postulate together with the corresponding functional
definition of the $T$-terms. For that definition, I claim, is given implicitly
when the postulate is presented as bestowing meanings on the $T$-terms
introduced in it. (1972, p. 254)
\end{quote}

Let us also sum up by returning to the central example of mind and body. So let $T$ be the folk psychology of mental states. It is a catalogue of platitudes about belief, desire, etc.,  connecting those mental states to public situations and behavioural dispositions, whose descriptions are given using the $O$-terms; and the $T$-terms  refer to the mental states. $T$ is taken to assert the \emph{existence} of inner states that fulfil the roles of appropriately mediating the connections between situations and behaviour. (Recall the detective-story parable.) The existence of such states is a fact \emph{beyond} the situation-behaviour correlations. $T$ may fall way short of giving necessary and sufficient conditions for them; and it is in general a contingent matter, varying from possible world to possible world,  \emph{which}  states they may happen to be.  Recall comments (i) and (ii) just before eq. \ref{ModRamSent}. This is \emph{functionalism} about mind and body.

 \section{Functionalist reduction}\label{DKL4}
 
 All the props are now on stage. We have developed a Nagelian account of reduction, taking in to account the problems of Faithlessness, Plenitude and Scarcity (Sections \ref{homered} and \ref{rednobstacle}); and Lewis' account of functional definition (Section \ref{homefunc}). We now put these together (again following Lewis) to get functionalist reduction: the second step of the Canberra Plan. As we announced in Section \ref{intro}, it is the {\em golden oldie} that, sadly, the literature on functionalism and reduction has forgotten. 
 
 Lewis' main idea is that thanks to Section \ref{homefunc}’s functional definitions of the $T$-terms, one can show that a Nagelian reduction of $T$ by some later or independent theory, $T^*$ say, has no need to {\em postulate} bridge laws (linking the vocabularies of $T$ and $T^*$)  as separate empirical hypotheses. For the functional definitions we extracted from $T$ mean that a version of the bridge laws---viz. with each $T$-term replaced by its functional definition using $O$-terms---can be a {\em theorem} of $T^*$. There is no bar against this, since these versions do not contain any vocabulary alien to $T^*$. (Lewis calls these versions the {\em definitionally expanded bridge laws}.) And if they {\em are} theorems of $T^*$, then: $T^*$ alone yields the bridge laws (by the transitivity of identity); and so $T^*$ alone reduces $T$. (In Section \ref{defext} we introduced the mnemonic labels $T_t$ and $T_b$ for the reduced and the reducing theories. But in this Section, it will be clearer to follow Lewis' notation of  $T$ and $T^*$.)

Of course, we already reviewed this idea in Section \ref{1111}’s example of pain and C-fibre firing. $T$ was the folk psychology of mental states from which functional definitions of mental terms like `pain' were extracted ; and $T^*$ was a neurophysiological theory, accepted later than or independently of folk psychology, which we took to include the theorem that C-fibre firing is the unique occupant of the pain-role. So $T^*$ asserts that  C-fibre firing is typically caused by tissue damage,  typically causes aversive behaviour etc.  So this theorem is the relevant definitionally expanded bridge law. Then as we said in Section \ref{1111}, it follows by transitivity of identity that pain = C-fibre firing.\footnote{`Transitivity of identity' is usually understood as identity of objects; and this accords with Lewis' taking all $T$-terms as names not predicates (1972, p. 253, quoted in Section \ref{DKLsentences}). But we can also understand it as co-extensiveness of predicates; cf. footnotes \ref{fnstateconcept} and \ref{nonchalant}.} 

 So our job now is to state this idea more precisely and generally. This will also make it clear how not only the bridge laws, but also the whole of $T$ is definitionally implied by $T^*$.

To state the idea precisely, it will be clearest to begin with the simplest scenario, where: (i) the $T$-terms have retained the interpretations they received when first introduced (i.e. as expounded in Section \ref{DKLsentences}) and (ii) `this is a reduction in which $T$ survives intact; not, what is more common, a simultaneous partial reduction and partial falsification of $T$ by $T^*$' (1970, p. 441). Afterwards, we can turn to the more common and complicated scenario, which Lewis' account of course admits just as Nagel and Schaffner do: (recall Sections \ref{Nagelmodif} and \ref{Faith}).

First, Lewis stresses that $T^*$ can be be miscellaneous; recall our disavowing reductionism, in (2) at the start of Section \ref{homered}. But all the terms of $T^*$ are taken as being understood (interpreted), just as the $O$-terms in Section \ref{homefunc} were. Besides,  one may as well use `$O$-term' to include also the `okay' terms of $T^*$ by which $T^*$'s newly-introduced, or `troublesome', terms---the $T^*$-terms---got their interpretations.  Then Lewis proposes to label as an $O^*$-term, whatever is either an $O$-term or $T^*$-term. 

So here, Lewis is at first registering what we dubbed the {\em ternary} contrast of vocabularies; but then he introduces the $O^*$ notation, so as to focus our attention on the binary contrast of greatest interest in reduction:  the contrast  between the $T$-terms of $T$, and the `rest', i.e. the  $O^*$-terms in the sense just stipulated.  For instance,   `tissue damage',  `aversive behaviour'  are $O$-terms, but they can enter the functional definition of the $T^*$-terms.  Thus Lewis writes   
\begin{quote}
The reducing theory $T^*$ need not be what we would naturally call a single theory; it may be a combination of several theories, perhaps belonging to different sciences. Parts of $T^*$ may be miscellaneous unsystematized hypotheses which we accept, and which are not properly called theories at all. Different parts of $T^*$ may have been proposed or accepted at different times, either before or after $T$ itself was proposed ... 
$T^*$, or parts of $T^*$, may introduce theoretical terms; if so, let us assume that these $T^*$-terms have been introduced by means of the same $O$-vocabulary which was used to introduce the theoretical terms of $T$. This is possible regardless of the order in which $T$ and $T^*$ were proposed. Any term that is either an $O$-term or a $T^*$-term may be called an $O^*$-term. (1970, p. 441)
\end{quote}

Now suppose that the following sentence is a theorem (a claim) of  $T^*$: $T(\rho_1,\ldots \rho_n)$, where $\rho_1, \ldots \rho_n$ are all $O^*$-terms, including, possibly, definite descriptions constructed using other $O^*$-terms. Here, $T(t_1, \ldots t_n)$ is of course the original postulate of $T$, just as at the start of Section \ref{DKLsentences}; and  $T(\rho_1,\ldots \rho_n)$ simply substitutes $\rho_i$ for each occurrence of $t_i$ therein.  Lewis calls $T(\rho_1,\ldots \rho_n)$  the \emph{reduction premise} for $T$ (1970, p. 441), or the \emph{weak reduction premise} for $T$ (1972, p. 255).

Notice that the original postulate $T(t_1, \ldots t_n)$  follows from the reduction premise, taken together with the \emph{bridge laws}, as usually formulated, viz.: $\rho_1 = t_1,\ \ldots \ \rho_n = t_n$. (Also, the bridge laws (thus formulated) follow from the reduction premise together with the postulate of $T$ and the functional definitions of the $t_1, \ \ldots \ t_n$.) {\em But where do we get the bridge laws?}

The traditional view is that the bridge laws are separate, empirical hypotheses.  In that case, says Lewis (p. 442), one may choose whether to posit $T^*$ and the bridge laws, and so derive $T$, or else to posit $T^*$ alone, in which case we will have to posit $T$ separately. 

But then Lewis asks us to consider the case in which $T^*$ has as theorems \emph{definitionally expanded bridge laws}:

$\rho_1 =_{df} \iota y_1 \exists y_2, \ldots y_n\forall x_1, \ldots x_n \left(T(x_1, \ldots x_n) \equiv (y_1 = x_1\ \&\ \ldots\ y_n = x_n)\right)$, etc.

\noindent whose right hand sides match the definitions in Section \ref{DKLdefns}, cf. eq. \ref{explicdefneqn}. Unlike the case for the original bridge laws, the definitionally expanded bridge laws contain only $O^*$-terms. So there is no problem of \emph{vocabulary} preventing these laws being theorems of $T^*$. 

And if they {\em are} theorems of $T^*$, then it follows by sheer logic (Leibniz's Law, the transitivity of identity) that $t_1 = \rho_1$, etc.  So there is no choice in the matter: $T^*$ reduces $T$, without the need for empirical hypotheses. Lewis sums up:
\begin{quote}
If $T^*$ yields as theorems a reduction premise for $T$, and also a suitable set of definitionally expanded bridge laws for $T$, then $T^*$---without the aid of any other empirical hypothesis---reduces $T$. ... The reduction of $T$ does not need to be justified by considerations of parsimony (or whatever) over and above the considerations of parsimony that led us to accept $T^*$ in the first place. (1970, p. 443; similarly, 1972, p. 255)\footnote{For brevity, we set aside: (i) Lewis' discussion of the \emph{auxiliary reduction premise} (1970, p. 443) and the \emph{strong reduction premise} (1972, p. 255); (ii) Lewis' examples (1970, p. 443-444; 1972, p. 256-258---which is mind and body); and (iii)  Lewis' discussion of adding definitionally expanded bridge laws to $T^*$ if they are not theorems already (1970, p. 444-445).}
\end{quote}

So much by way of summarising how bridge laws, and thereby the reduced theory $T$, might be deduced from the reducing theory $T^*$---in the simplest scenario, where (i) the $T$-terms have retained their original interpretations, and (ii) $T$ survives intact, rather than being partially reduced and partially falsified. Lewis ends by briefly discussing the more common and complicated scenario, where `the original theory is falsified while a corrected version is reduced. If $T$ is thus partially reduced and partially falsified, or revised for any other reason, do the $T$-terms retain their meanings?' (1970, p. 445). He considers two approaches. The first is prompted by Feyerabend's views. But we shall briefly report only the second: which we favour (as also Lewis seems to) and which leads us back to our previous topics of {\em causal descriptivism} (endorsed near the end of Section \ref{DKL1A}) and {\em near-realizations} (footnote \ref{nearrealize}). 

On this approach, we say that the $T$-terms name the nearest near-realization of the \emph{original} $T$.  Then so long as the corrections are not too extreme, the $T$-terms may denote, and we can make sense of the idea that adjustments to $T$ involve learning more about the objects thus denoted.  This meshes well with causal descriptivism. Thus Lewis:
\begin{quote}
 we permit the $T$-terms to name components of the nearest near-realization of $T$, even if it is not a realization of $T$ itself. For after $T$ has been corrected, no matter how slightly, we will believe that the original version of $T$ is unrealized. We will want the $T$-terms to name components of the unique realization (if any) of the corrected version of $T$. They can do so without change of meaning if a realization of the corrected version is also a near-realization of the original version.
 
According to this position, we may be unable to discover the meanings of theoretical terms at a given time just by looking into the minds of the most competent language-users at that time. ...  This situation is surprising, but it has precedent: a parallel doctrine about proper names has recently been defended.[a footnote cites Kaplan's famous 1968 paper `Quantifying in'] (1970, p. 446)
\end{quote}
Indeed, near-realizations and the causal theory of reference will be one of the themes in our discussion of Torretti ...

\section{Glimpsing the land of Torretti}\label{roberto}
As we announced in Section \ref{113}: now that we have developed our account of functionalist reduction, various projects in relation to Torretti's large and magisterial  {\em oeuvre} beckon us. Even once we postpone to our other papers the projects in the philosophy and history of geometry, there is much to do. For Torretti has written a lot about reduction, and related topics like scientific realism, the syntactic vs. semantic conceptions of theories, and structuralism (in its many senses): not just in geometry, but across all of physics. 

Besides, we already noted, from Section \ref{homered} through to \ref{DKL4}, several important topics we have had to postpone. For example, as regards reduction: defending the syntactic approach to theories; and as regards functionalism: comparing Lewis' and Carnap's treatment of theoretical terms, and assessing Newman's objection to the Ramsey-sentence approach, and to ``structural realism''. So we owe a discussion of these topics, especially in relation to Torretti's writings.

But for reasons of space, we must postpone these projects. So here,  as an appetizer for them and as a courtesy to Torretti, we will only discharge three small obligations to him and his work. First, we will say a little about the obvious project beckoning us: to compare functionalist reduction with Torretti’s discussions of reduction (Section \ref{peace}). Then (Sections \ref{implicit} and \ref{beltr}) we will exhibit two closely-related links between us and Torretti: both concern the philosophy and history of the axiomatic method. 

\subsection{Reduction: a peace-pipe}\label{peace}
We admit at the outset a major difference between Torretti's and our discussions of reduction. Like the majority of the literature, he uses a binary contrast between theory and observation, whereas we have advocated a ternary contrast. But when one looks at the details of his discussions, there is a good deal of convergence between him and us. 

At first, this may be surprising. For it is clear from the previous Sections that our intellectual outlook is close to figures such as Nagel and Lewis. These figures are more empiricist, more Humean, more scientific realist, and less holist about semantics than, we surmise, Torretti would like. (At least, so it seems to us, since after all, Torretti is also a profound Kant scholar.) But going beyond  vague `isms' and philosophical slogans: we find we agree with several of Torretti's main claims---and we surmise that figures like Nagel and Lewis would too. 

Let us briefly smoke this peace-pipe. As tobacco acceptable to both parties, we make the obvious choice: the rejection of a neutral observation language, and the issues to which it leads. We will begin with Torretti's writings, and then concur with him.

Torretti denies that from direct experience, unleavened by conceptualization---that `blooming, buzzing confusion', as William James put it---one can define worthwhile scientific concepts, let alone derive scientific knowledge. Expressed more positively: Torretti's view is that all observation subsumes its object under a general concept. This view is developed in several places in Torretti's {\em oeuvre}. For example, one finds it in his essay {\em Observation} (1986: p. 4, Section 5, and p. 21; labelled as the `principle of conceptual grasp') . It is announced as a leading theme at the start of  {\em Creative Understanding} (1990: pp. ix, 1, 5-7) whose Chapter 1 is a reworking of 1986, and whose Chapter 2 gives an extended critique of logical empiricists' distinction between theory and observation  (especially Section 2.4.3 about Carnap). It also appears in Torretti's  {\em Philosophy of Physics} (1999: pp. 402-404, 421), and of course also in his Kantian scholarship (e.g. 2008, pp. 81-82). 

In all these works, Torretti's development of this view is clear, gracious and often witty: as always with his writing. We cannot resist quoting a passage where Torretti goes one better than the famous metaphor invented by a kindred spirit:  viz. Quine's web of belief.
\begin{quote}
... if all our knowledge of physical objects is corrigible, it must be self-correcting, for there is no outside authority to which one could turn for help. Quine’s famous dictum that “our statements about the external world face the tribunal of sense experience not individually, but only as a corporate body”... is apt to be misleading. For in the trial of empirical knowledge the defendants are at once the prosecution, the witnesses, and the jury, who must find the guilty among themselves with no more evidence than they can all jointly put together. (1990, p. 7; cf. 1986, p. 9)
\end{quote}  

Of course, this view faces challenges, the obvious one being: without raw experience to adjudicate scientific claims, how can we secure the objectivity of scientific knowledge? In particular, how can we refute the  threat of incommensurability, urged by Kuhn and Feyerabend? Torretti answers these challenges, in various places within {\em Creative Understanding} and {\em Philosophy of Physics} (1990: pp. x-xi, 44; and 79-81, referring back to C.1-C.3 of pp. 32-33; 1999: 421-430). Broadly speaking, he gives various reasons for objectivity, even continuity, in scientific change, despite our lack of a theory-neutral observation language; among them the many ways that even our most arcane theoretical knowledge is rooted in our everyday world. We cannot resist quoting another passage where, again, he goes one better than the famous metaphor invented by a kindred spirit: in this case, Neurath's ship.
\begin{quote}
In Neurath's ship, the neat steel turrets of theory are built on and bridged by the wooden planks of common sense, which may be worn and musty but  are indispensible to keep afloat the enterprise of knowledge. Physicists  who advocate different theories do not `practice their trades in different worlds' (Kuhn 1962, p. 149), for there is but one world for them to wake up to, namely, the world they are in together with the persons they love and the goods they yearn for ...  (1999, p. 404; cf. also 421f. )
\end{quote}
  
With all this, we concur; (and we daresay Lewis, and to a large extent Nagel, would concur). Recall our previous scepticism about reductions of empirical knowledge to experience (cf. Section \ref{examp}), and more specifically, our joining Lewis in taking the $O$-terms to be, not `observational', but `old' or `okay' i.e. already-understood (cf. Section \ref{homefunc}). 

We also concur with much else in Torretti's discussions. Let us report two examples, from many that could be given. 
% {\jb STILL NEED TO THINK whether and how to include below a digest of / response to Torretti's 1990: Section 3.7-3-8, pp. 144-161; which give his positive views about reduction }\\
We choose them because they connect to previous Sections' discussions:\\
\indent \indent  (1): Torretti's rejection of `reference without sense', i.e. his rejection of Putnam's proposal in the mid-1970s to answer the challenge of incommensurability by adopting a causal theory of reference in which Fregean senses and-or Carnapian  intensions have no place (1990: pp. 51-70; echoed at 1999, p. 422); \\
\indent \indent  (2): Torretti's critique of the treatment by the Sneed-Stegm\"{u}ller school of structuralist analysis of scientific theories---to which, overall, he is sympathetic (1990, p. xi, 109f.; 1999, p. 412-417, 424)---of its ``problem of theoretical terms''; this critique is in (1990: pp. 129-130, 134-137: reviewed pp. 161-162)  and endorsed in (1999: p. 414, note 18).\\
As to (1), our agreement with Torretti is obvious. We recall that we joined Lewis in {\em causal descriptivism} (Section \ref{DKL1A}); and that this account of reference has since been taken up in detailed defence of scientific realism (Psillos 1999, 2012). 

As to (2), we shall give a bit more detail, since the proposals being criticised are less well known. The first thing to note (as Torretti does at the start of his discussion: 1990, p. 114) is that this ``problem of theoretical terms''  is {\em not} what goes by that name for the logical empiricists---and indeed, for the rest of us: viz. the semantic and epistemological questions, how theoretical terms get their meanings, and how we can be warranted in applying them. Instead, it is (roughly!) that for some terms in a theory, every way of establishing that they apply in a real situation presupposes that the theory `has an actual model', i.e. applies in that very situation or some other one. Sneed calls such terms `theoretical'. More precisely, since Sneed et al. adopt the semantic conception of theories: it is---not predicates, but---relations-in-extension in models, especially functions representing physical quantities, that are theoretical for Sneed. 

Sneed's problem about these ``terms'' is that although one might confirm the presupposition by finding an actual model (a successful real application of the theory), doing so will involve applying the term (measuring the function)---and so again presuppose some actual model. Thus a regress, or circularity, looms---prompting the question: `what empirical claim is being made by someone who holds such a theory?' (1990: p. 115). 

In both Sneed and the ensuing literature, classical mechanics (in a traditional point-particle formulation) is a frequent source of examples. Thus an apparent example of this problem is the oft-noted interdependence of Newton's laws and the concept of inertial frame: namely, that  the laws are to be true of motions only if they are described in inertial coordinates, while inertial frames get defined as those in which  Newton's laws hold---threatening a circle.  

Sneed's own answer to this question is that the empirical claim must involve only non-theoretical terms. This answer gets articulated in terms of models that assign extensions only to such terms, the theoretical terms being simply ignored: they are called `partial potential models' (1990: p. 116, 121). 
%{\jb ??ADD HEREABOUT Steven French and Da Costa?}

Torretti describes this answer in detail but---unsurprisingly, given his rejection of non-conceptual observation---rejects it as rooted in `foundationist scruples'. Indeed, he rejects the problem as a `pseudoproblem, stemming from a refusal to countenance a genuinely creative understanding of natural phenomena' (ibid. p. xi, p. 134).  Similarly, he describes in detail,  but also rejects, G\"{a}hde's alternative proposal for how to define theoretical terms (pp. 121-128). And he diagnoses the same defect, i.e. a foundationist belief in non-conceptual observation, in Ludwig's account of the interpretation of physical theories (pp. 131-137).  For both authors, the illustrations are again from classical mechanics, especially the interdependence of Newton's laws and inertial frames, noted above (pp. 117, 122-123, 136-137). 

 So far as we can judge, Torretti's critique is right. But we will not pursue details. We are just pleased to note a possible project for anyone sympathetic to the Sneed-Stegm\"{u}ller school of structuralism. For Sneed's problem of a regress, or circle, of presupposed successful real applications of a theory reminds us of the threat of logical circles of definition, which the functionalist---specifically Lewis---shows can be overcome. As we saw, there is nothing incoherent, or even suspicious, about the idea of simultaneous unique definition (cf. Section \ref{DKL2}).  So maybe the same idea can help with Sneed's problem.\footnote{ Certainly, it seems to fit the example of the interdependence of Newton's laws and inertial frames. That is: we propose that one can reasonably take classical mechanics (in a point-particle formulation of the sort Sneed et al. consider) as introducing simultaneously terms for time and length (and so: frame), and for mass: terms that are held by the theory to be uniquely realized in such a way as to make the theory  true---including Newton's laws, using these terms, being true, i.e. true for motions described in inertial coordinates.

Presumably, the starting point for this project would be Chapter III of Sneed (1979) and Chapter II.3 of Balzer et al. (1987). But the gap in the literature is wide. Sad to say: so far as we know, neither of the two sides---Lewis and other Canberra Planners, and the Sneed-Stegm\"{u}ller school---refers to the other.} 
%??ADD that we will return to this in connection with Huggett 2006 and Belot 2012 in our third paper??!!
%{\jb Agreed, disagreements between Torretti and us, about reduction and related topics, undoubtedly remain. By our lights, EXAMPLES : So complete peace has not broken out; and we cannot hope to establish it here. We can only look forward to continuing the debate with Torretti about these topics!}

But that is work for another day. Here, we must curtail our comparison of our and Torretti's general views about reduction and related topics. In the rest of this Section, we will turn to 
two more specific (and related) links between us and him. Both concern the philosophy and history of the axiomatic method, and what we have called the problem, or objection, of Faithlessness. The first link (Section \ref{implicit}) is general, and relates mostly to the functionalist idea of simultaneous unique definition. It leads to the second link (Section \ref{beltr}): which is specific, and relates mostly to reduction.

\subsection{Comparing functionalist definition with implicit definition}\label{implicit}
In this Section, we will compare the functionalist idea of simultaneous unique definition (or specification) with the idea that a mathematical axiom system defines its terms, or some of them---in some sense of `define'. This idea is often called {\em implicit definition}. But as Torretti and many authors point out, the word `definition' is misleading, since it suggests both uniqueness and Faithfulness to a given meaning---both features that in general fail in this context.

This idea was at the centre of discussions among mathematicians in the second half of the nineteenth century: especially about geometry, as mathematicians responded to the rise of non-Euclidean geometries. So---fortunately for us---it is discussed in detail by Torretti in his magisterial {\em Philosophy of Geometry from Riemann to Poincar\'{e}} (1978). This book is a treasure-trove, and we have learnt a lot from it: not just about this comparison, but about much else---especially the Helmholtz-Lie {\em Raumproblem} which will be our second paper’s first example of spacetime functionalism. But in this Section, we must stick to the comparison. (Section \ref{beltr} will focus on geometry and look ahead to our second paper.)

 More specifically, the idea was at the centre of the famous controversy between Frege and Hilbert, that was triggered by Frege's response to Hilbert's {\em Grundlagen der Geometrie} (1899). Thus their central dispute was `over the nature of axioms, i.e., over the question whether axioms are determinately true claims about a fixed subject-matter or re-interpretable sentences expressing multiply-instantiable conditions’,  as Blanchette's fine review puts it (2018, Section 3). Frege gave the first answer, Hilbert the second: and Torretti sides firmly with Hilbert.\footnote{One might similarly summarise  the dispute as about whether it is defect of an axiom system that it can be realized, i.e. made true, by very diverse interpretations (`models’ as we nowadays say in logic and model theory)--- with Frege saying `Yes’ and Hilbert saying `No’. (The dispute is often summarised as about whether axioms {\em implicitly define} their terms (with Frege saying `No’ and Hilbert saying `Yes’); but  as we mentioned, `implicit definition’ is a misleading term.) We recommend Potter's discussions (2000: 87-94; 2020, 124-132).}  As Hilbert put it to Frege:
\begin{quote}
Every theory is naturally only a scaffolding or schema of concepts, together with their necessary mutual relations, and the basic elements ({\em Grundelemente}) can be conceived in any way you wish. If I conceive my points as any system of things, e.g. the system {\em love, law, chimney-sweep}, ... and I just assume all my axioms as relations between these things, my theorems, e.g. the theorem of Pythagoras, will also hold of these things. In other words, every theory can always be applied to infinitely many systems of basic elements. It suffices to apply an invertible univocal transformation [i.e., a bijection] and to stipulate that the axioms hold correspondingly for the transformed things. [...] This property is never a shortcoming of a theory and is, in any case, inevitable. (Torretti, 1978, p. 251; also endorsed at Torretti 1999, p. 409)\footnote{In fact, Hilbert already took this view, some ten years before: witness the similar remark in 1891, made while waiting for a train in a station waiting-room: `one must be able to say at all times---instead of points, straight lines and planes---tables, chairs and beer mugs’: cf. Kennedy (1972, p. 133).}
\end{quote}
By 1900, Hilbert's view  of axioms (often called a `formal’ or `structural’ view) was already dominant. Not only did Hilbert's book and other writings deploy it to prove relative consistency and independence results, in unprecedented detail.  And not only was this work very influential, in the ensuing years, in moulding the modern conception of model theory and formal semantics. Also, as Torretti's discussions bring out: already in 1900, several contemporaries shared Hilbert's `formal’ or `structural’ view of axioms. For example, Torretti cites Padoa (a member of Peano's school) saying in 1900 that the undefined symbols of a deductive theory are `entirely devoid of meaning’ and that axioms `far from stating {\em facts}, i.e. {\em relations} between the {\em ideas} represented by the undefined symbols, are nothing but {\em conditions} with which the undefined symbols must comply’ (1978, p. 226).\footnote{{\label{threehistory}}{Below, we will mention another example, Pieri. We also note  three historical points. (1): Torretti (1999, p. 408-414) portrays this view as prompting the semantic or structural  (as against syntactic) view of theories that, as we saw in Section 6.1, he favours. (2): Gray (2008, Section 4.1, p. 176f.) makes an interesting case that this broad development represented a rise of modernism, in a sense analogous to that in art and literature. (3): The Peano school’s endorsement of the Hilbertian structural view of axiom systems also surfaced in the philosophy of arithmetic: where it leads us back to Russell's quip about `theft over honest toil'. Thus recall from Section \ref{221}’s discussion of Faithlessness, especially  footnote \ref{StrucPotter}, that Frege ``already played the role of Benacerraf’’. That is: he objected to his own definition of natural number; and then replied that the Faithlessness (specifically: over-shooting) was a price worth paying for the benefit of an otherwise successful definition. {\em A propos} of this, Alex Oliver points out to us (p.c.) that Peano also made this objection in 1901; and the reply that the price is right came from Russell, in his {\em Principles of Mathematics}. Writing with his usual verve (but lack of argument!), he in effect just outfaced Peano: `To regard a number as a class of classes must appear, at first sight, a wholly indefensible paradox. Thus Peano  remarks that “ … these objects have different properties”. He does not tell us what these properties are, and for my part I am unable to discover them’ (1903/2010: Section 111, p.115). Similarly in later writings: Russell had the `structuralism' of Peano, Dedekind and Hilbert in mind when he condemned the `method of postulating' as `theft': cf. Section \ref{223}, especially footnote \ref{toil}, and Linsky (2019, Section 3).}}

Against this background, we have two main points to make. The first (Section \ref{precurs}) will be that some nineteenth-century discussions of axiom systems contain striking precursors of simultaneous definition. Since we are by no means historians, we will here be wholly indebted to Torretti’s scholarship. Our second point (Section \ref{proFrege}) is about how one should understand {\em logical consequence}. Our view of this will imply that, although  Hilbert undoubtedly ``beat'' Frege as regards historical influence,  there was more right on Frege’s side than he is usually given credit for. Besides: more right than Torretti gives him credit for. So here, we regret to say, we have a philosophical disagreement with Torretti. At first sight, this may seem bad manners in a {\em Festschrift}. But we trust that with his liberal and gentlemanly outlook---and his relish for philosophical debate---Roberto will forgive us!

\subsubsection{Precursors of functionalist definition}\label{precurs}
Both our points, and our historical debt to Torretti, are best introduced by the closing passage (pp. 252-253) of Section 3.2.10 of Torretti (1978), which is entitled `Axioms and definitions. Frege's criticism of Hilbert' (Section 3.2 is entitled `Axiomatics'). For the passage itself will serve to establish our first point, about historical precursors of simultaneous definition. And Torretti's views, especially at the end of the passage, will be the springboard for our second point.

First, Torretti criticises the phrase `implicit definition', on the grounds that it is in general impossible to extract from an axiom system explicit definitions of its primitive terms. He also notes that this is disanalogous from the situation often cited as an analogue, viz. extracting the $n$ roots of $n$ simultaneous linear equations.\footnote{In Section \ref{DKL1A} we agreed that this is in general impossible, but argued that this is no objection to Lewisian functionalism.} On this second point, Torretti also cites in his support, Frege, who writes as follows; (it will be clearer in Section \ref{proFrege} why he does so): 
\begin{quote}
``If we survey the whole of Mr. Hilbert’s definitions and axioms, they will be seen to be comparable to a system of equations with many unknowns; for in each axiom you normally find several of the unknown expressions `point', `line', `plane', `lie on', `between', etc., so that only the whole, not particular axioms or groups of axioms, can suffice to determine the unknowns. But does the whole suffice? Who can say that this system is solvable for the unknowns, and that these are unambiguously determined?'' (note 148, p. 402).
\end{quote}

Then (p. 252), Torretti briefly mentions Pieri (who, like Padoa, was a member of Peano's school).  Torretti's previous passages about him (pp. 224-226, 250-251) describe how his papers of 1899 and 1900:\\
\indent \indent (a) urged the implicit/explicit contrast under the (respective) labels `real'/`nominal';\\
\indent \indent (b) were in the mathematical vanguard, in that Pieri wished to liberate geometry from any appeal to spatial intuition, and also sided (implicitly!) with Hilbert’s view of axioms, against Frege’s; although also Pieri:\\
\indent \indent (c) preferred to reserve the word `definition' for explicit/nominal definitions---a practice that logicians nowadays generally follow, and that Torretti endorses.\footnote{\label{BethRaum}{Two further points, which will be centre-stage in our second paper. (1): Beware: `Implicit definition’, as used {\em nowadays} by logicians, means something else. The idea is due to Padoa (Torretti 1978, pp. 226-227). It is essentially equivalent to metaphysicians’ notion of supervenience/determination (cf. footnote \ref{rigorlogic}). But it is utterly precise, and logically weaker than explicit definition (in general---though equivalent to it, for first-order languages, by Beth’s theorem: cf. Boolos and Jeffrey 1980, p. 245f., Hodges 1997, p. 149). Thus proving explicit undefinability by showing implicit undefinability is called {\em Padoa's method} (Hodges 1997, p. 58).  (2): For us, Pieri's own axiomatisation of geometry has the further interest that, following the tradition of the Helmholtz-Lie {\em Raumproblem}, it is based on the idea of rigid motions.}}

Then Torretti reports that the phrase `implicit definition', and the analogy with simultaneous equations (and so for us functionalists: the idea of simultaneous definition), seems to have been first used eighty years earlier, by the geometer Gergonne in his ``Essai sur la th\'{e}orie des d\'{e}finitions'' (1818). (Quine  (1964: p. 71) also cites Gergonne.) But Torretti ends by emphasising the critique of the idea of implicit definition with which he began on p. 252. Thus he writes 
\begin{quote}
Gergonne observes that a single sentence which contains an unknown word may suffice to teach us its meaning. Thus, if you know the words {\em triangle} and {\em quadrilateral} you will learn the meaning of {\em diagonal} if you are told that``a quadrilateral has two diagonals each of which divides it into two triangles''. [Now Torretti quotes Gergonne ...]

``Such phrases, which provide an understanding of one of the words which occurs in them by means of the known meaning of the others, might be called {\em implicit} definitions, in contrast with the ordinary definitions, which we would call {\em explicit}. There is evidently between the latter and the former the same difference as between solved and unsolved equations. One sees also that, just as two equations with two unknowns simultaneously determine both, two sentences which contain two new words, combined with other known words, can often determine their sense. The same can be said of a greater number of words combined with known words in a like number of sentences; but, in this case, one must perform a sort of elimination which becomes more difficult as the number of words in question increases.'' (1818, p. 23; Gergonne's italics).

[Then Torretti concludes ...] Gergonne has grasped well a familiar linguistic phenomenon and has given it an appropriate name. But his systems of simultaneous implicit definitions are something evidently very different from abstract axiom systems. In these, all designators and predicators behave, if you wish, as unknowns, and no process of elimination can lead to fix their meanings, one by one. We ought not to burden Gergonne with the paternity of the rather unfortunate description of axioms as implicit definitions. (1978, p. 253)
\end{quote}
The first thing to say about this passage is that the quotation from Gergonne is indeed a striking precursor of functionalists' idea of simultaneous definition. It is striking  for its contrast between known and unknown words (cf. Lewis' $O$-terms and $T$-terms), for its idea of simultaneous definition---and for its early date.  Thus ends our first point.  

\subsubsection{Logical consequence is not formal: Faithlessness and Frege}\label{proFrege}
As we announced, our second point is not historical, but philosophical; and marks a disagreement with Torretti. We think it worth displaying because it has a deep origin: in  different ways of thinking about {\em logical consequence}. In short: we  say that logical consequence is {\em not} a formal notion. It is {\em not} fully captured by the textbook definition of subset-hood among sets of models. But Torretti thinks it is thus captured. 

Thus in Section 3.2.2 of his (1978), entitled `Why are axiomatic theories naturally abstract?', he lays out a formal semantics of a fragment of English sufficient to state mathematical propositions (called `$m$-English': pp. 192-195). It is essentially the familiar logic-textbook semantics of (maybe higher-order) predicate languages; with the usual allowance of any set of objects as the domain $D$ of quantification, and any assignment of semantic values therein to the non-logical vocabulary (i.e. elements of $D$ to individual constants, subsets of $D$ to monadic predicates, sets of ordered pairs from $D$ to dyadic predicates etc.). Then he defines logical consequence in the usual way as subset-hood among sets of models.
\begin{quote}
A sentence $S$ is a logical consequence of a set of sentences $K$ if, and only if, every interpretation of $K \cup \{ S\}$ which satisfies $K$ also satisfies $\{ S \}$. We use the abbreviation $K \models S$ ... This relation does not depend on a particular interpretation of $K$ and $S$. Indeed, we can replace all interpretable words in $K$ and $S$ by meaningless letters   ... and it will still make sense to say that $K \models S$.  The mathematician who studies an axiomatic theory need not worry about the referents of its sentences ... The important thing is that, for every conceivable interpretation of the theory, if the axioms are true, then the theorems are also true. (p. 195)
\end{quote}

In reply to this, we propose a return to how we all think about logical consequence, before we study the modern textbook in logic class: a return, we daresay, to how everyone thought about it before the twentieth-century advent of formal semantics, and Hilbert’s ``beating’’ Frege in the way we reported in this Section's preamble. This means: logical consequence as a relation between a sentence $S$ and another $S'$ (or: a set of interpreted sentences $K$) that is {\em Faithful} to how they are understood (intended) in the language concerned. The relation is: if $S'$ (understood as intended) is or were true (or: all the sentences of $K$ are or were true), then $S$ (understood as intended) {\em must} be true. 

Our point is, of course, not about the {\em word} `logical consequence’. We are happy to give that to Torretti and the logic books; and to say instead `entailment', or `implication'. Our point is that this relation, entailment, does not involve {\em dis-interpreting} any words. In particular, it does not try to capture, or theorise about, the compulsoriness of the inference from $S'$ to $S$---what Wittgensteinians used to call the  ``force of the logical must''---by dividing the language’s vocabulary into two sets: non-logical vocabulary that is then to be interpreted (given semantic values) in an arbitrary way, irrespective of what the words in fact mean, and logical vocabulary whose interpretation is fixed.

Obviously, our point is not original. Countless discussions in philosophical logic, for example about the differences between natural and artificial languages, emphasise that when one is faced with the countless entailments (equivalently: valid arguments) in a natural language, one naturally tries to find patterns, every instance of which is an entailment (a valid argument). And one soon finds many such patterns that are:\\
\indent \indent (i) a matter of where in the entailment (the argument) a small number of  words---such as `and', `not', `all', `some'---occur, and yet:\\
\indent \indent (ii) wholly independent of what other words (`red', `tall',…) occur. \\
Thus is formal logic born. 

Besides, it is natural to make some sort of toy-model semantics: toy-models of how the world could have been different, to represent the phrase above `or were true’. And once we focus on the patterns just mentioned, it is natural to represent the entailment’s (the valid argument’s)  independence of the other words in (ii), by:\\
\indent \indent (a) letting the toy-models assign to the words in (ii) arbitrary semantic values; and then \\
\indent \indent (b) checking that whatever assignment a toy-model makes, $S$ comes out true in the toy-model provided that $S'$ does. \\
Thus are formal semantics and model theory born, with their textbook definition of logical consequence. 

The point now is: these toy-models, these arbitrary assignments, can only be expected to be appropriate for those entailments (valid arguments) that {\em are} instances of such patterns: i.e. for those that turn solely on the placing of the few words selected in (i). But countless entailments (valid arguments) are {\em not} such instances. For there are countless `meaning connections’ (`analytic connections’) between our words which cannot be plausibly claimed to be due to a discernible `logical form’, lurking just below `surface grammar’. Agreed, it is plausible that `all bachelors are unmarried’ is necessary, simply because `bachelor’ means `unmarried male’, so that it instantiates the textbook’s logically valid formula $(\forall x)((Ux \wedge Mx) \supset Ux)$. But such cases are exceptional. How, for instance, should we `analyse away’ the connection between the meanings of a specific colour predicate, and of `…is coloured’. Thus consider: $S’$ := `this pencil is blue’; $S$ := `this pencil is coloured’. $S’$ entails $S$. But predicate logic will render these sentences along the lines, with `this pencil’ as an individual constant $p$: $B(p)$ and $C(p)$; and the latter is by no means a logical consequence, in Torretti's sense, of the former:  $B(p) \nvDash C(p)$.  (We take this example from Coffa (1983), whom we discuss shortly.)

As indeed countless discussions say: this point has a long and varied history. Various authors have espoused various different conceptions of logical form, or `deep structure’, into which they hoped to analyse or `regiment’  natural language, so that all inferences are rendered formal---by the lights of their preferred notion of logical form. But there is no consensus about logical form, nor even agreement about the desiderata for it.\footnote{Indeed, the alleged contrast between logical form and `surface grammar’ is itself very questionable: cf. Oliver (1999).} We do not need details of this variety, let alone an assessment of the various authors’ proposals. All we need is to emphasise the gap---wide and poorly mapped---between entailment in natural language and these formal calculi. Let us briefly do so with two historically influential examples: Wittgenstein’s troubles with elementary propositions in 6.3751 of the {\em Tractatus}, and Davidson’s analysis of adverbs, and advocacy of events as genuine objects (particulars).

In short: Wittgenstein hoped to completely analyse all necessity as the unproblematic combinatorial necessity of truth-tabular tautologies, and so maintained that the elementary propositions in which analysis would end must be logically independent of each other (i.e. all truth combinations are genuinely possible). So he could never give an example of such a proposition, but only told us that, thanks to the mutual exclusion of `red and `green’, propositions about colour in a patch of space or one’s visual field were not elementary.\footnote{So this causes trouble for combining the doctrines of the {\em Tractatus} with the sort of phenomenalism Carnap espouses in the {\em Aufbau}. It is also, of course, an Achilles heel of the {\em Tractatus}: it prompted Wittgenstein to return to philosophy, and write his `Remarks on logical form’ in 1929; cf. e.g. Potter (2020, pp. 337-338, 409-410).}  

And again, in short: Davidson noticed (1967) that adverbs and adverbial phrases modifying descriptions of action give entailments (valid arguments), such as in the sequence: `John quickly buttered the toast with the knife’; so: `John quickly buttered the toast’; so: `John quickly buttered’. With ordinary objects, like John, the toast and the knife, as the referents of the singular terms, these entailments far outstrip the power of predicate logic to make them valid by a formal rule. For with these referents, predicate logic will render these sentences along the lines: `ButterQuick(John,the toast,the knife)'; and `ButterQuick(John,the toast)'; and `ButterQuick(John)’. But despite the typography, these predicates are distinct, indeed of different polyadicities; and predicate logic recognizes no compulsory `meaning connections’ between them.\footnote{Of course, Davidson, like Quine, advocates predicate logic as the preferred language for logical forms or `regimentations’. So he sees this situation as an argument for an ontology of events, in this example a human action. For then one has regimentations like, for our first two sentences: $(\exists x)((Butter(x) \wedge By(x,John) \wedge Quick(x)  \wedge Of(x,toast) \wedge With(x,knife))$, and $(\exists x)((Butter(x) \wedge By(x,John) \wedge Quick(x)  \wedge Of(x,toast))$. So our first inference becomes formally valid in predicate logic, i.e. an instance of dropping a conjunct in the scope of an existential quantifier.} 

These two examples are enough---more than enough for this paper!---to show that there is much more to entailment in natural language than one sees in formal logic; and that it is very hard to get a satisfactory theory of that `much more’.  And in that endeavour, we tend to endorse  traditional ideas of analyticity and synonymy: we side with Carnap, against Quine (cf. Stein 1992, 275-278, 282-285). So here ends our excursus in to philosophical logic: we can now return to Torretti.

As we have said, Torretti sides with Hilbert against Frege. He does not mince words: `To demand like Frege  that [...] shows a lack of understanding of logical consequence that is indeed astonishing in the founder of modern logic. ... Frege's obtuseness is truly baffling' (1978, p. 251: endorsed at his 1999, p. 408f). But as we said in this Section's preamble; our view is that there was more right on Frege’s side than this allows. 

And we have been pleased to find that long ago, Coffa gave a similar {\em apologia}  in his (otherwise laudatory) review (1983) of Torretti (1978). We shall quote Coffa liberally, since he makes the points clearly and wittily. And more important, his discussion lays the ground for our return to geometry in Section \ref{beltr}.  (Cf. also the discussion in Coffa’s posthumous book (1991: 48-57, 381-381).)

Coffa considers the work of Beltrami and Klein (principally in 1868 and 1871, respectively) on the independence of Euclid's fifth, parallels postulate, from the rest of Euclid's axioms: i.e. on providing a model of non-Euclidean geometry---principally what is now called the `Beltrami-Klein' model (Torretti 1978, pp. 125-137). Coffa starts with an incontrovertible point of philosophical logic about such theorems. Then he uses this point to make his historico-critical remark---favouring Frege against Hilbert, and Carnap against Quine: thereby lodging his, and our, criticism---courteous and minor!---of Torretti. 

The point of logic starts from the fact that an independence theorem, saying that an axiom A is independent of the other axioms, X say, must countenance non-isomorphic models: one (or some) making true both A and all of X, and another (or some others) making true not-A and X. So in accepting such a theorem, we cannot take the meaning of the terms in the sentences of X to be so constraining, so logically rich, that an interpretation making all the sentences of X true---a model of X---must be unique  (up to isomorphism, of course). Accepting the theorem as showing that A is not provable from X requires us to be liberal, undemanding,  about meanings. We must admit: `the models making true not-A and X are legitimate: in particular, no worse as models of X than the models making true A and X'. 

Of course, this point underlies the rise, in the years 1870 to 1900, of the Hilbertian `structural' view of axioms. And as we said at the start of Section \ref{implicit}: thereafter this view became well-nigh universal. For example, consider this quotation from Weyl in 1934: `a science can never determine its subject-matter except up to an isomorphic representation' (1934, p. 95-96). (Of course, nowadays quotations like this tend to remind philosophers of Putnam's model-theoretic argument.)  

But, says Coffa, it is also legitmate to {\em reject} such an independence theorem. An objector---call him `Gottlob'!---can object that to treat both sorts of model on a par, as equally legitimate, is to be {\em Faithless} to the intended meanings of one or more terms. Which terms, and in what way faithless, will of course vary from case to case: but Gottlob's strongest objections of Faithlessness  will probably be directed at the interpretations given to some terms in the axiom A, by models making true not-A and X.
   
Coffa's point here is not to urge that the objector is {\em right}; but only  to urge that he is not {\em silly}. He agrees that it is nowadays well-nigh universal practice to understand the enterprise of interpreting an axiomatic theory in the sense `famously advanced by Hilbert in his {\em Grundlagen} (in effect, the doctrine that axiomatics ignores the meanings of all non-logical signs)' (1983, p. 687). But he goes on: 
\begin{quote}
...[we should not ...]  conflate axiomatics with formalism; the former aims to identify within a class of claims C a manageable subclass that contains all of the information conveyed by C, the latter invites us to ignore the meanings of some words and to infer only in conformity with the remaining content. 

The point of this distinction may be clarified by recalling a doctrine of the celebrated Prof. Schmall, who argues that someone who accepts the claim \\
\hspace{1 truecm} (*) this pencil is blue \\
is still free to choose whether to assert or deny the claim \\
(**) this pencil is colored,\\
for, he says, (**) is no part (does not follow from, is not a consequence of) what (*) says. His reasoning is this: first we substitute `long' for `blue' in (*) and `short' for `colored' in (**); then we notice that the pencil in question is, in fact, long. QED.

Prof. Schmall's routine response to the routinely astonished faces of his listeners is that his reasoning in no way differs from Klein's widely acclaimed proof of the independence of Euclid's parallel postulate, so that his own claim stands or falls with Klein's.

On the face of it, Prof. Schmall seems to have a point. The centuries-old question concerning the parallel postulate was not, one may safely assume, whether once we ignore in Euclid's axioms the meanings of `point, `line' and all other geometric words, the logical skeleton of information surviving that semantic slaughter still contains one redundant claim---the surviving content of the parallel postulate. The question was, of course, whether what the parallel postulate says(about points, lines, etc.) is part of the information conveyed by the remaining axioms.

The standard move at this point is to draw a distinction between formal and material consequence. What Klein and Schmall deal with, we are told, is {\em formal} consequence; the other sort is not their concern. Perhaps the reason why Schmall's audience gets angry is because {\em they} are thinking of material consequence. End of solution.

The formal-material distinction is what one might call an anaesthetic
distinction. There is nothing technically wrong with it, but its normal tendency is to desensitize its users to philosophical problems. Material consequence is a fancy name for plain old consequence; B is a material consequence of A iff what B says is part of what A says. Formal consequence is what Klein and Schmall (should) have in mind. Far from solving the problem, the distinction only allows us to pose it in new terms: how come that a formal answer is being offered to what is clearly a material question?

In Schmall's case the answer is easy: he is silly. Was Klein silly? Was the entire geometric community silly? Frege, of course, thought so: Profs. Schmall and Hilbert were, in his opinion, guilty of precisely the same blunder. And one is bound to miss the force of his point if one jumps into Hilbert's arms as quickly as, say, Quine and Torretti. (p. 687-688).
\end{quote}
Thus Coffa argues that the mathematical community's well-nigh universal acceptance of Hilbert's viewpoint was a {\em decision}, not a rational necessity: albeit one made with good reasons. He goes on:
\begin{quote}
As I see it, the formalism that Hilbert eventually forced into the field of logic was not a discovery of the essence of axiomatics but an invention, a clever expedient inspired by a number of circumstances, prominent among them, the developments in 19th century geometry, which were hard to accommodate except within a formalist framework. In the course of that development it became clear that nothing was clear about geometric primitives except what the appropriate axioms implicitly stated about them, so that any extra-axiomatic considerations involving the meanings of these terms came to be seen as not only irrelevant but positively obtrusive. That Klein's proof was recognized as a proof of independence by the geometric community ... is no evidence of the fact that geometers had seen the essence of axiomatics but, rather, of the fact that they had tacitly reached a decision that Hilbert  
would make explicit three decades later: that as far as geometry is concerned, the meanings of geometric terms really does not matter, except to the extent that it is determined by the logical words in the axioms involved. The overall story of the episode leading to this decision, and its philosophical echoes, is told by Torretti much better than by anyone else I know; but I wish he had not put formalism at the beginning of his account, as the truth underlying the whole process, but at the end, as the result of a decision called for by the contingent course of geometric history. (p. 688-689)
\end{quote}
To conclude: although we admit to being {\em ingenus} about the history of geometry, we concur with Coffa's distinction between axiomatics and formalism, and  his suggestion that Frege's position was legitimate.

\subsection{Beltrami's model as an example of reduction---and an analogy}\label{beltr}
We turn to another link between Torretti's work and our views in this paper. Namely, between Beltrami's  Euclidean model for hyperbolic geometry (in 1868) as described by Torretti (1978), and our account of reduction, and the possible objection against reductions that we labelled {\em Faithlessness}. 

So this will be a specific historical example of the issues about definition, and the axiomatic method,  raised in Section \ref{proFrege}.  Indeed, we want to develop a striking analogy mentioned by Coffa in his  review of Torretti (1978); cf. also Coffa (1991, pp. 48-57). It is an analogy between:\\
\indent \indent  (i) Beltrami's conception of his project  in 1868; and \\
\indent \indent (ii) Frege's and Russell's logicism, i.e. their effort to reduce arithmetic to logic (in modern terms: to set-theory); which we discussed in Section \ref{examp}. \\
It is this analogy that will  give us the  vivid illustration of the {\em Faithlessness} objection.\footnote{\label{2ndanalogy}{Cf. Sections \ref{221} and \ref{Faith}. But our rationale for expounding this example is not just to illustrate this paper's themes. It also sets up another analogy, that we explore in our second paper: between Beltrami's construal  of hyperbolic geometry, and how a modern relativist who advocates a conformally flat spacetime would construe that paper's second example of spacetime functionalism: viz. Robb's axiom system for causal connectability. There is also another analogy hereabouts, which is both recent and mathematically deep. Beltrami’s embedding of 2-dimensional hyperbolic geometry in Euclidean space is a `baby version’ of the famous Nash embedding theorem, that any compact Riemannian manifold can be isometrically embedded in Euclidean space; cf. e.g. Tao (2016).}}  

Thus Coffa writes (1983, p. 684):
\begin{quote}
``An Essay of Interpretation of non-Euclidean Geometry''' [i.e. (Beltrami 1868)] contains what is now regarded as the first effort to produce a model of hyperbolic geometry. Although it is sometimes said that Beltrami's discovery was a death-blow to Kantianism, it emerges from Torretti's account that there is nothing in the``Essay'' that could have caused anything but glee to a Kantian.

To begin with, Beltrami believes that the only way to legitimize a geometric doctrine is to show that it can somehow be reduced to Euclidean elements. Indeed, {\em reduction} rather than {\em interpretation} is the decisive notion in Beltrami's work; for what he does to hyperbolic geometry is precisely what Frege and Russell aimed to do to arithmetic a few years later: to reduce a certain obscure and dubious doctrine to another one, which we understand and find well grounded.

The main conclusion of Beltrami's``Essay'' was that 2-dimensional hyperbolic geometry is no more than a fragment of Euclidean geometry in disguise; for it is, in fact, the geometry of a perfectly Euclidean surface of constant negative curvature named by Beltrami the `pseudosphere'. Moreover, since 3-dimensional hyperbolic geometry is, as far as Beltrami can tell, not reducible to Euclidean geometry, he concluded that no``real substratum'' underlies it. Therefore, even though, as Kant had implied, 3-dimensional hyperbolic geometry can be developed analytically (i.e., purely conceptually) without inconsistency, no real-intuitional geometric substratum for it could be offered. Beltrami's tacit premise, one may safely assume, is the Kantian doctrine that our intuitions necessarily conform to Euclidean laws.
\end{quote}

As we emphasised: we are not historians. But we concur  with Coffa's description of the case: which, as he says, matches Torretti's. And we of course endorse the end of Coffa's second paragraph, which mirrors our Section \ref{problegi}'s discussion of reducing the problematic to the unproblematic. So we will take it that Beltrami sees himself as showing a reduction of hyperbolic geometry to Euclidean geometry,  by giving as  a model of two-dimensional hyperbolic geometry  a `perfectly Euclidean surface of constant negative curvature': which he calls a `pseudosphere'. And here, `reduction to Euclidean geometry' connotes `legitimation from the viewpoint of Euclidean geometry'.

So according to this analogy: \\
\indent \indent (a) the advocate of hyperbolic geometry is like an advocate of arithmetic; and \\
\indent \indent  (b) the advocate of the reduction to Euclidean geometry----according to Torretti and Coffa: Beltrami himself, the displayer of the Euclidean model---is like the set-theorist reducer of arithmetic.

Let us compare the two cases, using our earlier discussion of reduction, especially the objection of Faithlessness (cf. Sections \ref{221} and \ref{Faith}). For convenience, we will call these advocates `H' and `E', respectively. So beware: Beltrami is E, not H. Rather, H is some gung-ho advocate of hyperbolic geometry as being well able to ``stand on its own two feet''. So H is a person who becomes historically possible only after 1868. Indeed: at a pinch, we can take `H' to stand for Helmholtz!\footnote{We take up Helmholtz in our second paper. Agreed, Gauss in his unpublished work envisaged that a non-Euclidean geometry could be the real geometry of space: so, cheekily, `H' could also stand for Gauss.} And E is not {\em any} Euclidean geometer, or philosopher of geometry (e.g. Kant): E is precisely the advocate of Beltrami's reduction (to Euclidean geometry).

The first point to make is the obvious one: that H and E are in danger of misunderstandings, i.e. of speaking at cross-purposes. For they are liable to mean different things by the same words such as `geodesic' or `straight'. For a hyperbolic straight line, i.e. geodesic in Beltrami's model (on the pseudosphere), is of course {\em not} a geodesic of the embedding Euclidean geometry. And a hyperbolic triangle is not a Euclidean triangle since its angles do not sum to 2$\pi$. And so on. 

So let us, more specifically, compare the cases in terms of {\em Faithlessness}. Thus we can envisage:
H will accuse E of Faithlessness to the meanings of his, H's, words:
\begin{quote} 
I, H, do not mean by `geodesic' a curve in Euclidean space that is not straight according to Euclidean geometry!
\end{quote}
And  H might well go on to accuse E of what Section \ref{221} called `over-shooting': just as Benacerraf  accused a set-theoretic reduction of arithmetic of over-shooting---yielding claims in the reduced theory (here: hyperbolic geometry) that are alien to it.\footnote{This is not to say the objection is right: recall footnotes \ref{StrucPotter} and \ref{threehistory}. Note that H might also accuse E of what Section \ref{222} called `Plenitude'. Thus the variety of ways to identify numbers with sets, which Benacerraf emphasised---and we called `Plenitude'---corresponds to e.g. different placements of the Beltrami pseudosphere within $\mathR^3$. So H might accuse E of having many equally good, and therefore equally bad, placements of the hyperbolic space, within $\mathR^3$. (We write $\mathR^3$, but the affine Euclidean space would be more accurate: no matter---nothing turns on this.)}

There is also a notable contrast between the two analogues: about the chronological order, and one might say `order of understanding', of the reduced and reducing theories. And this contrast will prompt another objection of Faithlessness---in the opposite direction,  from E to H.

Thus: In the reduction of arithmetic to set theory, arithmetic was used successfully for millennia before set theory was even formulated. (And that successful use surely means it was in some good sense `understood': albeit not analysed, and maybe also, problematic in ways that rightly prompted the logicists' enterprise of reduction.) In short: the reduced theory, arithmetic, `was there first’. 

But in the reduction of hyperbolic geometry to Euclidean geometry, it is the {\em reducing} theory that was used successfully for millennia (and again: surely in some good sense `understood', albeit not analysed, and maybe also, problematic), before the reduced theory, hyperbolic geometry, was even formulated. 

And this of course makes an objection of Faithlessness in the opposite direction, i.e. from E to H, also tenable. For the long history, the entrenchment, of Euclidean geometry makes E’s use of words such as ‘geodesic’  or `straight' much more natural for us. This is of course reminiscent of Frege's complaint against Hilbert: cf. the discussion in Section \ref{proFrege}.  Thus E might say:  
\begin{quote} 
H's use of ‘geodesic’ etc is faithless to my/Euclidean/the proper meaning. Agreed: Beltrami shows us how to define H's words ‘geodesic’ etc.---which for clarity we should really write as `geodesic(hyp)’ etc.---viz. as what I call an appropriately {\em curved} line in $\mathR^3$, in such a way that all H's claims are truths of  my/Euclidean/the proper geometry. In short, hyperbolic geometry is a part of Euclidean geometry. Namely: a small part---about a surface of constant negative curvature. But sadly, H presents his theory  with misleading words. For example, when H writes ‘geodesic’, he  really  should write e.g. `geodesic(hyp)’.
\end{quote}

\section{Conclusion }\label{concl}
We will not take the space to summarize {\em post facto} the main ideas---like simultaneous unique definition,  functionalism as reduction and the Canberra Plan---that this paper has argued for. Those ideas are clear enough from Section \ref{111}’s announcement of them. Instead, let us consider them in relation to spacetime, and so look ahead to our other papers.
 
Our complaint has been that these ideas---though standard undergraduate fare even forty years ago---are neglected in the recent literature on spacetime functionalism. Agreed, that might signal merely different concerns from the bulk of philosophers now writing about functionalism (and related topics like levels of description); and so it might merely signal a different use of the term, `functionalism'. Though the homonymy might be regrettably misleading, no real harm would be done. 

But we think some harm is done. For there are examples of spacetime functionalism, {\em stricto sensu}, to be had. Besides, these  examples are ``hiding in plain sight'', in either the philosophical literature (especially about relationism) or in the physics literature about relational or Machian traditions of dynamics: these literatures including some precise---indeed, impressive---results.  

Thus we think that someone, aware of the functionalist tradition we have reviewed, who sought for whatever reason to find examples of `spacetime functionalism', would in short order think of these examples. And since the examples are impressive, the recent literature's detaching the phrase `spacetime functionalism' from the tradition's concern with simultaneous unique definition and reduction, is regrettable. Hence our aim, in our other papers, to celebrate these examples of spacetime functionalism. \\ \\

{\em Acknowledgements}:---  We are grateful to: audiences at talks in Cambridge UK, Harvard (Black Hole Initiative), MIT, Munich, New York (MAPS), Oxford, and the `Quantum information structure of spacetime' Network. For conversations and comments on previous versions, we are very grateful: to David Chalmers, Grace Field, Sam Fletcher, Eleanor Knox, Dennis Lehmkuhl, James Read, Alex Roberts and Bobby Vos; especially to Erik Curiel, Sebastian De Haro, Josh Hunt, Alex Oliver and Bryan Roberts; and above all, to Adam Caulton for---as ever---such insight and generosity. We are also very grateful to Cristian Soto, not least for his patience.  \\ \\

\section{References}

\indent Austin, J.  (1962),  {\em Sense and Sensibilia}, ed. G.J. Warnock; Oxford University Press.  

Balzer, W., Moulines, C. and Sneed, J. (1987), {\em An Architectonic for Science}, Dordrecht: Reidel.

Beaney, M. (2004), Carnap’s Conception of Explication: From Frege to Husserl, in  S. Awodey and C. Klein (eds.), {\em Carnap Brought Home; the View from Jena}, Chicago: Open Court : p.117-150.

Benacerraf, P. (1965), `What numbers could not be', {\em The Philosophical Review} {\bf 74}, 47-73. 

Benacerraf, P. (1973),  `Mathematical Truth', {\em Journal of Philosophy} {\bf 70}, 661-679.

Blanchette, P. (2018), `The Frege-Hilbert controversy',  {\em Stanford Encyclopedia of Philosophy}, https://plato.stanford.edu/entries/frege-hilbert/

 Boolos, G. and Jeffery, R.  (1980),  {\em Computability and Logic}, second edition: Cambridge University Press.
 
 Braddon-Mitchell, D. and Nola, R. (2009),  {\em Conceptual Analysis and Philosophical Naturalism}, MIT Press: Bradford Books.

Braithwaite, R. (1953) {\em Scientific Explanation}, Cambridge University Press. 

Brown, H. (2006), {\em Physical Relativity}, Oxford University Press.

Butterfield, J. (2011), `Emergence, Reduction and Supervenience: A Varied Landscape',  {\em Foundations of Physics} {\bf 41}, 920–959. 

Butterfield, J. (2011a), `Less is Different: Emergence and Reduction Reconciled', {\em Foundations of Physics} {\bf 41}, 1065–1135

Butterfield, J. (2014),  `Reduction, emergence and renormalization',  {\em Journal of Philosophy} {\bf 111}, 5-49.

Butterfield, J. (2018), `On Dualities and Equivalences Between Physical Theories', http://philsci-archive.pitt.edu/14736/. Forthcoming (abridged) in {\em Philosophy Beyond Spacetime} (Oxford University Press), edited by N. Huggett, B. Le Bihan and C. W\"{u}thrich.
  
Button, T. (2013), {\em The Limits of Realism}, Oxford University Press. 

 Button, T. and Walsh S. (2018), {\em Philosophy and Model Theory},  Oxford University Press.  

Carnap, R. (1936), `Testability and meaning', {\em Philosophy of Science} {\bf 3}: 419-471.

Carnap, R. (1963), `Intellectual Autobiography', in P. A. Schilpp (ed.) {\em The Philosophy of Rudolf Carnap: The Library of Living Philosophers volume XI}, Open Court.
%quoted in Friedlman 2007 , `The {\em Aufbau} and the rejection of metaphysics', in Friedman, M. and Creath R. (eds.) {\em The Cambridge Companion to Carnap}\\

Coffa, J. A. (1983),  Review of Torretti (1978), {\em Nous} {\bf 17}, pp. 683-689.

Coffa, J. (1991), {\em The Semantic Tradition from Kant to Carnap: to the Vienna Station}, (ed. L. Wessels), Cambridge University Press.   

Coffey, K. (2014), `Theoretical equivalence as interpretative equivalence', {\em British Journal of Philosophy of Science} {\bf 65}, 821–844.

Davidson, D. (1967), `The logical form of action sentences', in N. Rescher (ed.), {\em The Logic of Decision and Action}. University of Pittsburgh Press.

De Haro, S. (2020), `Theoretical equivalence and duality', {\em Synthese}; https://doi.org/10.1007/s11229-019-02394-4

Dewar, N. (2019), `Supervenience, reduction and translation', {\em Philosophy of Science} {\bf 86}, 942-954.

Dizadji-Bahmani, F., Frigg R. and Hartmann S., (2010), `Who's afraid of Nagelian reduction?' {\em Erkenntnis}, {\bf 73}, 393-412.

Dummett, M. (1991) {\em Frege: Philosophy of Mathematics}, Duckworth.

Field, H. (1980), {\em Science Without Numbers}, Blackwell.

Fletcher, S. (2019), `Counterfactual reasoning within physical theories’, {\em Synthese}; published online at https://doi.org/10.1007/s11229-019-02085-0

Frege, G. (1884), {\em The Foundations of Arithmetic}, translated and edited by J. Austin 1974, Blackwell. 

Gray, J. (2008) {\em Plato's Ghost}, Princeton University Press.

Halvorson, H. (2019), {\em Logic in the Philosophy of Science}, Cambridge: University Press.  

Harman, G. (1977), {\em The Nature of Morality: an introduction to ethics}, Oxford University Press.

Hellman, G., Thompson, F. (1975),  `Physicalism: ontology, determination and reduction', .{\em Journal of Philosophy}  {\bf 72}, 551– 564.

Hempel, C. (1965), {\em Aspects of Scientific Explanation}. Free Press, New York. 

Hempel, C. (1966), {\em Philosophy of Natural Science}. Prentice-Hall, New York.

  Hodges, W. (1997), {\em A Shorter Model Theory}, Cambridge University Press. 
  
Hudetz, L. (2019), The semantic view of theories and higher-order languages, {\em Synthese} {\bf 196}: 1131–1149.
 
Hudetz, L. (2019a), Definable categorical equivalence, {\em Philosophy of Science} {\bf 86}: 47-75.

Hurley, S. (1989), {\em Natural Reasons: Personality and Polity}, Oxford University Press.

Jackson, F. (1998), {\em From Metaphysics to Ethics: A Defence of Conceptual Analysis}, Oxford University Press.

Janssen-Lauret, F. and  MacBride, F. (2020), `Lewis’s Global Descriptivism and Reference Magnetism’, {\em Australasian Journal of Philosophy} {\bf 98}, 192-198, DOI: 10.1080/00048402.2019.1619792

Kennedy, H. (1972), `The Origins of Modern Axiomatics: Pasch to Peano', {\em The American Mathematical Monthly} {\bf 79}, 133-136.

Kroon, F. (1987), `Causal descriptivism', {\em Australasian Journal of Philosophy}, {\bf 65}, pp. 1-17. 

Lewis, D.~(1966), `An Argument for the Identity Theory', \emph{Journal of Philosophy} \textbf{63}, pp.~17-25.  

Lewis, D. (1969), Review of `Art, Mind, and Religion', {\em Journal of Philosophy} {\bf 66}: 23-25.

 Lewis, D.~(1970), `How to Define Theoretical Terms', \emph{Journal of Philosophy} \textbf{67}, pp.~427-446.  
 
  Lewis, D.~(1972), `Psychophysical and theoretical identifications', \emph{Australasian Journal of Philosophy}, \textbf{50}: 3, pp.~249-258.  
   
Lewis, D. (1983) `New Work for a Theory of Universals', {\em Australasian Journal of Philosophy}, {\bf 61}, pp. 343-77. 

Lewis D. (1984) `Putnam's paradox', {\em Australasian Journal of Philosophy}, {\bf 62}, pp. 221-236.
  
Lewis, D. (1989), 'Dispositional Theories of Value', {\em Proceedings of the Aristotelian Society,} Supplementary  Volume. {\bf LXIll}:  113-137.

Lewis, D. (1993) `Many, but almost one', in Keith Campbell, John Bacon, and Lloyd Reinhardt (eds.), {\em Ontology, Causality and Mind: Essays on the Philosophy of D. M. Armstrong}, Cambridge: Cambridge University Press, pp. 23–38.

Lewis, D. (1994), Reduction of mind, in S. Guttenplan (ed.) {\em A Companion to the Philosophy of Mind}, pp. 412-431, Blackwell.

Linsky, B. (2019), `Logical construction', {\em Stanford Encyclopedia of Philosophy},\\
 https://plato.stanford.edu/entries/logical-construction

Lutz, S. (2017), `What Was the Syntax-Semantics Debate in the Philosophy of Science About?' {\em Philosophy and Phenomenological Research}, {\bf 95}, 319-352. 

 Lutz, S. (2017a), `Newman's objection is dead. Long live Newman's objection!',\\
  https://philsci-archive.pitt.edu/13018/
 
 Moore, G. (1903), {\em Principia Ethica}, Cambridge University Press.
 
Nagel, E. (1961), {\em The Structure of Science: Problems in the Logic of Scientific Explanation}, Harcourt.

Nagel, E. (1979), 'Issues in the logic of reductive explanations', in his {\em Teleology Revisited and other essays in the Philosophy and History of Science}, Columbia University Press; reprinted in Bedau and Humphreys (2008); page reference to the reprint.

Niebergall, K-G. (2000), `On the logic of reducibility: axioms and examples’, {\em Erkenntnis} {bf 53}: 27–61.

Niebergall, K-G. (2000), `Structuralism, model theory and reduction’, {\em Synthese} {\bf 130}: 135–162.

Oliver, A. (1996), `The metaphysics of properties', {\em Mind} {\bf 105} pp. 1-80.

Oliver, A. (1999), `A few more remarks on logical form', {\em Proceedings of the Aristotelian Society} {\bf 99}: 247-272.

Oliver, A. and Smiley, T. (2013), `Zilch', {\em Analysis} {\bf 73}: 601-613.

Oliver, A. and Smiley, T. (2016), {\em Plural Logic}, Oxford University Press.

Potter M. (2000), {\em Reason's Nearest Kin}, Oxford University Press.

Potter, M. (2020) {\em The Rise of Analytic Philosophy, 1879-1930}, Routledge.

Psillos, S. (1999), {\em Scientific Realism: how science tracks truth}, Routledge.

Psillos, S. (2012),  `Causal descriptivism and the reference of theoretical terms', in A. Raftopoulos and P. Machamer (eds.) {\em Perception, Realism, and the Problem of Reference},  Cambridge: University Press.  

Quine, W. (1960), {\em Word and Object}, MIT Press (new edition: 2013).

Quine, W. (1964), `Implicit definition sustained',  {\em Journal of Philosophy}, \textbf{61}: 71-74.

Russell, B.(1903/2010), {\em The Principles of Mathematics}, Allen and Unwin; 2010 reprint by Routledge.

Russell, B. (1918), `The Philosophy of Logical Atomism' in {\em The Monist} {\bf 28} (Oct. 1918): 495–527, {\bf 29} (Jan., April, July 1919): 32–63, 190–222, 345–80. Page references to {\em The Philosophy of Logical Atomism}, D.F. Pears (ed.), La Salle: Open Court, 1985, 35–155.

Russell, B. (1924), `Logical Atomism', in {\em The Philosophy of Logical Atomism}, D. F. Pears (ed.), La Salle: Open Court, 1985, 157–181: also in {\em The Collected Papers of Bertrand Russell:  vol. 9, Essays on Language, Mind and Matter: 1919–1926}, J.G. Slater (ed.), pp. 160–179; 2001: London and New York.

Russell, B. (1919), {\em Introduction to Mathematical Philosophy}, New York and London.

Ryle, G. (1949), {\em The Concept of Mind}, London: Hutchinson.

Schaffner, K. (1967), `Approaches to reduction' {\em Philosophy of Science}, {\bf 34}, 137-147.

Schaffner, K. (1976). `Reductionism in biology: Prospects and problems', In R. Cohen, et al. (Eds.), {\em PSA 1974}, pp. 613–632. Dordrecht: Reidel.

Schaffner, K. (2006), `Reduction: the Cheshire cat problem and a return to roots', {\em Synthese} {\bf 151}, 377–402.

Schaffner, K. (2012), `Ernest Nagel and reduction, {\em Journal of Philosophy} {\bf 109}, 534-565.

Shapiro, S. (2000), {\em  Thinking About Mathematics} Oxford University Press.

Sklar, L. (1982), `Saving the noumena', {\em Philosophical Topics} {\bf 13}, 89-110

Smith, P. and Jones, O. (1986), {\em The Philosophy of Mind}, Cambridge University Press.

Sneed, J. (1979), {\em The Logical Structure of Mathematical Physics}, Dordrecht: Reidel, Pallas paperbacks.

Sober, E. (1999),  `The multiple realizability argument against reductionism', {\em Philosophy of Science}, {\bf 66}, 542-564.

Stein, H. (1992), Was Carnap entirely wrong, after all?, {\em Synthese} {\bf 93}, 275-295.

Tanney, J. (2015), `Gilbert Ryle', {\em Stanford Encyclopedia of Philosophy}, \\
https://plato.stanford.edu/entries/ryle/

Tao, T. (2016), `Notes on the Nash embedding theorem’: on his blog at: https://terrytao.wordpress.com/2016/05/11/notes-on-the-nash-embedding-theorem/

Taylor, B. (1993), `On Natural Properties in Metaphysics', {\em Mind} {\bf 102}, pp. 81-100.

Torretti, R. (1978), {\em Philosophy of Geometry from Riemann to Poincar{\' e}}, Reidel.

Torretti, R. (1986), `Observation', {\em British Journal for the Philosophy of Science}, {\bf 37}, 1-23.

Torretti, R. (1990), {\em Creative Understanding: philosophical reflections on physics}, University of Chicago  Press.

Torretti, R. (1999), {\em Philosophy of Physics}, Cambridge University  Press.

Torretti, R. (2008), `Objectivity; a Kantian perspective', {\em Royal Institute of Philosophy: Supplement}, {\bf 63}, 81-94.

van Fraassen, B. (1980) {\em The Scientific Image}, Oxford University Press. 

van Fraassen, B. (1991) {\em Quantum Mechanics: an empiricist view}, Oxford University Press.

van Fraassen, B. (2008) {\em Scientific Representation}, Oxford University Press.

van Riel, R. and van Gulick, R. (2019), `Scientific reduction',  {\em Stanford Encyclopedia of Philosophy}, https://plato.stanford.edu/entries/scientific-reduction/

Weatherall, J. (2018), Why not categorical equivalence?, https://arxiv.org/abs/1812.00943. 

Weatherall, J. (2018a), Theoretical equivalence in physics, {\em Philosophy Compass} {\bf 14} e12592 and e12591: and at https://arxiv.org/abs/1810.08192. 

Weyl, H.  (1934), `Mind and nature', in {\em Mind and Nature: selected writings in philosophy, mathematics and physics} (2009) ed. P. Pesic, Princeton University Press.

\end{document}